\documentclass[fleqn,usenatbib]{mnras} 

\usepackage[T1]{fontenc}

\DeclareRobustCommand{\VAN}[3]{#2}
\let\VANthebibliography\thebibliography
\def\thebibliography{\DeclareRobustCommand{\VAN}[3]{##3}\VANthebibliography}

\usepackage{color}
\usepackage{xcolor}
\usepackage{hyperref}
\usepackage{ulem}
\usepackage{graphicx}	
\usepackage{amsmath}	
\usepackage{amssymb}	

\usepackage{subcaption}

\newcommand{\kms}{{\rm km\,s^{-1}}}

\usepackage{tikz,xcolor,hyperref}

\definecolor{lime}{HTML}{A6CE39}
\DeclareRobustCommand{\orcidicon}{
	\begin{tikzpicture}
	\draw[lime, fill=lime] (0,0) 
	circle [radius=0.16] 
	node[white] {{\fontfamily{qag}\selectfont \tiny ID}};
	\draw[white, fill=white] (-0.0625,0.095) 
	circle [radius=0.007];
	\end{tikzpicture}
	\hspace{-2mm}
}

\foreach \x in {A, ..., Z}{\expandafter\xdef\csname orcid\x\endcsname{\noexpand\href{https://orcid.org/\csname orcidauthor\x\endcsname}
			{\noexpand\orcidicon}}
}

\usepackage{newtxtext,newtxmath}

\title{Rotation curves and dynamical masses of MaNGA barred galaxies}

\author[E. O. Schmidt et al.]{
Eduardo O. Schmidt \orcidA{},$^{1,2,3}$\thanks{E-mail: eduardo.schmidt@unc.edu.ar}
Damián Mast \orcidB, $^{1,2}$
Gaia Gaspar \orcidD, $^{1,2}$
and Walter Weidmann \orcidC, $^{1,2}$
\\
$^{1}$Observatorio Astronómico de Córdoba, Universidad Nacional de Córdoba, Laprida 854, Córdoba, Argentina.\\
$^{2}$Consejo de Investigaciones Cient\'{i}ficas y T\'ecnicas de la Rep\'ublica Argentina, Buenos Aires, Argentina.\\
$^{3}$Instituto de Astronomía Teórica y Experimental (IATE), CONICET-UNC, Laprida 854, Córdoba, Argentina
}

\date{Accepted XXX. Received YYY; in original form ZZZ}

\pubyear{2022}

\begin{document}
\label{firstpage}
\pagerange{\pageref{firstpage}--\pageref{lastpage}}
\maketitle

\begin{abstract}
   
   {In this paper we analyze a sample of 46 barred galaxies of MaNGA. Our goal is to investigate the stellar kinematics of these galaxies and obtain their rotation curves. Additionally, we aim to derive the total stellar and dynamical masses, as well as the maximum rotation velocity, in order to examine their distributions and scaling relations.}

    {Using the Pipe3D dataproducts publicly available we obtained the rotation curves, which were fitted considering two components of an axisymmetric Miyamoto–Nagai gravitational potential.}
  
   {We found a wide range of the maximum rotation velocities (117-340 $\kms$), with a mean value of 200 $\kms$. In addition we found that the total stellar and dynamical masses are in the range of log(M$_{star}$/M$_{\odot}$)$=$ 10.1$-$11.5, with a mean value of log(M$_{star}$/M$_{\odot}$)$=$ 10.8, and log(M$_{dyn}$/M$_{\odot}$)$=$ 10.4$-$12.0, with a mean value of log(M$_{dyn}$/M$_{\odot}$)$=$ 11.1, respectively. We found a strong correlation between dynamical mass and maximum velocity, between maximum velocity and magnitude, and between stellar mass and maximum velocity. According to these results, barred galaxies exhibit similar behaviour to that of normal spiral galaxies with respect to these relations, as well as in terms of the distribution of their dynamical mass and maximum rotation velocity. However, we found that the distribution of stellar masses of barred galaxies is statistically different from other samples including non-barred galaxies. Finally, analyzing the galaxies that show nuclear activity, we find no difference with the rotation curves of normal galaxies.}

\end{abstract}




\section{Introduction}
\label{sec:intro}

Rotation curves (RCs) are the main tracers of the distribution of mass in spiral galaxies \citep[e.g.,][]{Ashman1992,Persic1995,Sofue2001, Kalinova2017a}. As such, they represent the major tool for determining the mass distribution, providing fundamental information for understanding the evolution, dynamics, and formation of spiral galaxies \citep[e.g.,][]{Persic1995, Sofue1999, Sofue2001, Martinez-Medina2020}.

There are typical kinematic tracers for galaxies which include stars \citep[e.g.,][]{Bekeraite2016a,Bekeraite2016b,Kalinova2017a,Kalinova2017b,Yoon2021ApJ} and ionised, atomic or molecular gas \citep[e.g.,][]{Levy2018}. In this context, astronomers have at their disposal several observing techniques and methods for determining rotation curves and velocity fields for both gas and stars \citep{Falcon-Barroso2017AA,Garcia-Lorenzo2015AA}. For example, traditional long slit spectroscopy, Fabry-Perot spectrographs, or integral field instruments are often used for these purposes \citep[see for example the review of][]{Sofue2001}. With the advent of large integral-field spectroscopic (IFS) surveys such as CALIFA \citep[Alto Legacy Integral Field Area,][]{Sanchez2012AA}, SAMI \citep[Sydney-AAO (Australian Astronomical Observatory) Multi-object IFS,][]{Croom2012MNRAS}, or MaNGA \cite[Mapping Nearby Galaxies at Apache Point Observatory,][]{Bundy2015}, the possibility of carrying out detailed studies with samples covering wide ranges of parameters has become a reality. Rotation curves can be obtained by measuring spectroscopic emission lines such as H$\alpha$, [OIII], [NII], [SII], HI, and CO lines and stellar absorption lines such as the generated by Ca, H, He, K, Mg, Na, among other chemical elements, and techniques vary from single line to full spectrum fitting \citep{Cappellari2004PASP,Cappellari2017MNRAS,Sanchez2016a}.

Each of the different tracers has its difficulties when studying and analyzing them \citep[e.g.,][]{Leung2018, Martinez-Medina2020}, in addition to preferentially tracing different components in the galaxy. For example, the gas kinematic has an intrinsic turbulent nature and therefore gas velocity fields are complex \citep[e.g.,][]{Barrera-Ballesteros2014AA}. Particularly, the CO molecule emission is considered to be a good tracer of the circular velocity of disc galaxies, however the typical presence of bars and spiral arms clearly influences the molecular gas orbits, making them turbulent \citep[e.g.,][]{Laine1999, Shetty2007}. In this context, although stellar kinematics can present higher velocity dispersion, stellar velocity fields are less distorted than gas fields \citep[e.g.,][]{Adams2012, Bekeraite2016a}. However, stellar kinematics presents its complications when it comes to deriving properties. \cite{Leung2018} studied and analyzed three dynamical models used to calculate masses from stellar dynamics, which are the Jeans method, the asymmetric drift correction method, and the Schwarzschild model method. These authors found that the three models can reproduce the CO circular velocity at 1 effective radius to within 10\%. The scatter in the inner regions can be of $\sim$20\%, because some assumptions may break down \citep[see][ for more details]{Leung2018}. These authors found that all three models can recover the dynamical mass at 1 effective radius better than 20\%. For their part, \cite{Martinsson2013} studied [OIII]$\lambda$5007 and stellar kinematics of 30 nearby spiral galaxies and found that the [OIII] and stellar rotation curves exhibit signatures of asymmetric drift with a rotation difference that is 11\% of the maximum rotation speed of the galaxy disk.

It is clear that a sufficiently complete view of the mass distribution in a galaxy can only be achieved through the use of different tracers. Moreover, understanding the scope and drawbacks, not only of the tracers but also of the limitations of the observational techniques used for each of the measurements, is a fundamental step in our understanding of the mass distribution in galaxies. Thus, having independent mass determinations for homogeneous samples of galaxies that allow us to compare analysis methods and even observational regimes (such as low spatial resolution and spectroscopic observations) is a fundamental building block in the study of galaxy dynamics and evolution in this new era of large integral field spectroscopy galaxy surveys.

In this work, we study the stellar kinematics of barred galaxies to obtain stellar rotation curves that are less affected by the turbulent motions more characteristic of gas kinematics. We intend to avoid the volatility inherent to gas curves that are generated by any perturbation. This way, we aim to analyze the stellar kinematic and derive the stellar rotation curves of a sample of MaNGA barred galaxies. 

Several papers in the literature investigate and present RCs using different methods. For example, \cite{Mathewson1992} study the RCs of 965 southern spiral galaxies through H$\alpha$ and HI emission. \cite{Makarov2001} present RCs of 135 edge-on galaxies through H$\alpha$ and \cite{Vogt2004a, Vogt2004b} study rotation curves of 329 spiral galaxies using H$\alpha$ and HI emission. In addition, \cite{Marquez2002} derive rotation curves of 111 galaxies through H$\alpha$ and \cite{Spano2008} study rotation curves of 36 galaxies through HI emission. For their part, \cite{Lang2020} present RCs of 67 galaxies (including 45 barred galaxies) through CO emission. Moreover, \cite{Martinsson2013} determined the ionized-gas ([OIII$\lambda$5007]) and stellar rotation curves of a sample of 30 nearby spiral galaxies and \cite{Hernandez2005} present the H$\alpha$ gas kinematics of 21 barred spiral galaxies. In addition, \cite{Bekeraite2016a} study the stellar rotation curves of 199 rotating galaxies and \cite{Kalinova2017b} present the stellar rotation curves of 238 galaxies. Moreover, \cite{DiTeodoro2023} present the HI rotation curves of 15 massive spiral of the local universe and \cite{Kalinova2017a} study the stellar rotation curves of 18 late type spirals. \cite{Yoon2021ApJ} analysed the shape of the RCs of the MaNGA data release. In general, the aims of those works vary from testing the cold dark matter (CDM) models of galaxy formation to analyzing scaling relations. In any case, it is important to have homogeneously observed and analyzed dataset. In this context, one of the main goals of this work is to study the stellar kinematic and obtain the stellar rotation curves of a sample of 46 barred galaxies, and then, by fitting these RCs, derive the dynamical masses of an homogeneous galaxy sample.

There are different methods for determining dynamical masses of spiral galaxies. For example, \cite{Aquino-Ortiz2020} use the stellar kinematics for galaxies from MaNGA to explore a universal fundamental plane and the relation between the dynamical and stellar masses. In this scenario, RCs consitute an independent method for determining the dynamical masses of spiral galaxies \citep[e.g.,][]{Burbidge1959,Burbidge1961,Rubin1964,Krumm1977,Pence1981,Aguero2004,Jalocha2010,Daod2019}. As we mentioned above, obtaining the total dynamical mass of the sample of barred spiral galaxies by analyzing their stellar rotation curves, represents a valuable contribution to the study of galaxy dynamics.

Even though barred galaxies constitute $\sim$50\% of all disk galaxies \citep[e.g.,][]{Marinova2007, Sheth2008, Aguerri2009}, they have been observed and studied less than non-barred galaxies in part due to their kinematic complexity. However, although the kinematic study of barred galaxies can be more difficult than for non-barred galaxies because their gas tracers are very in-homogeneously distributed on small scales \citep[e.g.,][]{Bosma1996}, the large-scale properties of the rotation curves of  barred  galaxies  are  generally  similar  to those of non-barred galaxies \citep[e.g.,][]{Sofue2001}.

This paper is organized as follows. Sect. \ref{sec: sample} presents the sample of barred spiral galaxies and the data we worked with. Sect. \ref{sec: procedure} describes the procedure by which the velocity fields and radial velocity curves were obtained. In Sect. \ref{sec: results} we present our results, i.e., the process of obtaining and fitting the rotation curves, the derivation of the dynamical and stellar masses of the galaxies, and the study of the scalling relations. In Sect. \ref{sec:comparison} we make a comparison between our results and those of the literature. Finally, in Sect. \ref{sec:final_remarks} we provide a brief summary of the work and make final remarks.

\section{Sample}
\label{sec: sample}

For our study, we selected the galaxy sample analysed by \cite{Guo2019}, which consists of barred galaxies from MaNGA. This survey investigates the internal structure and composition of gas and stars in a sample of $\sim$ 10000 galaxies of the nearby universe \citep[for more details of the project see][]{Bundy2015}. We used the integral field unit (IFU) spectroscopic data taken at the 2.5 m Sloan Telescope \citep{Gunn2006}, which has 1423 fibres with 2\arcsec\, core diameters over a 3º diameter field of view.

The sample studied by \cite{Guo2019} is made up of 53 plate-IFU observations. Note that plates IFU 8256$-$6101 and 8274$-$6101 correspond to the galaxy SDSS J105456.35+412954.3 and plates 8588-3701 and 8603-12701 correspond to SDSS J163233.73+390751.7. In addition, we excluded from the analysis the galaxies SDSS J154702.69+542027.1, SDSS J162200.73$+$493114.8, SDSS J164029.54$+$391406.3 and SDSS J162907.83$+$403954.1 (plate-IFU: 8481-12701, 8482-12703, 8601-12705, and 8603-12703, respectively) which have a noisy velocity map. Moreover, as we will see below (Sect. \ref{sec: results}) one galaxy (SDSS J090641.14$+$412154.3, plate-IFU 8247$-$3701) was rejected from the analysis because the dispersion of velocities dominates its entire curve compared with the rotation velocity. This way, our final sample consists of 46 barred galaxies. 
Table \ref{table:sample} lists the main parameters of the galaxy sample, such as the MaNGA ID, the galaxy name, right ascension (J2000), declination (J2000), morphological type, absolute r-band magnitude from SDSS-DR9 \citep[][]{Ahn2012}, effective radius in the r-band from SDSS-DR9, redshift, galaxy inclination, galaxy kinematic PA and systemic velocity. All data were taken from \cite{Guo2019}, except the galaxy name (column 2), which has been taken from the Sloan Digital Sky Survey (SDSS) IV DATA Release 13 \citep[][]{Albareti2017} and the systemic velocity (column 11), which has been obtained in this work (Sect. \ref{sec: procedure}). The galaxy inclination was measured from the ellipticity radial profile using ellipse fitting of r-band SDSS image, while the galaxy kinematic PA was measured from the velocity map using fit\_kinematic\_pa.py\footnote{https://www-astro.physics.ox.ac.uk/~cappellari/software/} code \citep[see][for more details]{Guo2019}. In addition, we have made use of the Nasa Extragalactic Database (NED)\footnote{The NASA/IPAC Extragalactic Database (NED) is operated by the Jet Propulsion Laboratory, California Institute of Technology, under contract with the National Aeronautics and Space Administration.} to verify galaxy parameters such us the radial velocity. 

\begin{figure*}
   \includegraphics[width=0.45\textwidth]{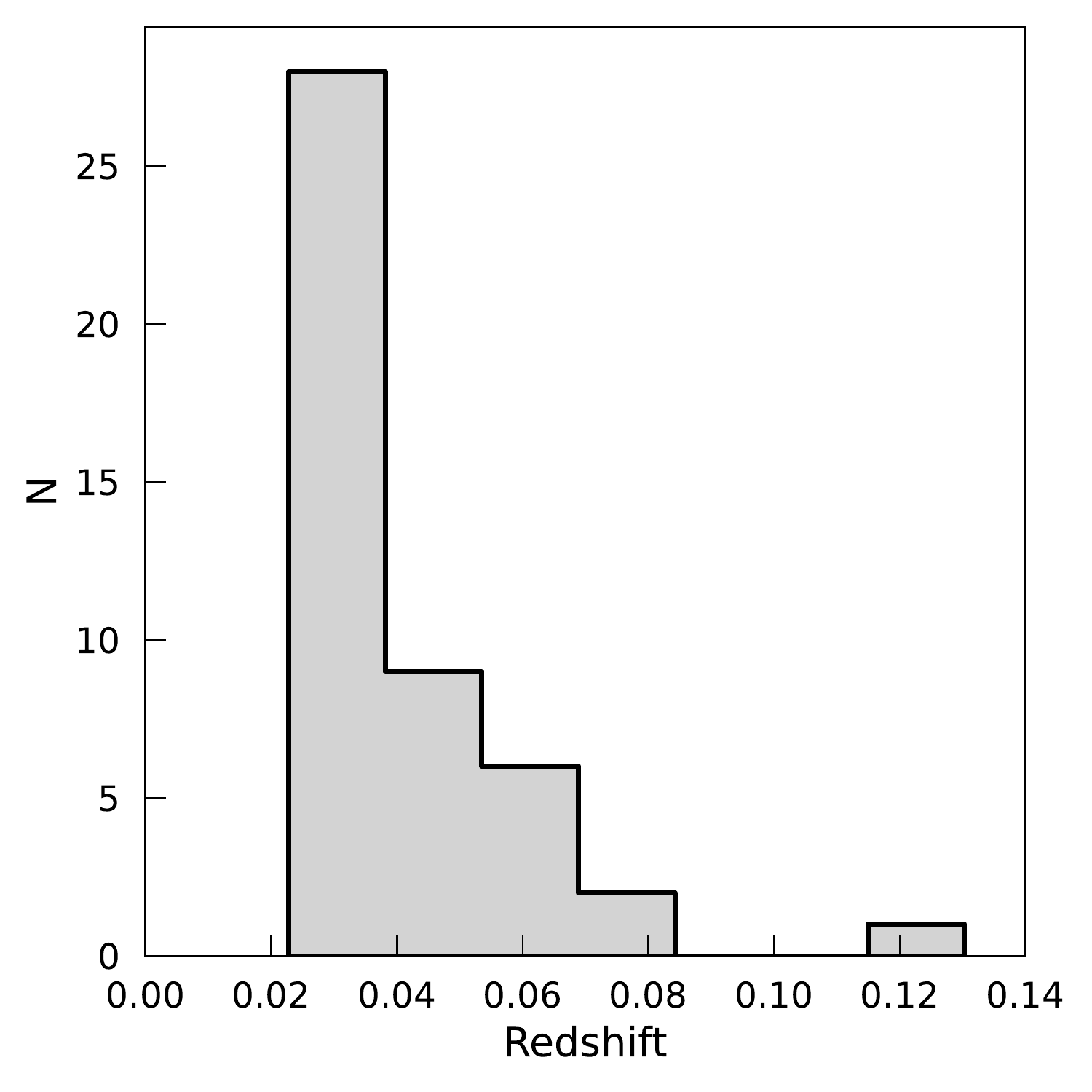}
   \includegraphics[width=0.45\textwidth]{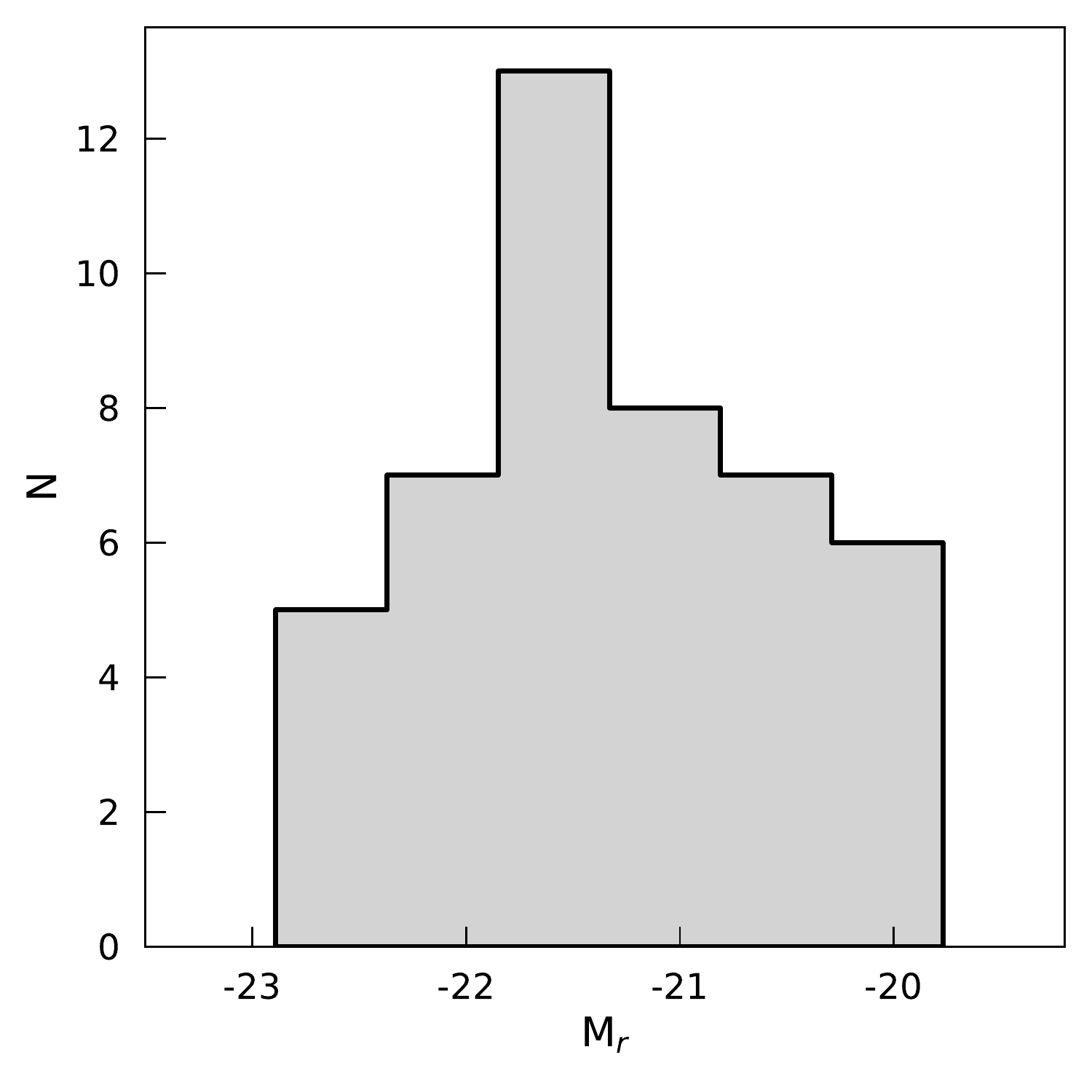}
    \caption{Redshift and r-band magnitude distributions for the galaxies of our sample are shown in the left and right panels, respectively. The data of both plots were taken from \citet{Guo2019}.}
     \label{fig:sample}
\end{figure*}

In order to make the characterisation of the sample more comprehensive, in Fig. \ref{fig:sample} we present the redshift and r-band magnitude distributions for our sample of galaxies (left and right panels, respectively). The redshift of the entire sample ranges from 0.0228 to 0.1303 and has a distribution with a mean value of z $=$ 0.0399, a standard deviation of 0.019 and an interquartile range (IQR) of 0.021. There is one object with $z \sim$ 0.13, while the remaining galaxies (98 \% of the sample) have redshifts in the range 0.0228 $-$ 0.0753. By removing this object from the analysis, the redshift distribution would have a standard deviation of 0.0138. Regarding the magnitude, it ranges from $-$22.89 to $-$19.77, with a mean value of $-$21.33 and a distribution with a standard deviation of 0.82 and an IQR of 1.16. Both distributions show that the galaxy sample is homogeneous in both redshift and magnitude.

\begin{table*}
	\centering
\caption{\label{table:sample} The sample of 46 barred galaxies. Columms: (1) MaNGA ID; (2) galaxy name; (3) right ascension (J2000); (4) declination (J2000); (5) morphological type; (6) absolute r-band magnitude from SDSS-DR9; (7) effective radius in the r-band from SDSS-DR9; (8) redshift; (9) galaxy inclination; (10) galaxy kinematic PA; (11) systemic velocity. All data were taken from \citet{Guo2019}, except the galaxy name (from the SDSS DR13) and the systemic velocity (Sect. \ref{sec: procedure}).}
\begin{tabular}{c c c c c c c c c c c}
\hline
Plate IFU   &	     Galaxy name         &         RA	  &      DEC  &  Morph.  &	  Mr          &	$R_{e}$  &	$z$      &	$i$	&        $PA_{kin}$    & $V_{syst}$\\ 
	    &	                         &         (°)	  &      (°)  &		     &  (mag)         &	(“) &	       &	(°)	&        (°)    &  ($\kms$)  \\ 
  (1)	    &             (2)	         &         (3)	  &      (4)  &	  (5)        &    (6)         &	(7) &	(8)    &	(9)     &	 (10)   &  (11)    \\ 
7495$-$12704 &	SDSS J134145.21$+$270016.9  &	205.4384  &	27.0048	&  SBbc &	    $-$21.40 &  8.65 &	0.0289 &	52.2$\pm$0.6 &	173.0 $\pm$0.6 &	8768  \\ 
7962$-$12703 &	SDSS J172452.14$+$280441.8  &	261.2173  &	28.0783	&  SBab	&       $-$22.33 &	8.34 &	0.0477 &	61.2$\pm$1.2 &	37.0 $\pm$0.9  &	14452 \\ 
7990$-$3704  &	SDSS J172817.97$+$564629.2  &	262.0749  &	56.7748	&  SB0	&       $-$20.15 &	3.83 &	0.0291 &	39.4$\pm$1.4 &	15.2 $\pm$3.4  &	8800  \\ 
7990$-$9101  &	SDSS J171901.32$+$571024.6  &	259.7555  &	57.1735	&  SBc	&       $-$19.77 &  4.51 &	0.0280 &	71.8$\pm$0.2 &	20.0 $\pm$3.8  &	8488  \\ 
7992$-$6104  &	SDSS J170107.08$+$644036.6  &	255.2795  &	64.6769	&  SBc	&       $-$20.31 &	8.78 &	0.0271 &	46.7$\pm$1.8 &	6.0 $\pm$2.8   &	8188  \\ 
8082$-$6102  &	SDSS J031947.01$+$003504.4  &	49.9459   &	0.5846	&  SB0	&       $-$21.46 &	6.91 &	0.0242 &	41.3$\pm$0.5 &	99.0 $\pm$0.9  &	7346  \\ 
8083$-$6102  &	SDSS J032427.59$-$000510.8  &	51.1150	  & $-$0.0863& SBa	&       $-$21.62 &	4.70 &	0.0365 &	70.4$\pm$0.2 &	62.8 $\pm$0.9  &	11028 \\ 
8083$-$12704 &	SDSS J032247.22$+$000857.6  &	50.6968   &	0.1494	&  SBbc	&       $-$21.03 &	13.32& 	0.0228 &	41.7$\pm$0.9 &	167.0 $\pm$1.4 &	6946  \\ 
8133$-$3701  &	SDSS J072819.02$+$431807.5  &	112.0793  &	43.3021	&  SBb	&       $-$20.10 &	2.39 &	0.0437 &	44.6$\pm$1.1 &	102.8 $\pm$3.5 &	13159 \\  
8134$-$6102  &	SDSS J073941.88$+$455445.4  &	114.9245  &	45.9126	&  SB0a	&       $-$21.40 &	5.98 &	0.0320 &	53.8$\pm$0.9 &	93.0 $\pm$0.8  &	9670  \\ 
8137$-$9102  &	SDSS J074809.26$+$433526.4  &	117.0386  &	43.5907	&  SBb	&       $-$21.07 &	6.68 &	0.0311 &	43.3$\pm$2.2 &	132.8 $\pm$1.9 & 	9417  \\ 
8140$-$12701 &	SDSS J074743.27$+$412310.9  &	116.9303  &	41.3864	&  SBa	&       $-$20.61 &	5.69 &	0.0286 &	37.8$\pm$1.3 &	62.8 $\pm$1.8  &	8674  \\ 
8140$-$12703 &	SDSS J075135.62$+$425248.3  &	117.8985  &	42.8801	&  SBb	&       $-$21.87 &	9.85 &	0.0320 &	55.0$\pm$0.6 &	28.0 $\pm$1.1  &	9652  \\ 
8243$-$6103  &	SDSS J083641.96$+$534338.0  &	129.1749  &	53.7272	&  SB0	&       $-$21.65 &	4.75 &	0.0315 &	59.1$\pm$0.6 &	9.8 $\pm$0.6   &	9526  \\ 
8244$-$3703  &	SDSS J084758.26$+$513603.6  &	131.9928  &	51.6010	&  SB0	&       $-$21.03 &	2.50 &	0.0483 &	46.1$\pm$1.1 &	71.5 $\pm$1.6  &	14561 \\  
8249$-$6101  &	SDSS J091014.98$+$461735.8  &	137.5625  &	46.2933	&  SBc	&       $-$20.27 &	4.64 &	0.0267 &	48.7$\pm$1.4 &	63.5 $\pm$1.6  &	8092  \\ 
8254$-$9101  &	SDSS J104502.80$+$434217.0  &	161.2617  &	43.7048	&  SBa	&       $-$21.78 &	8.00 &	0.0253 &	44.1$\pm$1.6 &	27.2 $\pm$0.8  &	7698  \\ 
8256$-$6101  &	SDSS J105456.35$+$412954.3  &	163.7348  &	41.4985	&  SBa	&       $-$20.79 &	6.06 &	0.0246 &	51.4$\pm$2.6 &	134.0 $\pm$0.9 & 	7492  \\ 
8257$-$3703  &	SDSS J110637.35$+$460219.6  &	166.6557  &	46.0388	&  SBb	&       $-$20.34 &	4.03 &	0.0250 &	58.3$\pm$0.6 &	155.2 $\pm$1.2 &	7543  \\ 
8257$-$6101  &	SDSS J110102.71$+$445317.7  &	165.2613  &	44.8882	&  SBc	&       $-$20.86 &	5.77 &	0.0294 &	45.0$\pm$2.2 &	159.2 $\pm$1.5 &	8904  \\ 
8312$-$12702 & 	SDSS J162105.00$+$395502.6  &	245.2709  &	39.9174	&  SBc	&       $-$21.24 &	7.23 &	0.0320 &	42.9$\pm$1.1 &	95.2 $\pm$1.8  &	9698  \\ 
8312$-$12704 &	SDSS J162912.98$+$410903.1  &	247.3041  &	41.1509	&  SBb	&       $-$21.00 &	7.47 &	0.0296 &	46.1$\pm$0.7 &	34.0 $\pm$1.8  &	8977  \\ 
8313$-$9101  &	SDSS J155847.40$+$415617.1  &	239.6975  &	41.9381	&  SBb	&       $-$21.87 &	6.76 &	0.0387 &	38.6$\pm$0.7 &	110.5 $\pm$1.0 &	11685 \\ 
8317$-$12704 &	SDSS J125448.95$+$440920.1  &	193.7040  &	44.1556	&  SBa	&       $-$22.68 &	7.14 &	0.0543 &	69.2$\pm$0.3 &	101.8 $\pm$0.9 &	16355 \\ 
8318$-$12703 &	SDSS J130455.76$+$473013.0  &	196.2324  &	47.5036	&  SBb	&       $-$22.21 &	9.09 &	0.0393 &	61.8$\pm$0.9 &	53.8 $\pm$0.9  &	11869 \\ 
8320$-$6101  &	SDSS J134630.60$+$224221.6  &	206.6275  &	22.7060	&  SBb	&       $-$20.37 &	5.22 &	0.0266 &	50.0$\pm$0.6 &	5.0 $\pm$1.1   &	8046  \\ 
8326$-$3704  &	SDSS J141924.05$+$455402.8  &	214.8502  &	45.9008	&  SBa	&       $-$20.25 &	3.83 &	0.0265 &	50.4$\pm$1.1 &	159.8 $\pm$3.0 &	8039  \\ 
8326$-$6102  &	SDSS J142004.29$+$470716.8  &	215.0179  &	47.1213	&  SBb	&       $-$22.06 &	2.95 &	0.0704 &	51.9$\pm$0.9 &	145.8 $\pm$1.6 &	21156 \\ 
8330$-$12703 &	SDSS J133329.90$+$403146.8  &	203.3746  &	40.5297	&  SBbc	&       $-$20.67 &	7.51 &	0.0269 &	45.0$\pm$0.5 &	68.5 $\pm$1.9  &	8143  \\ 
8335$-$12701 &	SDSS J142134.86$+$402129.1  &	215.3953  &	40.3581	&  SBb	&       $-$21.66 &	4.39 &	0.0633 &	67.0$\pm$0.5 &	78.2 $\pm$1.4  &	19096 \\  
8439$-$6102  &	SDSS J093106.75$+$490447.1  &	142.7782  &	49.0797	&  SBab	&       $-$21.64 &	4.54 &	0.0339 &	49.3$\pm$0.5 &	45.5 $\pm$1.1  &	10229 \\ 
8439$-$12702 &	SDSS J092609.43$+$491836.7  &	141.5393  &	49.3102	&  SBa	&       $-$21.57 &	8.10 &	0.0269 &	55.1$\pm$0.4 &	31.5 $\pm$0.5  &	8167  \\ 
8440$-$12704 &	SDSS J090434.15$+$412352.1  &	136.1423  &	41.3978	&  SBb	&       $-$21.12 &	4.56 &	0.0270 &	57.9$\pm$0.4 &	150.0 $\pm$0.8 &	8201  \\ 
8447$-$6101  &	SDSS J134431.98$+$401424.0  &	206.1333  &	40.2400	&  SBb	&       $-$22.89 &	4.48 &	0.0753 &	63.9$\pm$0.8 &	178.2 $\pm$1.2 &	22661 \\ 
8452$-$3704  &	SDSS J103009.35$+$471642.2  &	157.5390  &	47.2784	&  SBc	&       $-$19.97 &	4.34 &	0.0251 &	59.7$\pm$0.3 &	72.0 $\pm$2.5  &	7597  \\ 
8452$-$12703 &	SDSS J102713.35$+$481441.2  &	156.8057  &	48.2448	&  SBb	&       $-$22.83 &	8.13 &	0.0610 &	45.7$\pm$2.4 &	65.0 $\pm$1.2  &	18356 \\  
8482$-$9102  &	SDSS J102713.35$+$481441.2  &	242.9559  &	49.2287	&  SBb	&       $-$21.59 &	3.54 &	0.0580 &	62.6$\pm$0.6 &	63.8 $\pm$1.9  &	17493 \\ 
8482$-$12705 &	SDSS J161652.00$+$501655.8  &	244.2167  &	50.2822	&  SBb	&       $-$22.06 &	7.39 &	0.0417 &	63.0$\pm$1.0 &	117.0 $\pm$1.0 &	12583 \\  
8486$-$6101  &	SDSS J155209.48$+$461911.2  &	238.0396  &	46.3198	&  SBc	&       $-$21.57 &	3.56 &	0.0589 &	40.4$\pm$1.2 &	113.5 $\pm$1.4 &	17734 \\ 
8548$-$6102  &	SDSS J162205.36$+$463726.9  &	245.5224  &	46.6242	&  SBc	&       $-$20.83 &	3.85 &	0.0478 &	54.1$\pm$0.4 &	58.8 $\pm$3.6  &	14386 \\ 
8548$-$6104  &	SDSS J162259.35$+$464031.2  &	245.7474  &	46.6753	&  SBc	&       $-$20.47 &	2.73 &	0.0480 &	62.2$\pm$1.6 &	120.2 $\pm$5.0 &	14465 \\ 
8549$-$12702 &	SDSS J160505.14$+$452634.7  &	241.2714  &	45.4430	&  SBb	&       $-$22.03 &	6.72 &	0.0433 &	54.3$\pm$2.6 &	100.8 $\pm$1.0 &	13084 \\ 
8588$-$3701  &	SDSS J163233.73$+$390751.7  &	248.1406  &	39.1310	&  SBb	&       $-$22.88 &	4.43 &	0.1303 &	40.4$\pm$1.7 &	136.2 $\pm$1.9 &	39164 \\ 
8604$-$12703 &	SDSS J163103.40$+$395018.5  &	247.7642  &	39.8385	&  SBab	&       $-$21.67 &	9.08 &	0.0305 &	48.8$\pm$1.0 &	97.8 $\pm$1.0  &	9254  \\ 
8612$-$6104  &	SDSS J170001.67$+$384857.7  &	255.0069  &	38.8160	&  SBb	&       $-$21.83 &	8.60 &	0.0356 &	42.4$\pm$2.3 &	153.5 $\pm$1.8 &	10749 \\ 
8612$-$12702 &	SDSS J165547.13$+$391837.8  &	253.9464  &	39.3105	&  SBc	&       $-$22.60 &	8.26 &	0.0631 &	52.3$\pm$1.0 &	44.0 $\pm$1.4  &	19004  \\ 

\hline
	\end{tabular}
 \end{table*}


\section{Procedure}
\label{sec: procedure}

The data included in this paper are from Sloan Digital Sky Survey (SDSS) IV DATA Release 13 \footnote{\url{https://data.sdss.org/sas/dr13/manga/spectro/redux/v1_5_4/}} \citep{Albareti2017}. We used the publicly available dataproducts obtained by Pipe3D \citep{Pipe3d,Pipe3dII}. From these cubes we extracted the V-band pseudo-image, the stellar radial velocity field, and the error in the stellar radial velocity determination for each spaxel. Using MARVIN software \citep{Marvinsoft2019} we also obtained the SDSS color image with the indicated IFU MaNGA FoV for comparison with the cube maps. The pseudo V-image was used to determine the peak emission and the systemic velocity. From the kinematic position angle (PA$_{Kin}$) determined by \cite{Guo2019} we carried out an extraction of the radial velocities in a 1 spaxel width pseudo-slit. This allowed us to construct the stellar radial velocity curves. Errors were estimated from the error plane in MaNGA datacubes. We used the value of the radial velocity error indicated in the Pipe3D maps for each spaxel. A typical error bar is indicated in the plots. In the upper panels of Fig.~\ref{fig:curves} we show the SDSS images, the pseudo V-images, and the stellar radial velocity fields for each galaxy.

For the choice of the systemic velocity ($V_{syst}$) for each galaxy, the velocities corresponding to the peak of the pseudo-image obtained in the V-band were compared with the velocity that best symmetrizes the radial velocity curve. This comparison is performed to analyze possible off-centering of the velocity fields with the luminosity distribution, but in all cases the displacements were less than 2 arcsec, which implies that they are within the spatial resolution element. Thus the $V_{syst}$ adopted are those that best symmetrize the radial velocity curves. In all cases these $V_{syst}$ are in agreement with the values found in the literature and in the NED.

\section{Results}
\label{sec: results}

\subsection{The rotation curves}
\label{sub:rotation_curves}

Once the radial velocity curves are determined, the rotation velocity (V$_{rot}$) can be calculated as:

\begin{equation}
\label{eq:vel}
V_{rot}= \frac{V_{obs}-V_{syst}}{sin(i)}
  \end{equation}
  
where V$_{obs}$ is the observed radial velocity, V$_{syst}$ is the systemic radial velocity and $i$ is the galaxy inclination.\\

In order to fit all the RCs, we made use of \href{https://www.galpy.org/}{galpy}\footnote{\url{http://github.com/jobovy/galpy}} \citep{Bovy2015}, which is a Python package for galactic dynamics. For this, we considered two mass components corresponding to axisymetric Miyamoto$-$Nagai gravitational potentials \citep{Miyamoto-Nagai1975}:

\begin{equation}
\label{eq:pot}
    \Phi~(R,z) = - \frac{G M}{\sqrt{R^2 + [a + \sqrt{z^2 + b^2}]^2}} 
    \end{equation}

where $\Phi (R,z)$ is the Miyamoto$-$Nagai potential at $(R, z)$ position, $a$ and $b$ are shape parameters and $M$ is the total mass. By way of example, the ratio b/a $\sim$ 0.4 corresponds to a flattened disk distribution, b/a $=$ 1 to an ellipsoidal distribution and b/a $\sim$ 5 to a spherical distribution  \citep[e.g.,][]{Binney}.

Given its simplicity, the Miyamoto$-$Nagai potential is one of the most widely used models for fitting rotation curves \citep[e.g.,][]{Binney, Sofue2017,Granados2017,Ciotti2021, Ciotti2022}. It is a generalisation of the Plummer's spherical potential (when a $=$ 0) and the Kuzmin's potential of a razor-thin disk (when b $=$ 0). This way, depending on the values of the two parameters a and b, the Miyamoto$-$Nagai model can represent the potential of anything from an infinitesimally thin disk to a spherical distribution \citep[see for example][for more details]{Binney}.

From the centrifugal-equilibrium condition, the rotation velocity in the equatorial plane is related to the gravitational potential $\Phi$(R, z$=$0) in the form:

\begin{equation}
\label{eq:vel1}
    V_{rot}~(R)=\sqrt{-R (\partial \Phi / \partial R)}
    \end{equation}

Taking into account Eq. \ref{eq:pot} and Eq. \ref{eq:vel1}, the rotation velocity in the galactic plane at z $=$ 0 is given by:

\begin{equation}
\label{eq:vel2}
    V_{rot}~(R) = R \sqrt{\frac{G M}{[R^2 + (a + b)^2]^{3/2}}} 
    \end{equation}


The fitting of the rotation curves was carried out in a two-stage procedure. The first thing we did was to perform a least squares fit to the data in order to have a first approximation. However, an automatic fit may consider regions where there are non-circular motions generated by nonaxisymmetric disturbances and thus affect the global fit that takes into account rotation velocities. Because of this, we carefully checked each curve individually, and using the least square fit as a reference, we made used of galpy improving each fit by making it take into account only the observed rotation motions, which are representative of the gravitational potential. This procedure was carried out varying the input parameters mass, a, and b and considering each galaxy individually to ensure the best fit, which has been carefully inspected by eye. To determine the uncertainties of the fits, we repeated the measurement procedure 15 times for all galaxies. We assumed that any errors are given by the dispersion in the distribution of these measurements, considering them at 1$\sigma$. Concerning the fits, relative errors in the velocity are in the range  3\% $-$ 8\%. These errors take into account only our measurements procedures, and do not involve other factors such as the observation process, reduction or calibrations.

All galaxies of the sample were fitted considering the presence of two galactic subsystems: a central ellipsoidal component and a flattened disk, in agreement with previous works that also have fitted rotation curves of barred spiral galaxies \citep[e.g.,][]{Diaz1999,Aguero2004,Schmidt2019}. 

Given the morphological characteristics of this sample, a significant fraction of the galaxies are late type objects (see Table \ref{table:sample}). Related to this and considering the rotation curve fitting, the central component is almost absent or negligible compared to the disk component in all but a few galaxies (see for example the fits of 7495$-$12704 and 7962$-$12703, among others, in Fig. \ref{fig:curves}).
Although for consistency in all the fittings the two components were kept. In none of the fits a third component corresponding to the halos was used, on the one hand because the fit was satisfactory with two components and on the other hand because the scale radius of the halo is larger than the radii involved in our data \citep[e.g.,][]{Sofue2016}. Related to this, it has to be noted that the possible gravitational influence of a dark matter halo may be implied by the disc component. However, in this work we do not intend to physically interpret the different components separately, but consider the total fit of the rotation curve that will allow us to robustly calculate the dynamical masses of the galaxies of the sample (see Sect. \ref{sub:Mdyn}). 


In order to check for the reliability of the rotation curves as dynamical mass estimators, we follow the procedure used by \citet{Kalinova2017a} and \citet{Bekeraite2016a}. This procedure considers that when the ordered-over-random motion V$\phi/\sigma_{R}$ < 1.5, the rotation curve might not be a reliable tracer. To do this, we made use of the dispersion maps available as part of the Pipe3D dataproducts to construct the dispersion velocity radial profiles to evaluate the V$\phi/\sigma_{R}$ ratio. We performed this test for all galaxies in our sample. As we said in Sect. \ref{sec: sample}, we rejected one galaxy (SDSS J090641.14+412154.3, plate-IFU 8247-3701) whose entire extension presents an important contribution of the stellar velocity dispersion compared with the rotation velocity. This way, our whole sample is constituted by galaxies whose dynamic is dominated by rotation. Considering that practically all the galaxies in our sample present a negligible bulge component, and in order to make an homogeneous treatment to our data and to avoid any potentially incorrect assumption, we will study the stellar kinematic of the galaxies neglecting the correction methods mentioned in Sect.\ref{sec:intro} \citep[e.g.,][]{Bureau2002, DiTeodoro2021, DiTeodoro2023}.

\begin{figure}
    \centering
    \includegraphics[width=0.45\textwidth]{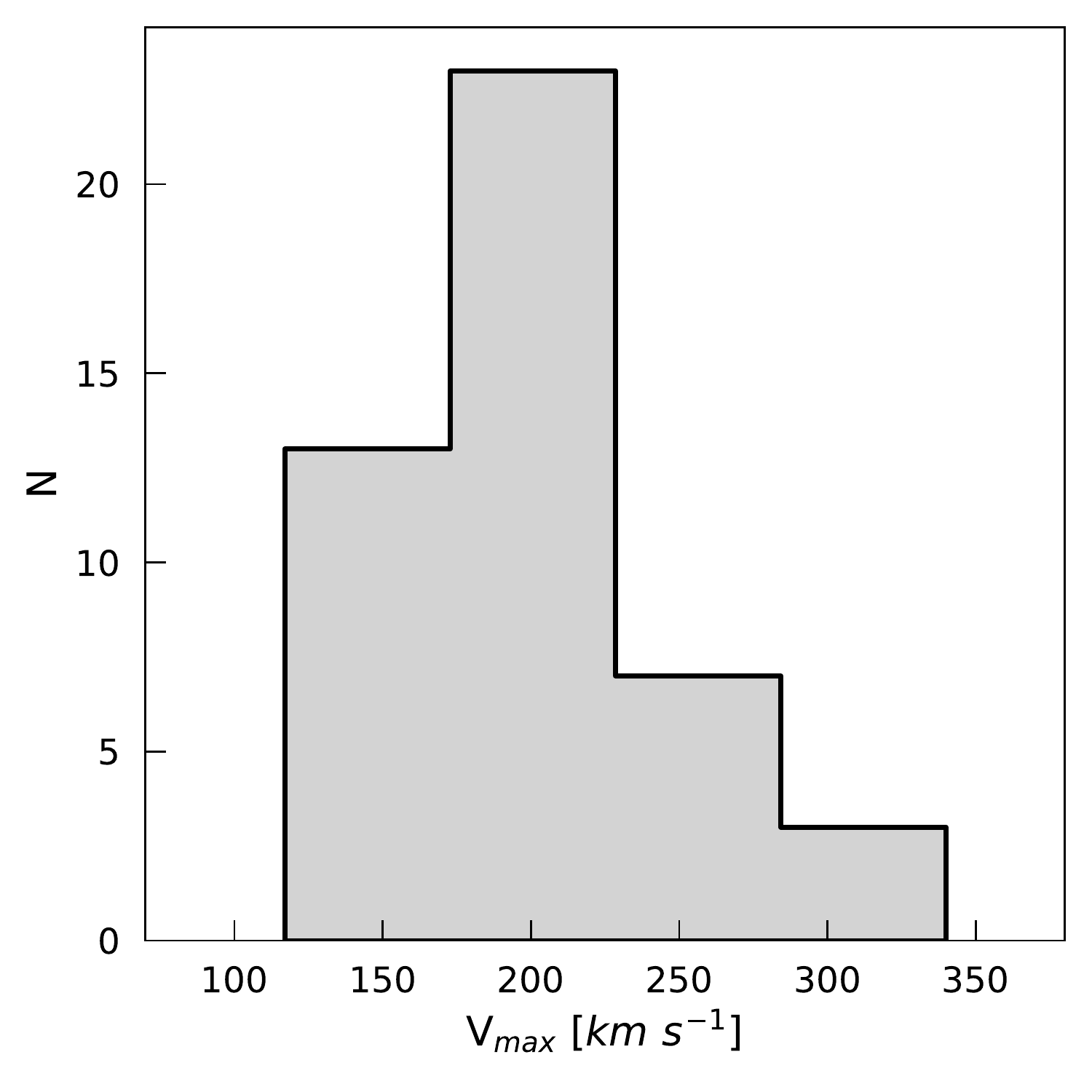}
   \caption{Distribution of the maximun rotation velocity in units of $\kms$ for our sample of barred galaxies.}
              \label{fig:hist_velocidad}%
    \end{figure}

Table \ref{table:params} lists the obtained parameters of our fits, such as the scale parameters, the mass of each galactic subsystem and the total mass.
Fig. \ref{fig:curves} presents all the fitted rotation curves of the galaxy sample. In addition, Fig. \ref{fig:hist_velocidad} shows the distribution of the maximun rotation velocity, i.e., the highest rotation velocity measured over each rotation curve. According to our results, the maximun rotation velocities show a wide range, with values from 117 to 340 $\kms$, with most of the objects showing a maximun rotation velocity in the range $\sim$120$-$250 $\kms$. The distribution has a mean value of 200 $\kms$, a standard deviation of 50 $\kms$ and an IQR of 57. These values are in agreement with previous results on spiral galaxies \citep[e.g.,][among others]{Persic1995, Sofue2001, Aguero2004, Sofue2017, Schmidt2019, Martinez-Medina2020}. There are 11 galaxies from our sample (7495-12704, 8137-9102, 8257-6101, 8318-12703, 8330-12703, 8335-12701, 8439-12702, 8452-12703, 8482-12705, 8549-12702 and 8601-12705) that were also studied by \cite{Pilyugin2019}, who obtained rotation curves through the analysis of the H$\alpha$ velocity field. In general, the rotation velocities determined in this work are in good agreement with the results reported in the mentioned paper. There are, however, some differences in the rotation velocities of a few galaxies. These differences may probably be due to the procedure used to construct the rotation curves, the adopted galaxy inclination, and to the fact that these authors used H$\alpha$ kinematics instead of the stellar velocity fields.

There are some galaxies in our sample that present nuclear activity: 8257$-$6101 \citep[][]{Edelson1987, Toba2014}, 8312$-$12704 \citep[][]{Toba2014, Rembold2017}, 8317$-$12704 \citep[e.g.,][]{Rembold2017, Bing2019}, 8320$-$6101 \citep[e.g.,][]{Stern2012}, 8326$-$6102 \citep[e.g.,][]{Greene2007, Stern2012}, 8452$-$12703 \citep[e.g.,][]{Rembold2017}, 8549$-$12702 \citep[e.g.,][]{Greene2007, Stern2012}, 8588$-$3701 \citep[e.g.,][]{Rembold2017}, 8604$-$12703 \citep[e.g.,][]{Rembold2017} and 8612$-$12702 \citep[e.g.,][]{Rembold2017, Ilha2019}. We did not find any specific peculiarity in the rotation curves of these galaxies, which is to be expected due to the distances involved. Related to this, nuclear activity should not affect the kinematics at large distances from the center \citep[e.g.,][]{Sofue1999,Schmidt2016, Sofue2017,YU2022AA}. This would be in accordance with the hypothesis that nuclear activity is generated by a more local cause rather than by a global dynamical mass distribution \citep[e.g.,][]{Sofue1999, Sofue2017}, although a study with a larger sample size would be necessary to delve deeper into this aspect \citep{Sanchez2018}. 

On the other hand, as mentioned above, there are galaxies that present evidence of non-circular motions which are manifested through strong departures from the fit. These galaxies are 7962-12703, 7992-6104, 8312-12704, 8317-12704, 8330-12703, 8452-12703, 8482-9102, 8486-6101, 8548-6102 and 8549-12702. This behaviour is expected in this kind of galaxies \citep[e.g.,][]{Vaughan1989,Regan1994,Schmidt2019} and may be caused by the presence of nonaxisymmetric structures such as spiral arms and bars \citep[e.g.,][]{Binney1991, Athanassoula1992, Burton1993, Sofue2017}.

\subsection{Dynamical masses}
\label{sub:Mdyn}

\begin{figure}
    \centering
    \includegraphics[width=0.45\textwidth]{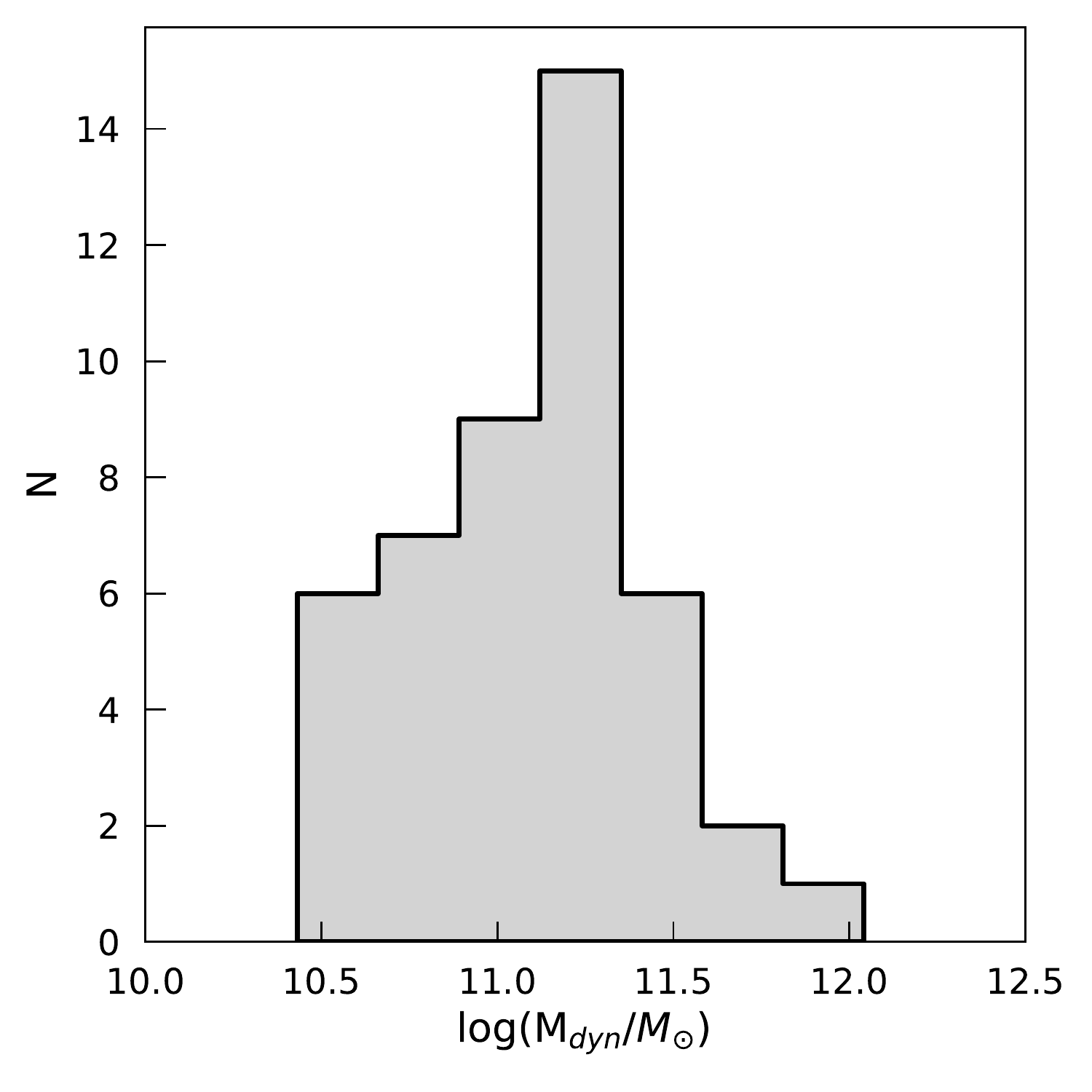}
   \caption{Total dynamical mass distribution for our sample of barred galaxies.}
              \label{fig:total_mass}%
    \end{figure}

The derived dynamical masses for all barred galaxies of the sample are presented in column 6 of Table \ref{table:params}. In addition, Fig. \ref{fig:total_mass} shows the distribution of this parameter. The masses are in the range of log(M$_{dyn}$/M$_{\odot}$) $=$ 10.4$-$12.0, with a mean value of log(M$_{dyn}$/M$_{\odot}$) $=$ 11.1. These values are in agreement with previous results considering masses of spiral galaxies \citep[e.g.,][among others]{Burbidge1959,Burbidge1961,Rubin1964,Krumm1977,Pence1981,Aguero2004, Salucci2008b, Jalocha2010,Daod2019,Schmidt2019}. The distribution has a standard deviation of 0.36 and an IQR of 0.47. This result would suggest that barred galaxies present dynamical masses similar to those of the general spiral galaxies population.

Considering the uncertainties given by the fitting process of the rotation curves (see Sect. \ref{sub:rotation_curves}), the typical relative error in the determination of the dynamical mass is in the range $\sim$7\% $-$ 15\% (Table \ref{table:params}). As mentioned above, these errors take into account only our measurement process.\\

There are 11 galaxies in our sample which were also studied by \citet{Garma-Oehmichen2020}. These objects are 7495-12704, 7962-12703, 8256-6101, 8257-3703, 8312-12704, 8313-9101, 8317-12704, 8318-12703, 8439-6102, 8439-12702 and 8453-12701. These authors present the stellar mass and the molecular gaseous mass derived from the Pipe3D analysis. In the mentioned work, the molecular gas mass was estimated adopting the dust-to-gas ratio based on the dust attenuation obtained by Pipe3D. As expected, the dynamic masses obtained in this work are higher than the sum of the masses obtained in \citet{Garma-Oehmichen2020}. This difference is in general $\sim$0.3 dex. The observed difference between our dynamically determined mass and the sum of stellar and molecular mass estimates may be attributed to the contribution of other components, such as ionized gas and dark matter \citep[e.g.,][]{Sofue2001}. Indeed, it is well-known that the dynamical mass estimates of galaxies account for all gravitationally bound matter within their extent, including components beyond the visible stellar and molecular matter \citep[e.g.,][]{Lelli2016}. In this context, the contribution of ionized gas, which can comprise a substantial fraction of the total gas content in galaxies, can be particularly relevant. Moreover, the presence of dark matter, which is thought to constitute an important fraction of matter in the universe, is expected to impact the dynamical mass of galaxies and, thus, the observed discrepancy between different mass estimates \citep[e.g.,][]{Persic1995, Sofue2001, Lelli2016}. Therefore, our finding of a greater dynamical mass relative to the sum of stellar and molecular mass estimates, while expected, is noteworthy as it underscores the importance of knowing the dynamical masses of these objects as they consider the contribution of all elements to the gravitational potential. To our best knowledge, this is the first direct determination of dynamical masses through rotation curves in this sample of galaxies.

\subsection{Stellar masses}
\label{sub:Mstar}

\begin{figure}
    \centering
    \includegraphics[width=0.45\textwidth]{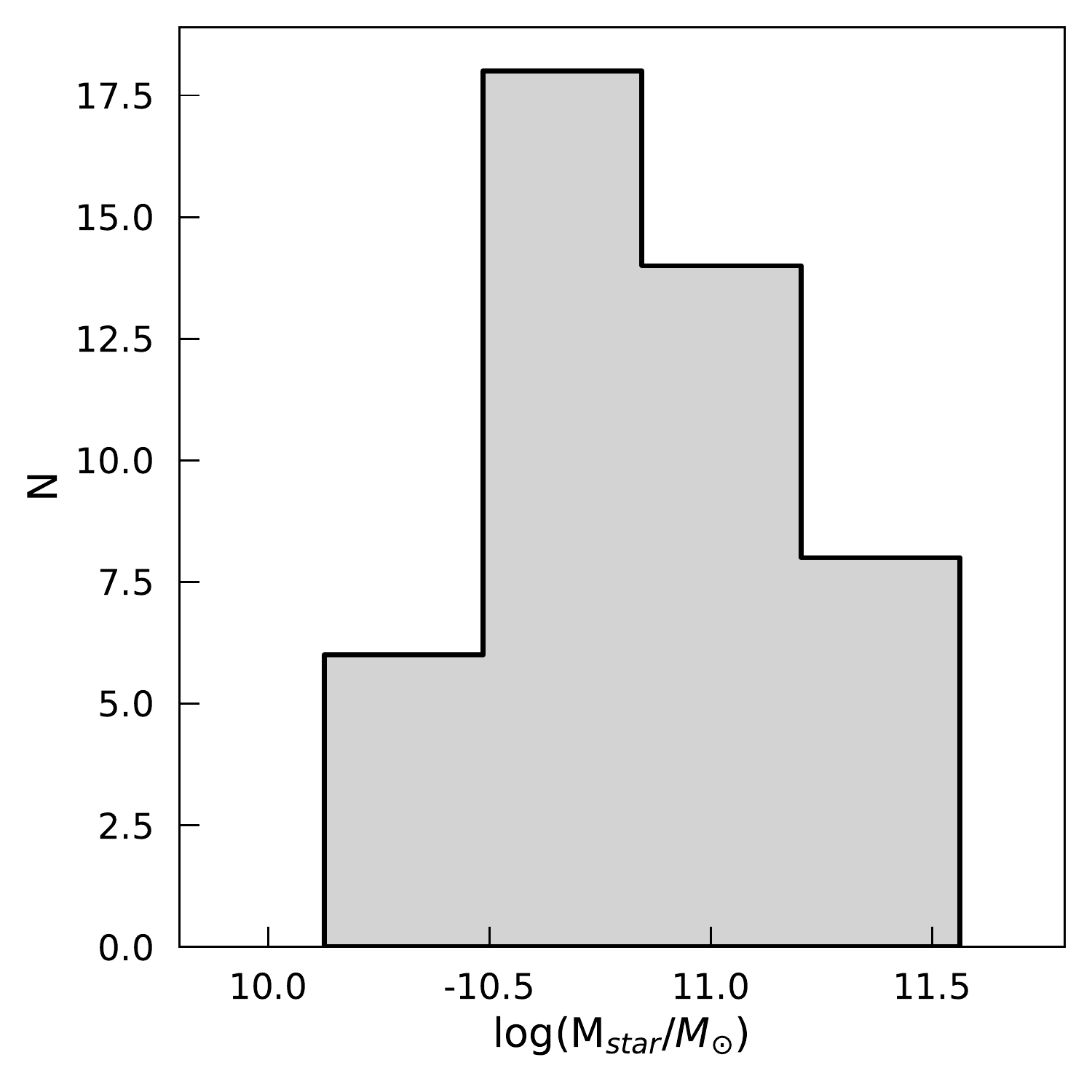}
   \caption{Stellar mass distribution for our sample of barred galaxies.}
              \label{fig:hist_mstar}%
    \end{figure}

The relation between the dynamical and stellar mass of galaxies gives us clues to understand their structure and evolution, since it links the mass already contained in stars with the total mass of the galaxy, i.e. gas, dust, and dark matter. The integrated stellar masses reported in Table \ref{table:params} were determined from the absorption-corrected stellar mass density maps as indicated in \citet{Sanchez2016a,Sanchez2016b}. The errors of stellar masses are of the order of $\sim$ 1\% and were obtained through error propagation considering the mass determination of each spaxel. These errors refer to our measurement process and do not include errors in observations, calibrations, etc. Fig. \ref{fig:hist_mstar} shows the distribution of stellar mass for the whole sample, which is in the range of log(M$_{star}$/M$_{\odot}$) $=$ 10.13 $-$ 11.53, with a mean value of log(M$_{star}$/M$_{\odot}$) $=$ 10.8. According to our results, this distribution has a standard deviation of 0.35 and an IQR of 0.54. These values are in agreement with previous
results on spiral galaxies \citep[e.g.,][]{Williams2010, Davis2018}.
As mentioned in Sec.\ref{sub:Mdyn},  \citet{Garma-Oehmichen2020} studied the stellar mass of 11 galaxies that are also part of our sample and as expected the results are in good agreement.

\subsection{Scaling relations}
\label{sub:scaling}

Scaling relations are empirical relationships between different parameters that can be used to study the properties of galaxies \citep[e.g.,][]{DiTeodoro2021, Krut2023}. They are a fundamental tool for understanding the structure, dynamics, and evolution of galaxies \citep[e.g.,][]{Aquino-Ortiz2018,Ferrero2021AA}. For example, in the context of rotation curves, scaling relations provide insights into the underlying physical processes that govern the dynamical properties of galaxies. These relations quantify the relationships between different observable properties and are widely used to study the formation and evolution of galaxies \citep[e.g.,][]{Lapi2018, DiTeodoro2023} . As such, scaling relations provide a powerful framework for connecting observations with theoretical models and for testing our understanding of galaxy formation and evolution. In this context, we analyse and study the scaling relations taking into account the magnitude and the parameters we have measured namely dynamical mass, stellar mass, and maximum rotation velocity.

\begin{figure}
    \centering
    \includegraphics[width=0.47\textwidth]{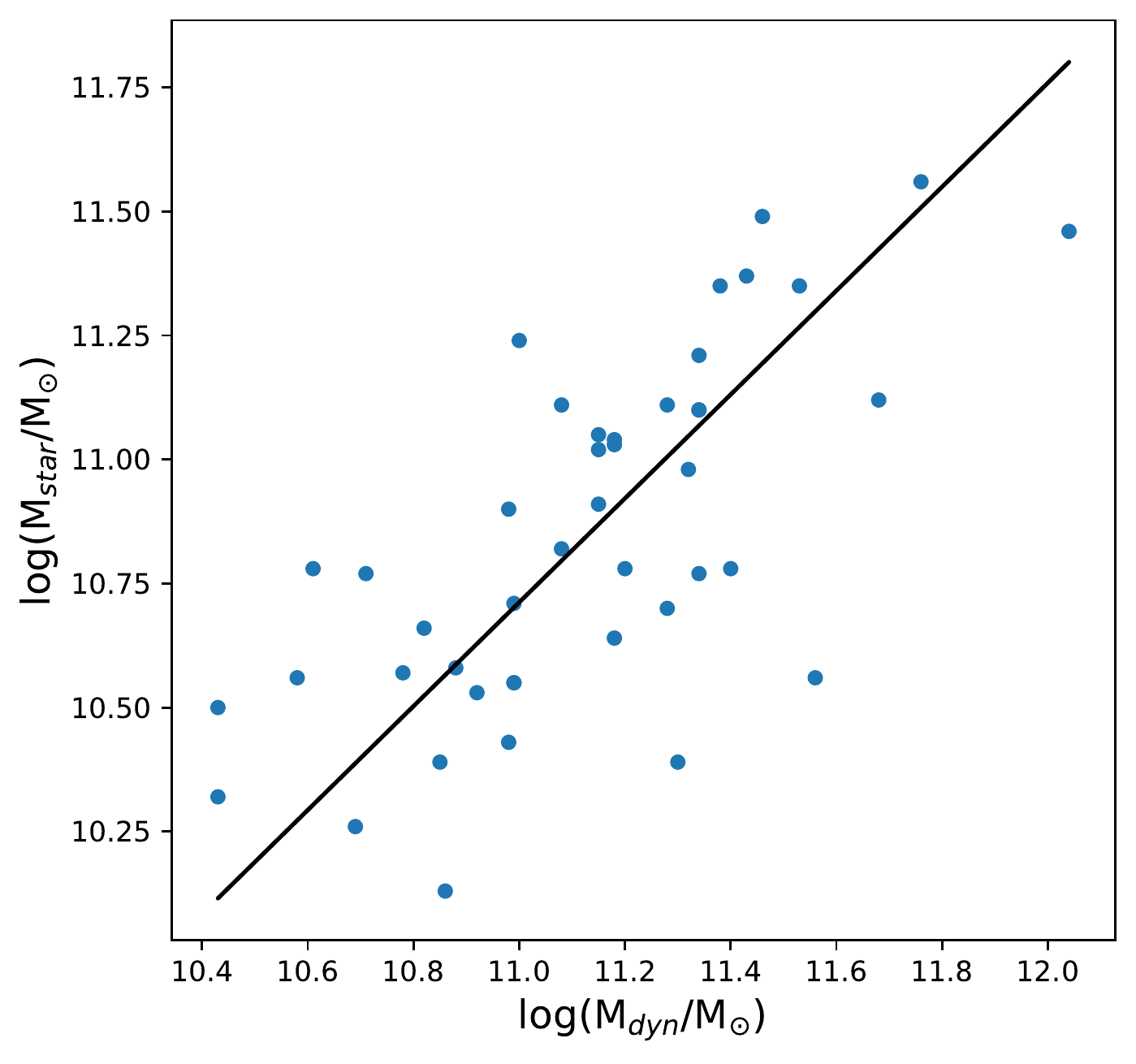}
   \caption{Relation between the dynamical mass and the stellar mass. The galaxies 8082-6102 and 8318-12703 are not shown in the plot and are not considered in the analysis. The black solid line represents the OLS bisector fit for our data. }
              \label{fig:Mdyn_Mstar}%
    \end{figure}

Firstly, we study the relation between the dynamical mass and the stellar mass. In Fig. \ref{fig:Mdyn_Mstar} we show this correlation, which has a Pearson coefficient r$_{p}=$ 0.70 and a p-value of 1.4 $\times$10$^{-7}$. As expected, this result implies that as the stellar mass of a galaxy increases, so too does its total dynamical mass. Although the relation is clear, it can be seen that there is some dispersion in the data. It is worth noting that there are many different factors that can influence the dynamics of spiral galaxies, including the distribution of dark matter, the presence of gas and dust, and the possible effects of mergers and interactions with other galaxies. These factors may interact in complex ways and can be different in each galaxy. We performed an OLS (Ordinary Least Squares) bisector fit for our data, which has a slope of 1.0 $\pm$ 0.1 and an offset of -0.80 $\pm$ 1. 
Due to the deviation, the galaxies 8082$-$6102 and 8318$-$12703 are not shown in the plot and are not considered for this particular analysis. These objects, however, are considered for the rest of the analysis regarding the other parameters.
Additionally, it has to be noted that a few objects present higher stellar mass compared to the dynamical mass. These galaxies are: 8083-6102, 8249-6101, 8257-3703, 8326-6102, 8548-6104 and 8612-12702 (see Fig. \ref{fig:Mdyn_Mstar} and Table \ref{table:params}). While these differences fall within the errors, caution should be exercised when dealing with these particular values.

\begin{figure}
    \centering
    \includegraphics[width=0.47\textwidth]{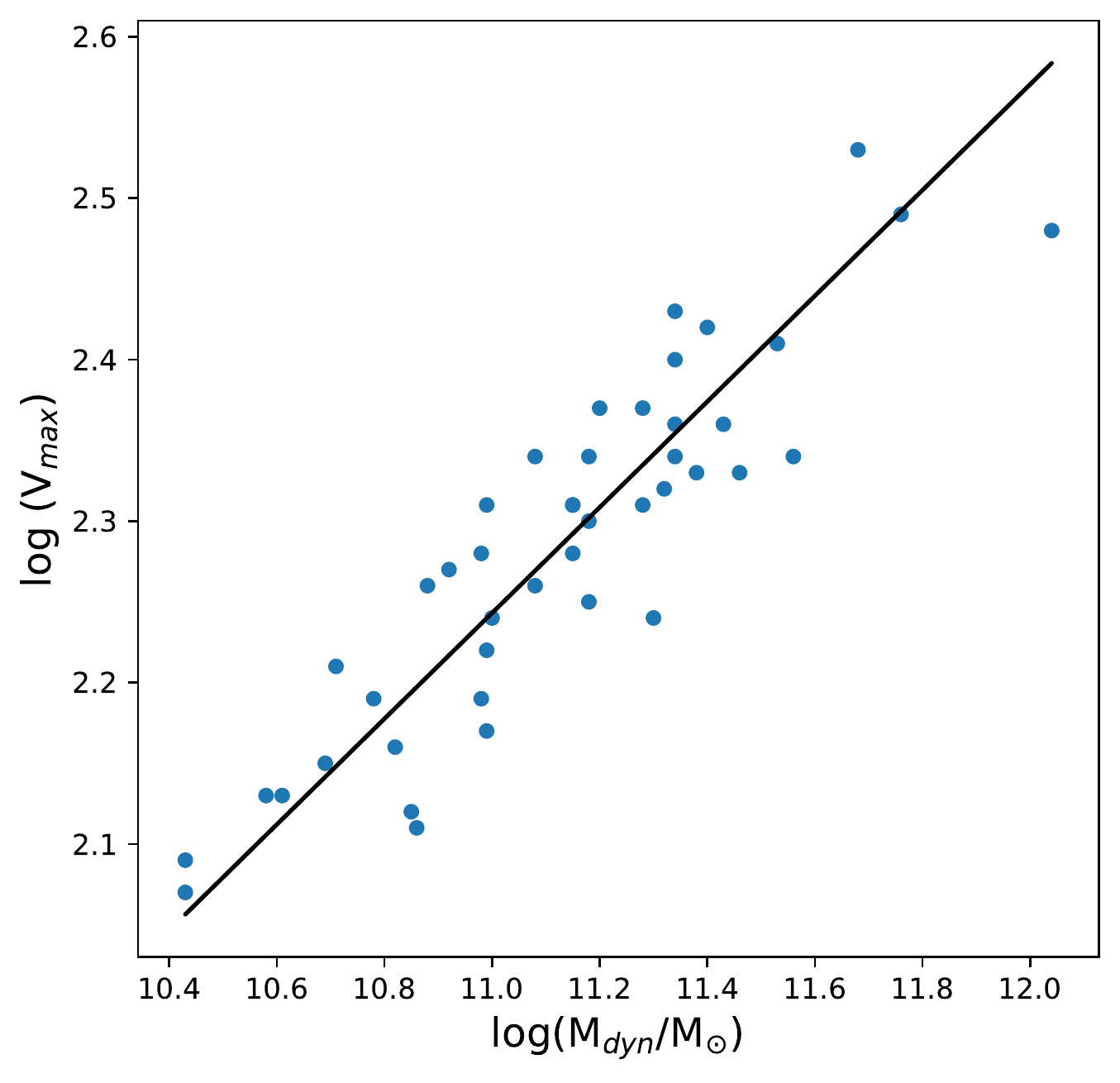}
   \caption{Relation between the dynamical mass and the maximun rotation velocity (km s$^{-1}$). The galaxies 8082-6102 and 8318-12703 are not shown in the plot and are not considered in the analysis. The black solid line represents the OLS bisector fit for our data.}
              \label{fig:Mdyn_Vmax}%
    \end{figure}

In addition, we study the correlation between dynamical mass and maximum rotation velocity, which is a well-known relationship in the study of spiral galaxies \citep[e.g.,][]{Rubin1978}. This correlation implies that as the mass of a galaxy increases, so does its maximum rotation velocity. Physically, this can be understood as a result of the balance between gravity and centrifugal forces within the galaxy \citep[e.g.,][]{Binney}. Fig. \ref{fig:Mdyn_Vmax} shows the relation between the dynamical mass and the maximun rotation velocity. This correlation has a Pearson coefficient r$_{p}=$ 0.90 and a p-value of 7.9 $\times$10$^{-17}$, indicating the close relationship between the two parameters. As in the previous case, the galaxies 8082$-$6102 and 8318$-$12703 are not shown in the plot and are not considered for this particular analysis. These objects, however, are considered for the rest of the analysis. We performed an OLS bisector fit for our data, which has a slope of 0.33 $\pm$ 0.02 and an offset of -1.36 $\pm$ 0.27.

\begin{figure}
    \centering
    \includegraphics[width=0.47\textwidth]{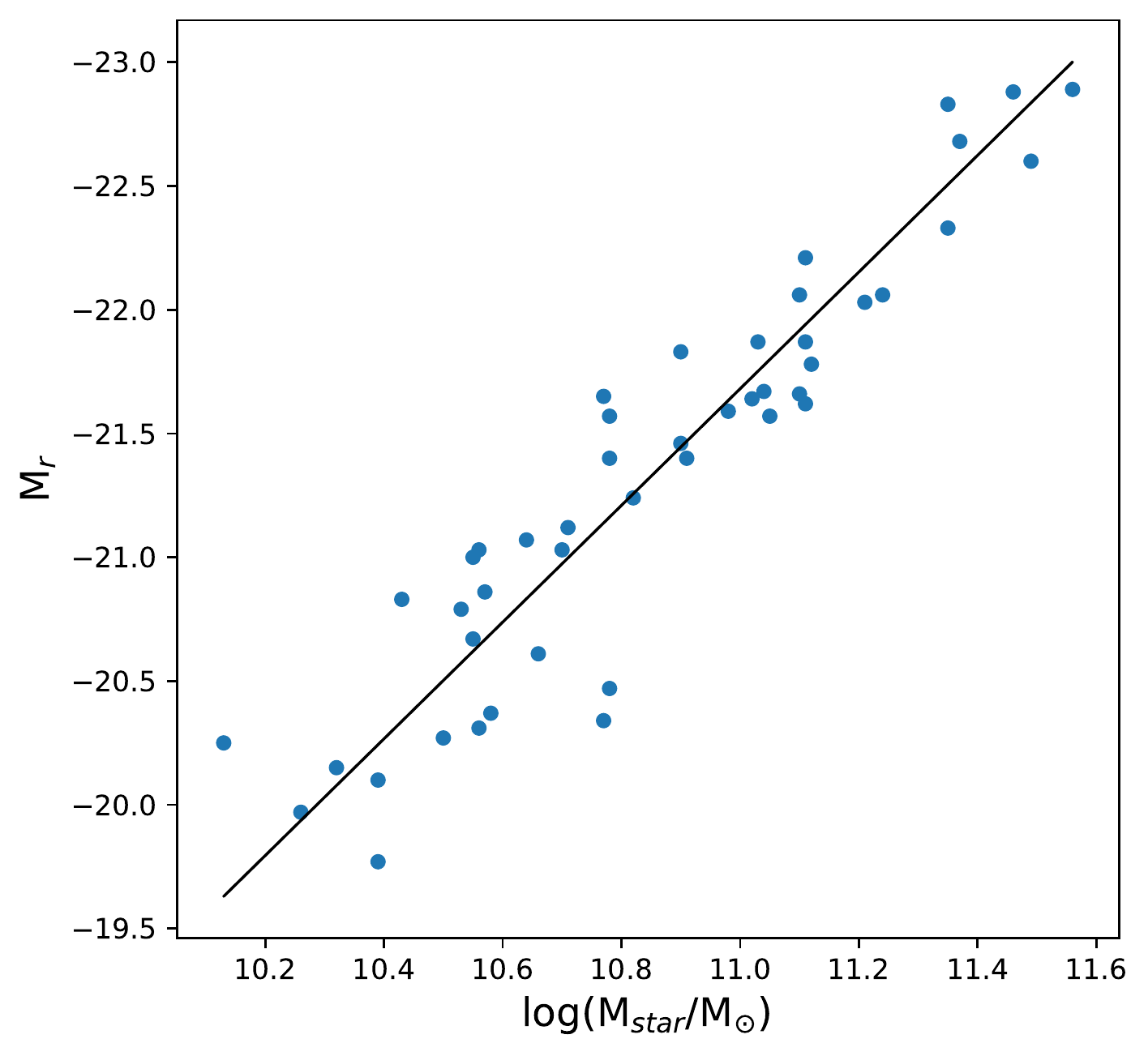}
   \caption{Relation between stellar mass and magnitude M$_{r}$. The black solid line represents the OLS bisector fit for our data. }
              \label{fig:Mstar_Mr}%
    \end{figure}

Moreover, in Fig. \ref{fig:Mstar_Mr} we show the relation between the stellar mass and the magnitude M$_{r}$. As expected, a close correlation is seen, with a Pearson coefficient r$_{p}=$ -0.93 and a p-value of 3.4 $\times$10$^{-21}$. The OLS bisector fit has a slope of -2.36 $\pm$ 0.12 and an offset of 4.2 $\pm$ 1.4. As expected, galaxies with higher stellar masses tend to have lower magnitudes, indicating that they are brighter.

\begin{figure}
    \centering
    \includegraphics[width=0.47\textwidth]{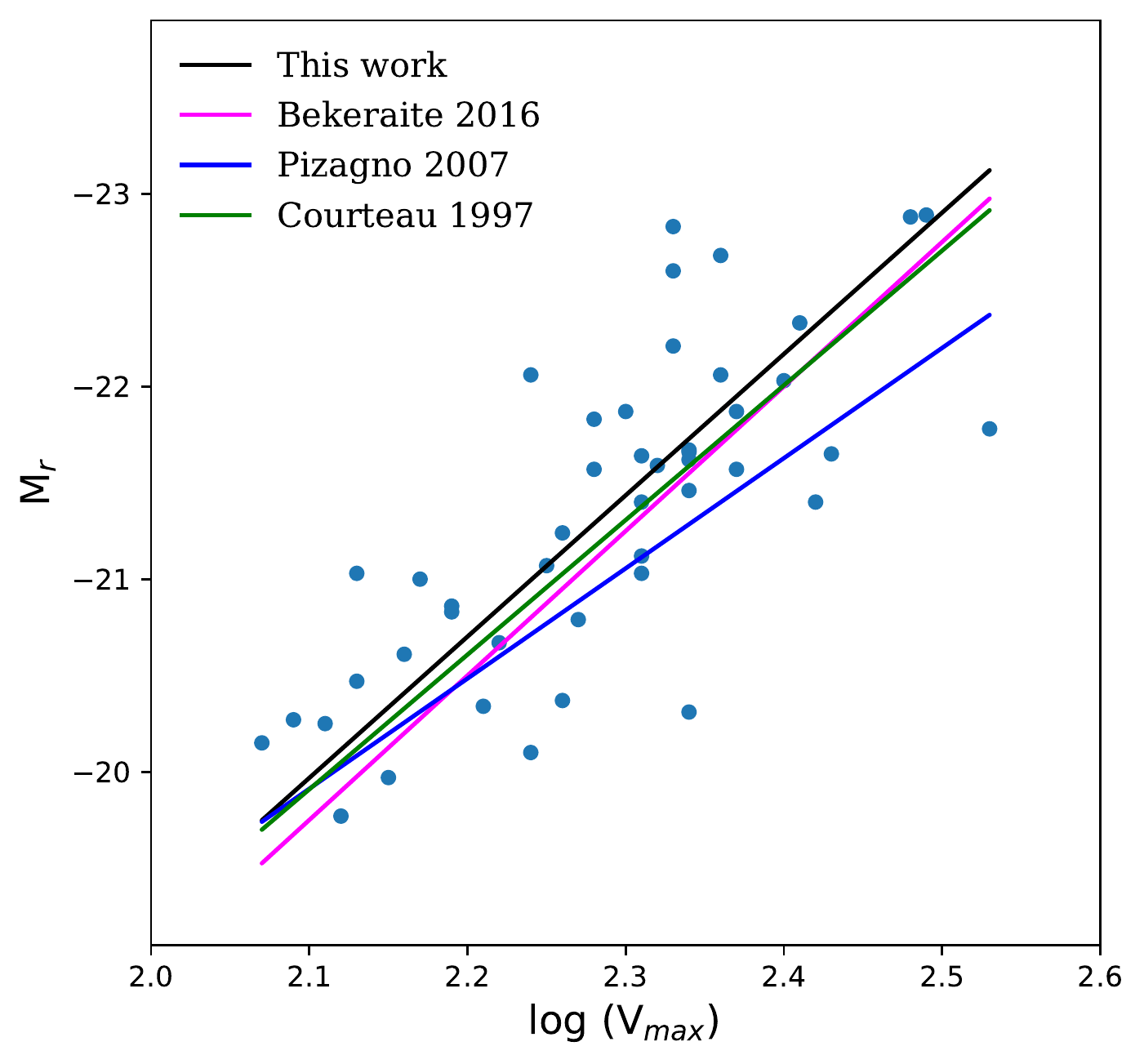}
   \caption{T-F relation between the maximun rotation velocity (km s$^{-1}$) and the magnitude M$_{r}$. The black solid line represents the OLS bisector fit for our data. For comparison we also show the \citet{Courteau1997}, \citet{Pizagno2007} and \citet{Bekeraite2016a} T-F fits.}
              \label{fig:Vmax_Mr}%
    \end{figure}

In the context of scaling relations, the Tully-Fisher (TF) relation is a powerful tool for understanding the intrinsic properties of rotationally supported galaxies \citep[][]{Tully1977,Corteau1997AJ}. While the primary use of the TF relation has been to determine distances in extragalactic studies, it also provides fundamental insights into the mechanisms of disk assembly and evolution \citep[e.g.,][]{Bekeraite2017}. In one of its best-known forms, this relation provides a link between two key properties of galaxies: their circular rotation velocities and their luminosities \citep[e.g.,][]{Pizagno2007,Bekeraite2016a}. We study the TF relation in our sample through the maximun rotation velocity and the magnitude M$_{r}$ (Fig. \ref{fig:Vmax_Mr}). Considering our data, this relation has a Pearson coefficient r$_{p}=$ -0.76 and a p-value of 1.1 $\times$10$^{-9}$. An OLS bisector fit yields a slope of -7.3 $\pm$ 0.7 and an offset of -4.6 $\pm$ 1.6. Our fitting results were compared with \cite{Courteau1997}, \cite{Pizagno2007} and \cite{Bekeraite2016a} findings. The first two studies investigated the TF relation through rotation curves using H$\alpha$ emission and considering log(v) as independent variable. On the other hand, \cite{Bekeraite2016a} explored the TF relation in a sample of galaxies from CALIFA using stellar dynamics and considering the magnitude as the independent variable. Our bisector fit's slope and offset are presented in Table \ref{tab:TF}, along with the corresponding values reported by \cite{Courteau1997}, \cite{Pizagno2007} and \cite{Bekeraite2016a} (see Table 1 of this last work). As can be seen, the slope and offset reported in this work are within the range obtained by \citet{Bekeraite2016a} and \citet{Courteau1997}. This suggests that the behaviour of the sample of barred galaxies considering this relation does not differ from that of other studies involving galaxies of different morphological types including non-barred galaxies.

\begin{table}
 \caption{Fit parameters of the Tully-Fisher relation (V$_{max}$ and magnitude M$_{r}$) and literature values.}
 \label{tab:TF}
 \begin{tabular}{lcc}
  \hline
  Work & Slope & Offset\\
  \hline
  This work & -7.3 $\pm$ 0.7 & -4.6 $\pm$ 1.6 \\
  \citet{Bekeraite2016a} & -7.5 $\pm$ 0.5 & -4.0 $\pm$ 1.0\\
  \citet{Pizagno2007} & -5.72 $\pm$ 0.19 & -7.90 $\pm$ 0.03 \\
  \citet{Courteau1997} & -6.99 $\pm$ 0.33 &  -5.23 $\pm$ 0.46\\
  \hline
 \end{tabular}
\end{table}









\begin{figure}
    \centering
    \includegraphics[width=0.47\textwidth]{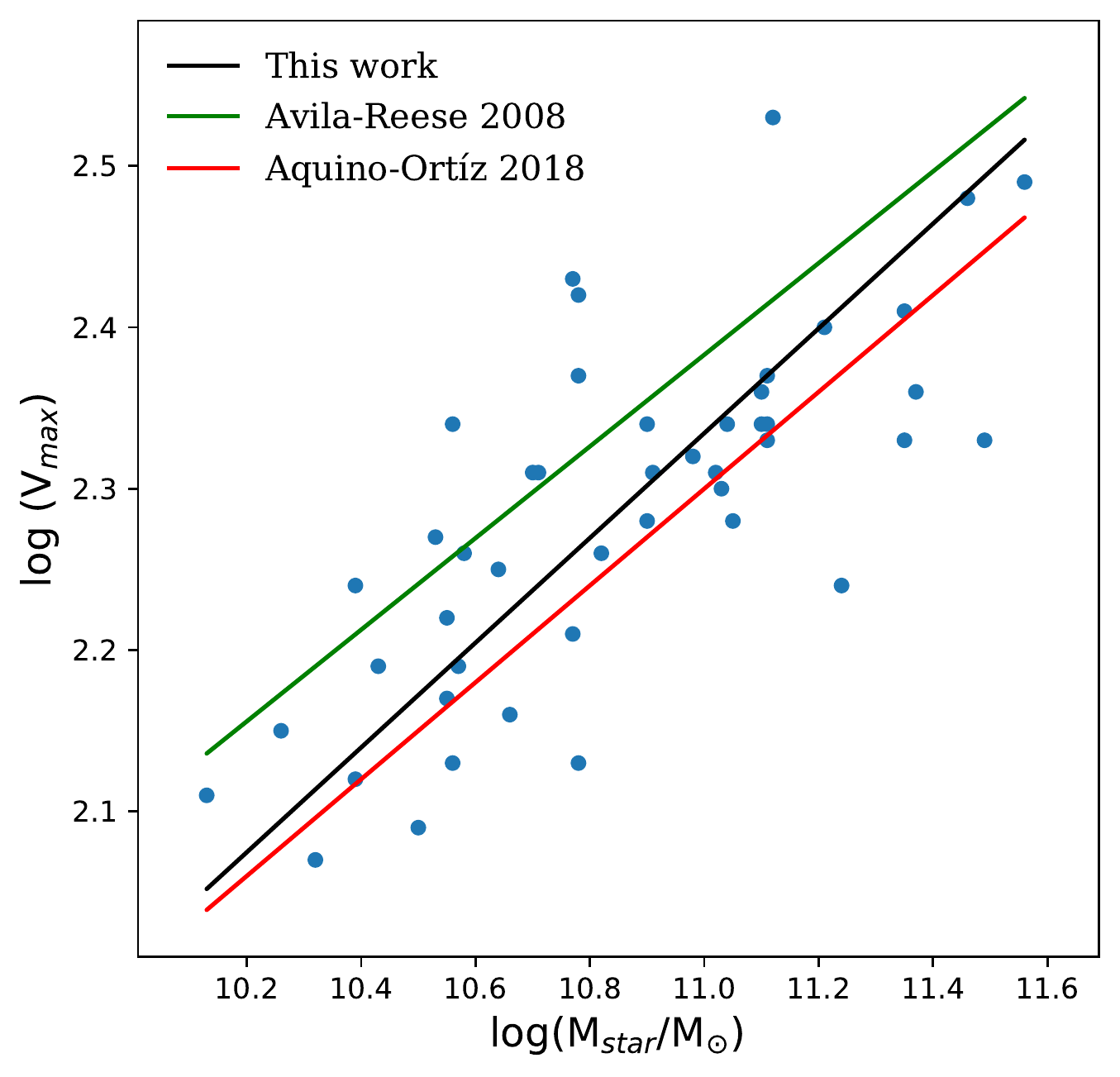}
   \caption{T-F relation between the stellar mass and the maximun rotation velocity (km s$^{-1}$). The black solid line represents the OLS bisector fit for our data. For comparison we also show the \citet{Aquino-Ortiz2018} and \citet{Avila-Reese2008} T-F fits.}
              \label{fig:mstar_vmax}%
    \end{figure}

A known alternative form of the TF relation is expressed in terms of the stellar mass and the maximum rotation velocity \citep[e.g.,][]{Avila-Reese2008}. We examine this relation in Fig. \ref{fig:mstar_vmax}, which has a Pearson coefficient r$_{p}=$ 0.74 and a p-value of 3.3 $\times$10$^{-9}$. The OLS bisector fit has a slope of 0.32 $\pm$ 0.03 and an offset of -1.2 $\pm$ 0.3. We compared our fitting results with those found by \cite{Avila-Reese2008} and \cite{Aquino-Ortiz2018}. \cite{Avila-Reese2008} study the TF relation in a compiled sample of 76 normal disk galaxies of all morphological types and performed, among others, a bisector fit (their table 1), while \cite{Aquino-Ortiz2018} study the TF relation in galaxies from CALIFA performing an orthogonal linear fit considering, among others, the stellar dynamics of only spiral galaxies and using the stellar mass as independent variable (their table 1). In Table \ref{tab:TF2} we present our values of slope and offset for our bisector fit of the TF relation.  This table also shows the parameters provided by \cite{Avila-Reese2008} and \cite{Aquino-Ortiz2018}. The slope value we obtain is higher than the values obtained by both works, however, within error it is similar to the one reported by \citet{Aquino-Ortiz2018}. In Fig. \ref{fig:mstar_vmax} it can be seen that our fit is between the fits of these two mentioned works, what indicates that the behaviour of this sample of barred galaxies does not differ from samples of varying morphologies.\\

\begin{table}
 \caption{Fit parameters of the Tully-Fisher relation (M$_{star}$/M$_{\odot}$ and V$_{max}$ in km s$^{-1}$) and literature values.}
 \label{tab:TF2}
 \begin{tabular}{lcc}
  \hline
  Work & Slope & Offset\\
  \hline
  This work & 0.32 $\pm$ 0.03 & -1.2 $\pm$ 0.3 \\
  \citet{Aquino-Ortiz2018} & 0.30 $\pm$ 0.02  & -1.00 $\pm$ 0.02\\
  \citet{Avila-Reese2008} & 0.284 $\pm$ 0.012 & -0.741 $\pm$ 0.127  \\
  \hline
 \end{tabular}
\end{table}

\section{Comparison with literature}
\label{sec:comparison}

\begin{figure}
    \centering
    \includegraphics[width=0.49\textwidth]{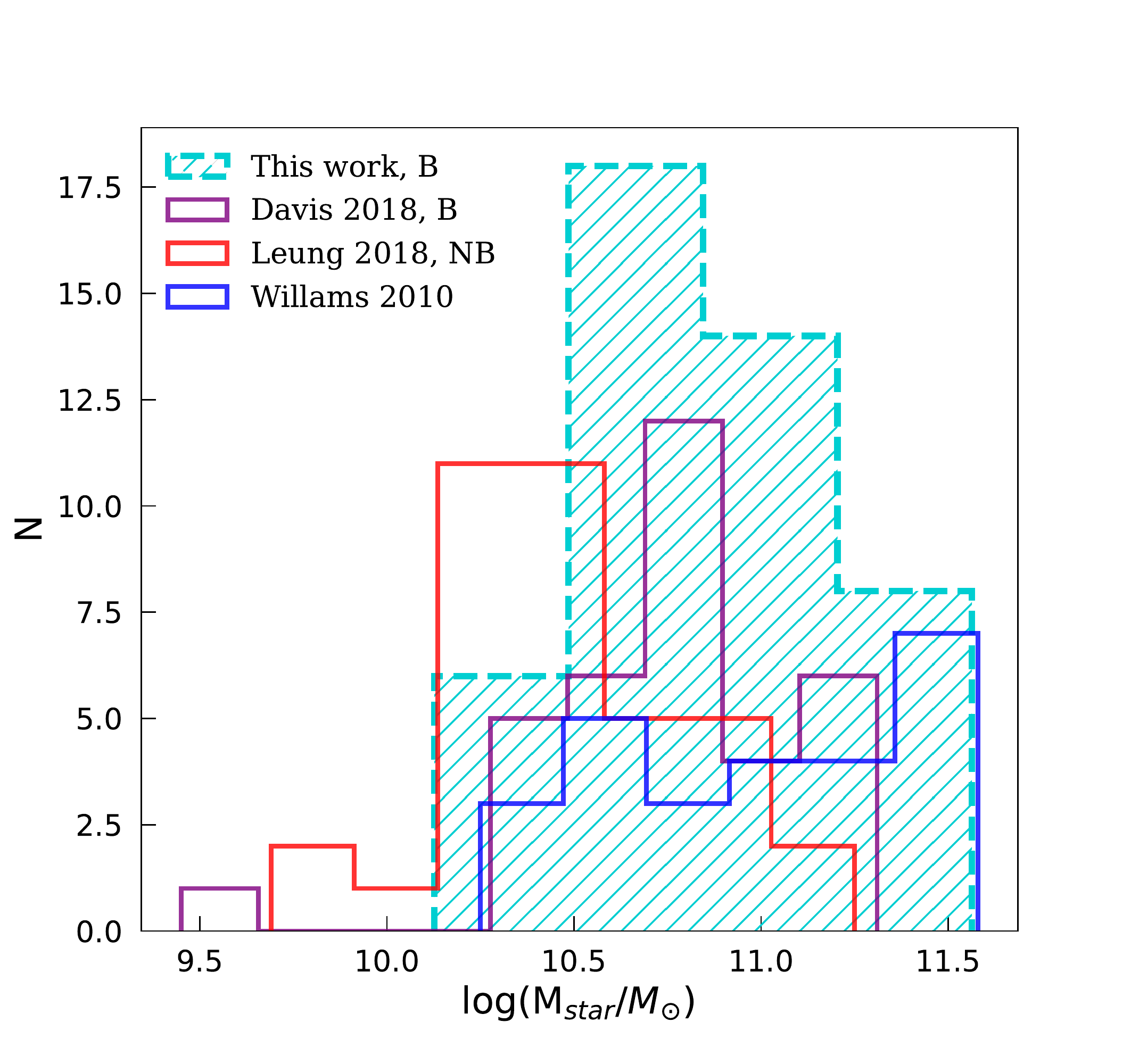}
    \includegraphics[width=0.49\textwidth]{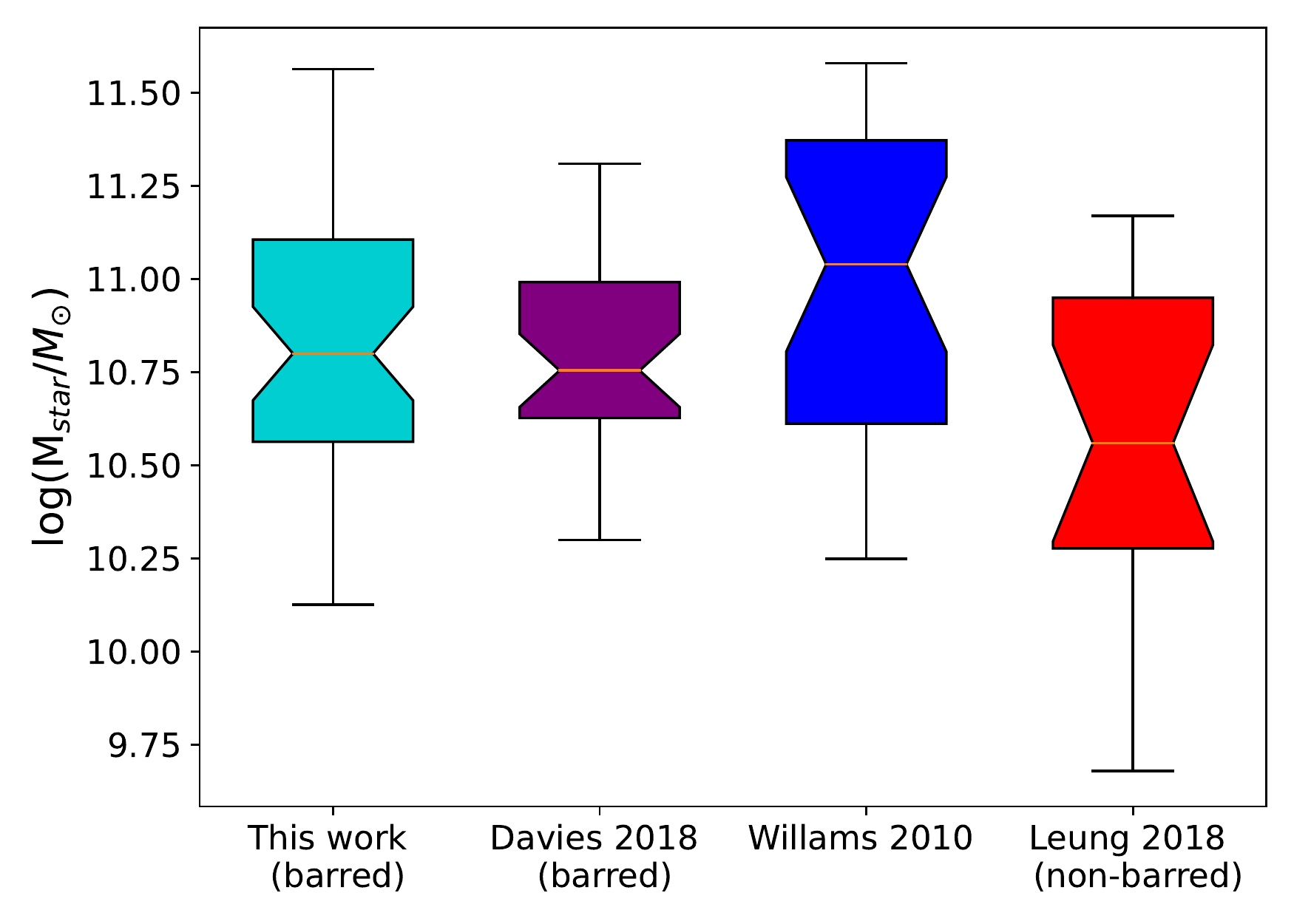}
   \caption{Comparison of distributions of stellar mass for our sample and other three spiral galaxy samples (see text for details). Top panel shows histograms for each sample with the barred and non-barred samples marked with "B" and "NB" labels. The bottom panel displays box plots with the same color scheme. The orange horizontal line within each box plot represents the median value. The height of the boxes indicate the interquartile range (IQR) for each sample, while the whiskers extending from the boxes indicate the range of non-outlier values. Additionally, the notches in the boxes provide a visual estimate of the statistical significance of differences between groups. If the notches of two boxes do not overlap, it suggests that there is a significant difference in the medians of the two groups at a 95\% confidence level.}
              \label{fig:comp_Mstar}%
    \end{figure}

This section presents a comparison of the parameters obtained for our sample of galaxies with those of other samples found in the literature. The primary objective is to examine the results in relation to other studies on barred galaxies and to investigate potential differences between barred and non-barred galaxies in terms of their stellar mass, total mass, and V$_{rot}$. By conducting this analysis, we aim to gain a deeper understanding of the properties and behavior of barred galaxies and how they relate to the broader population of galaxies.

We compared the stellar mass distribution of our sample with three additional samples of spiral galaxies by generating histograms and box plots (see Fig. \ref{fig:comp_Mstar}). Box plots are an effective tool for summarising the distribution of data based on quartiles and offer several advantages, including the ability to visualize central tendency, dispersion, and outliers of each sample, even for small samples (n $>$ 5). In addition, we have included the histograms of the samples as they provide a quick and direct visualization of their distributions. Barred and non-barred galaxies were distinguished for each sample using the morphological type reported in the NED, and we only included in this analysis sub-samples with a sufficient number of galaxies to conduct a meaningful comparison.

Of the sub-samples examined, only the sample presented in \citet{Davis2018} contained a sufficient number of barred galaxies for a valid comparison. The medians of our sample and the \citet{Davis2018} sample are statistically similar, as shown in the bottom panel of Fig. \ref{fig:comp_Mstar}, taking into account the uncertainties represented by the notches of the boxes. Furthermore, we used the Kolmogorov-Smirnov (KS) test to compare the underlying distributions of the two samples and obtained a statistic of 0.2123 and a p-value of 0.2895. As the p-value is significantly higher than the limit value of 0.05, there is no strong evidence to reject the null hypothesis. Therefore, based on the KS test, it is likely that the two samples were drawn from the same underlying distribution suggesting that our sample and the sample of \cite{Davis2018} may follow a similar stellar mass distribution.

The non-barred sub-sample we used to compare our results came from \cite{Leung2018} where the stellar masses were measured for the EDGE-CALIFA survey \citep{Bolatto2017}, which mainly includes face-on galaxies. The box plot in Fig. \ref{fig:comp_Mstar} reveals some overlapping between our sample and the sample from \citet{Leung2018} but the medians of both samples are clearly different. The KS test yielded a statistic of 0.3849 and a p-value of 0.0008 for the comparison between them, indicating that the two samples likely have different underlying distributions. This suggests that there could be a statistically significant difference in the stellar mass distributions of barred and non-barred galaxies.

Finally, we compared our stellar masses with a third sample of edge-on spiral galaxies compiled by \cite{Williams2010}. We have selected only the non-barred galaxies based on their classification in the NED. However, we note that despite the NED classification, \citet{Williams2010} report that 75\% of the galaxies in this sample exhibit boxy structures that may be interpreted as bars viewed side-on. Furthermore, given that all galaxies in this sample are edge-on and that bars in such galaxies may be harder to observe, it is likely that this sample is strongly contaminated with barred galaxies. As such, care must be taken when comparing our results to this sample.
The box plot in Fig. \ref{fig:comp_Mstar} shows a significant difference between the median of the stellar mass for this sample and ours and even more with respect to the non-barred sample from \cite{Leung2018}. For the comparison with our sample the KS test yielded similar results as the obtained when comparing with the barred sample of \citet{Davis2018}: a statistic of 0.2876 and a p-value=0.1016, implicating that we can not rule out the null hypothesis where both samples, ours and the one from \citet{Williams2010}, have been drawn from the same stellar mass distribution. This reinforces the idea that the galaxies in \citet{Williams2010} sample could be mainly barred.

\begin{figure}
    \centering
    \includegraphics[width=0.49\textwidth]{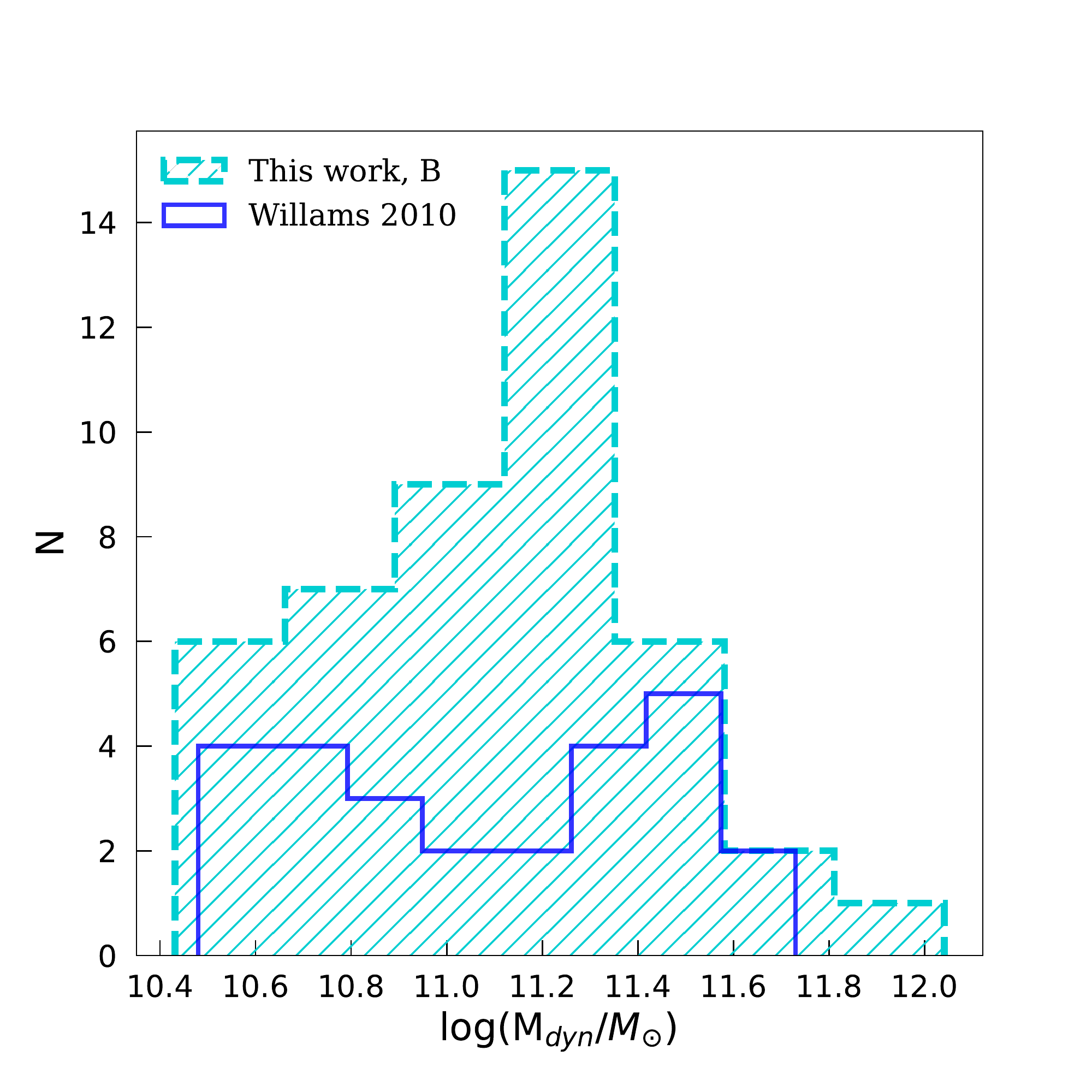}
    \includegraphics[width=0.49\textwidth]{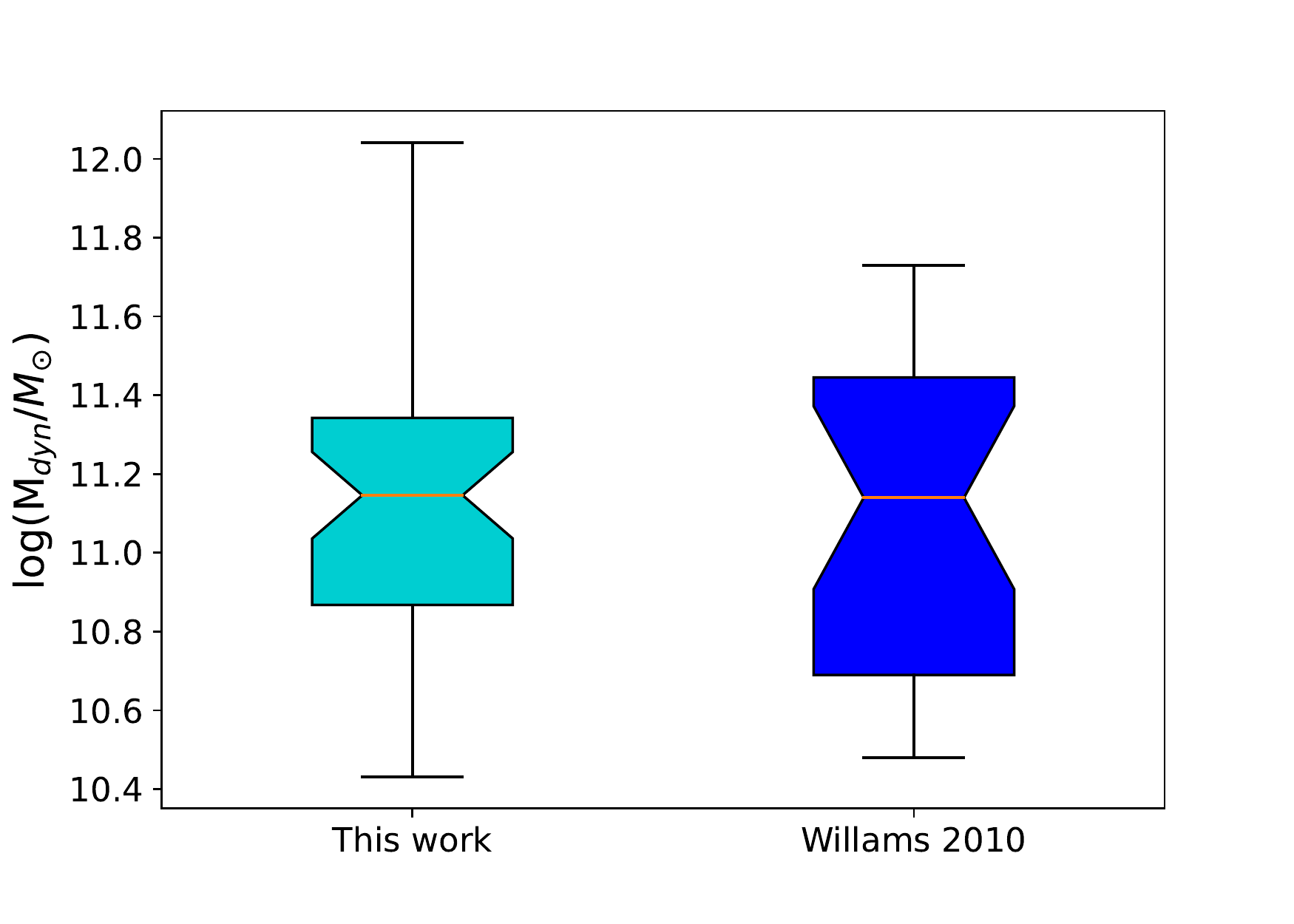}
   \caption{Same as Fig. \ref{fig:comp_Mstar} but for the total dynamical mass.}
              \label{fig:comp_Mtot}%
    \end{figure} 

Regarding the total dynamical mass we found that \cite{Williams2010} used the $K_s$ total magnitude to estimate the total mass of the galaxies in their sample. 
To evaluate the agreement between their result and our own, we present a comparison in Fig. \ref{fig:comp_Mtot}, where we used histograms and box plots to illustrate the distributions of both samples. Despite the relatively small sample size in \cite{Williams2010}, our analysis indicates significant agreement between the two datasets. Results from a KS test yield a statistic of 0.1572 and a corresponding p-value of 0.7364, indicating that the two distributions are statistically indistinguishable within the limits of these samples.

\begin{figure}
    \centering
    \includegraphics[width=0.49\textwidth]{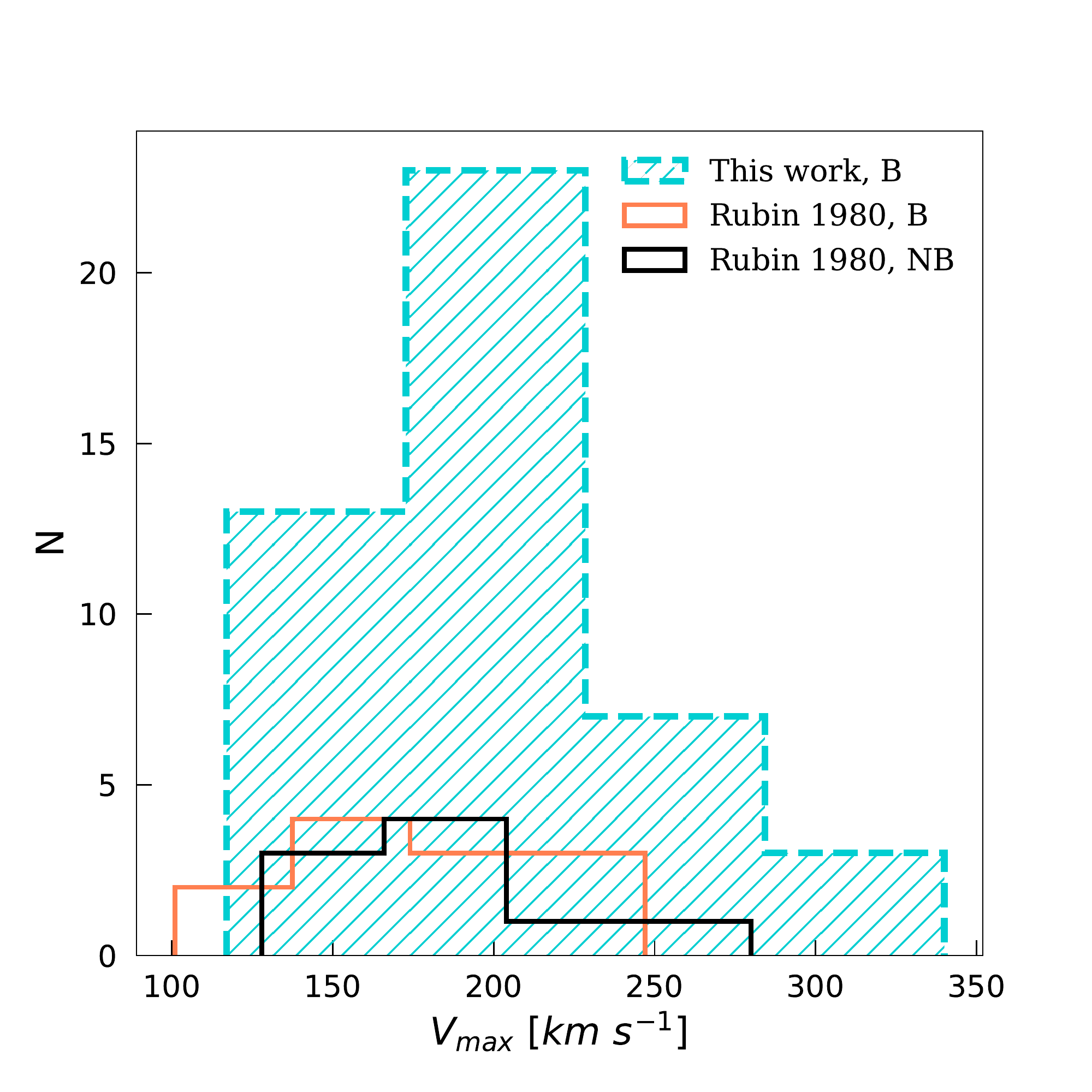}
    \includegraphics[width=0.49\textwidth]{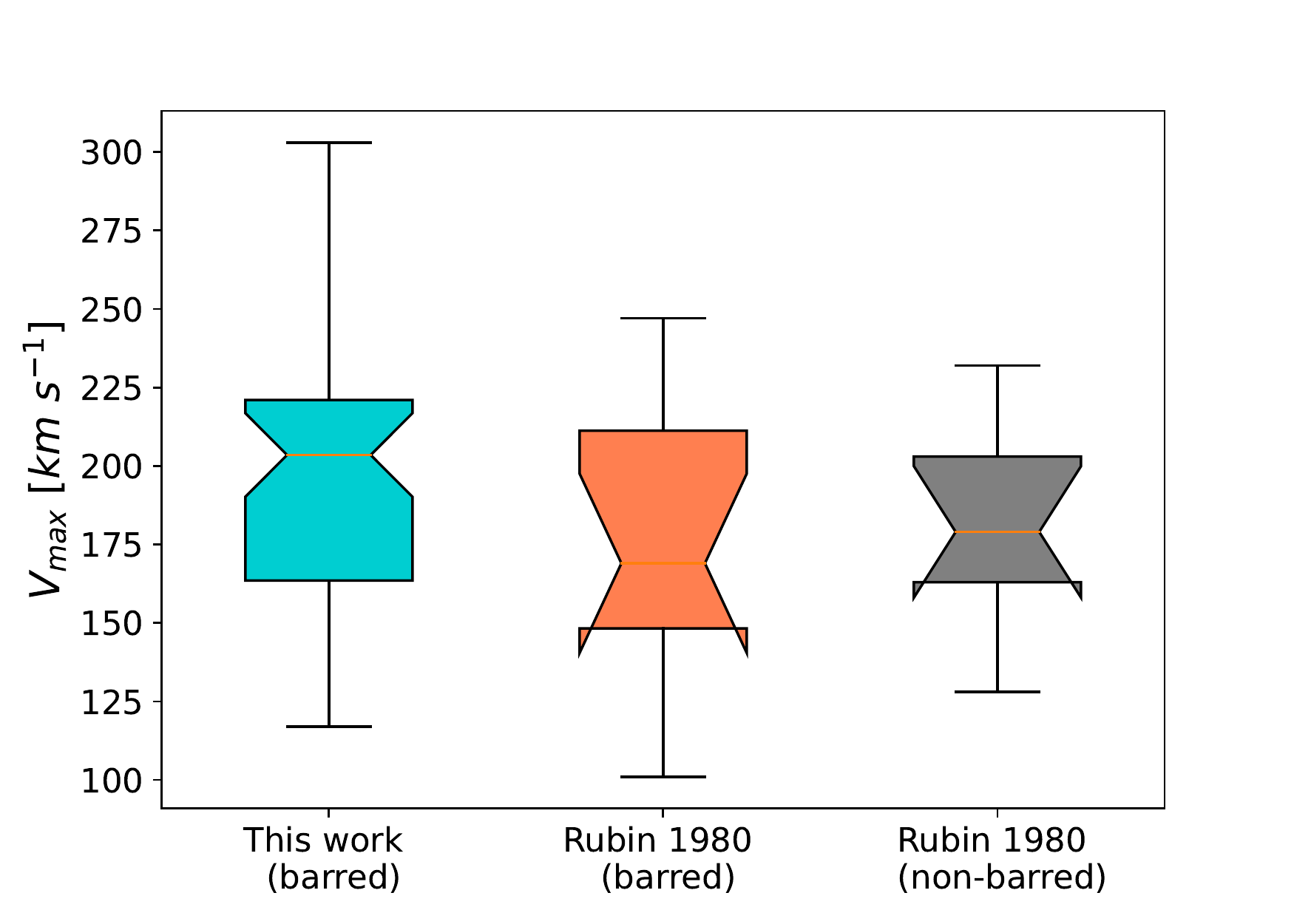}
   \caption{Same as Fig. \ref{fig:comp_Mstar} but for the maximum velocity.}
              \label{fig:comp_vrot}%
    \end{figure}

In order to compare the $V_{max}$ values of our sample with those of other studies, we used a sample of spiral galaxies presented by \cite{Rubin1980}. We employed the NED to classify the galaxies into barred and non-barred subsamples, which resulted in 12 and 9 galaxies, respectively. Although the sample size is relatively low, we conducted a statistical analysis to investigate the differences between the samples. The histograms and box-plots depicted in Fig. \ref{fig:comp_vrot} suggest that the three samples are statistically indistinguishable.

Furthermore, we implemented the Kolmogorov-Smirnov test to evaluate the significance of the similarities and differences between the samples. The test yielded a statistic of 0.2608 and a p-value of 0.4612 for the comparison between our sample and the barred sample of \citet{Rubin1980}. Additionally, we obtained a statistic of 0.2777 and a p-value of 0.5121 for the comparison with their non-barred sample. In both cases, the null hypothesis of the samples corresponding to the same distribution cannot be rejected. Thus, based on the KS test, we can conclude that the two samples are statistically similar to ours in their distribution, even though caution should be exercised due to the low number of galaxies in each of the sub-samples from \citet{Rubin1980}. A larger sample of non-barred galaxies is necessary in order to study if bars have some impact in the $V_{max}$ of galaxies.

\section{Final remarks}
\label{sec:final_remarks}

We have obtained and presented the rotation curves of 46 barred galaxies of MaNGA (Fig. \ref{fig:curves}). By intrinsic quality of the individual curves, size, and homogeneity, this sample constitutes one of the best samples of RCs available for barred galaxies to date. As such, it will offer a very good opportunity for investigating and analyzing the dynamical properties of this kind of objects.\\
We have fitted all the RCs considering two components of an axisymmetric Miyamoto–Nagai gravitational potential, the parameters of which are presented in Table \ref{table:params}. Our main results can be summarised as follows:

\begin{itemize}
    \item We have found a wide range of rotation velocities (117 $-$ 340  $\kms$), with most galaxies showing values in the range 120 $-$ 250  $\kms$. These velocities are consistent with previous results in barred galaxies and no difference have been found with a sample of non-barred galaxies. Nevertheless a larger sample of galaxies is necessary to confirm this trend.
    \vspace{0.5cm}
    
    \item The total dynamical masses of our sample lie in the range of log(M$_{dyn}$/M$_{\odot}$)$=$ 10.1$-$12.0, with a mean value of log(M$_{dyn}$/M$_{\odot}$)$=$ 11.0. These values are in agreement with previous results considering total masses of spiral galaxies. According to our results, the distribution has a standard deviation of 0.39 and an IQR of 0.55.
    \vspace{0.5cm}
    
    \item Stellar masses of our sample of barred galaxies are in the range log(M$_{star}$/M$_{\odot}$)$=$ 10.13$-$11.53 with a standard deviation of 0.35 and an IQR of 0.54. The comparison with other sample of barred galaxies yielded no statistical difference while comparison with a non-barred galaxies sample yielded that both distributions are statistically different. This difference between the stellar mass of barred and non-barred galaxies could imply that the bar plays a role in the assembly of stellar mass of galaxies.
    \vspace{0.5cm}
    
    \item The behaviour of our sample in the scaling relations was investigated. We found a Pearson coefficient r$_p$ = 0.70 and a p-value of 1.4 x 10$^{-7}$ for the relation between M$_{dyn}$ and M$_{star}$ while for the relation between M$_{dyn}$ and V$_{max}$ the statistic yielded r$_p$= 0.9 and a p-value of 7.9 x 10$^{-17}$. We also explored the Tully-Fisher relation in two ways: by considering the relationship between V$_{max}$ and M$_r$ (r$_p$= -0.76 and a p-value of 1.1 x 10$^{-9}$), and separately by examining V$_{max}$ and M$_{star}$ (r$_p$= 0.74 and a p-value of 3.3 x 10$^{-9}$). Our analysis revealed that the Tully-Fisher relation in barred galaxies is consistent with reported relations in the literature for samples of varying morphologies, indicating that barred galaxies do not deviate from this scaling relation commonly observed in other galaxy types.

\end{itemize}

Summing up, we studied the dynamical masses, stellar masses and maximun rotation velocities of barred galaxies, investigating their distributions and scaling relations and comparing them with other studies. Our results suggest that stellar masses could be different in barred galaxies compared with non-barred galaxies. The dynamical mass and maximun rotation velocities as well as scalling relations such as the Tully-Fisher relation show no difference with other samples involving galaxies with different morphological types including non-barred galaxies.

It has to be noted the importance of knowing the dynamical masses of galaxies as they consider the contribution of all constituent elements to the gravitational potential. To our best knowledge, this is the first time that the dynamical masses of these galaxies have been determined in a direct and independent way through their rotation curves. We hope that the work presented here will constitute a useful tool to increase our understanding of the dynamical behaviour of barred spiral galaxies.

\section*{Acknowledgements}
We want to thank our referee for constructive comments and suggestions that improved this work. We are also grateful to Rubén Díaz for fruitful discussions. This work was partially supported by grant PICT 2017-3301 awarded by Fondo para la Investigaci\'on Científica y Tecnol\'ogica (FonCyT). G.G is a postdoctoral fellow of Consejo de Investigaciones Científicas y T\'ecnicas (CONICET), Argentina. W.W., E.S., and D.M. are members of the Carrera del Investigador Científico of CONICET, Argentina.  
This research has made use of the NASA/IPAC Extragalactic Database (NED) which is operated by the Jet Propulsion Laboratory, California Institute of Technology, under contract with the National Aeronautics and Space Administration. We made an extensive use of the following Python libraries and packages: \href{https://www.galpy.org/}{galpy}, \href{https://pandas.pydata.org/}{pandas}, \href{http://www.matplotlib.org/}{Matplotlib}, \href{http://www.numpy.org/}{NumPy} and \href{http://www.scipy.org/}{SciPy}. This project makes use of the MaNGA-Pipe3D dataproducts. We thank the IA-UNAM MaNGA team for creating this catalogue, and the ConaCyt-180125 project for supporting them.
Funding for the Sloan Digital Sky Survey IV has been provided by the Alfred P. Sloan Foundation, the U.S. Department of Energy Office of Science, and the Participating Institutions. SDSS acknowledges support and resources from the Center for High-Performance Computing at the University of Utah. The SDSS web site is \url{www.sdss.org}.
SDSS is managed by the Astrophysical Research Consortium for the Participating Institutions of the SDSS Collaboration including the Brazilian Participation Group, the Carnegie Institution for Science, Carnegie Mellon University, Center for Astrophysics | Harvard \& Smithsonian (CfA), the Chilean Participation Group, the French Participation Group, Instituto de Astrofísica de Canarias, The Johns Hopkins University, Kavli Institute for the Physics and Mathematics of the Universe (IPMU) / University of Tokyo, the Korean Participation Group, Lawrence Berkeley National Laboratory, Leibniz Institut für Astrophysik Potsdam (AIP), Max-Planck-Institut für Astronomie (MPIA Heidelberg), Max-Planck-Institut für Astrophysik (MPA Garching), Max-Planck-Institut für Extraterrestrische Physik (MPE), National Astronomical Observatories of China, New Mexico State University, New York University, University of Notre Dame, Observatório Nacional / MCTI, The Ohio State University, Pennsylvania State University, Shanghai Astronomical Observatory, United Kingdom Participation Group, Universidad Nacional Autónoma de México, University of Arizona, University of Colorado Boulder, University of Oxford, University of Portsmouth, University of Utah, University of Virginia, University of Washington, University of Wisconsin, Vanderbilt University, and Yale University.

\section*{Data Availability}

All data used and analysed in this article are available in \url{www.sdss.org/dr17/}



\bibliographystyle{mnras}
\bibliography{Bibliography} 

\begin{thebibliography}{}
\makeatletter
\relax
\def\mn@urlcharsother{\let\do\@makeother \do\$\do\&\do\#\do\^\do\_\do\%\do\~}
\def\mn@doi{\begingroup\mn@urlcharsother \@ifnextchar [ {\mn@doi@}
  {\mn@doi@[]}}
\def\mn@doi@[#1]#2{\def\@tempa{#1}\ifx\@tempa\@empty \href
  {http://dx.doi.org/#2} {doi:#2}\else \href {http://dx.doi.org/#2} {#1}\fi
  \endgroup}
\def\mn@eprint#1#2{\mn@eprint@#1:#2::\@nil}
\def\mn@eprint@arXiv#1{\href {http://arxiv.org/abs/#1} {{\tt arXiv:#1}}}
\def\mn@eprint@dblp#1{\href {http://dblp.uni-trier.de/rec/bibtex/#1.xml}
  {dblp:#1}}
\def\mn@eprint@#1:#2:#3:#4\@nil{\def\@tempa {#1}\def\@tempb {#2}\def\@tempc
  {#3}\ifx \@tempc \@empty \let \@tempc \@tempb \let \@tempb \@tempa \fi \ifx
  \@tempb \@empty \def\@tempb {arXiv}\fi \@ifundefined
  {mn@eprint@\@tempb}{\@tempb:\@tempc}{\expandafter \expandafter \csname
  mn@eprint@\@tempb\endcsname \expandafter{\@tempc}}}

\bibitem[\protect\citeauthoryear{{Adams}, {Gebhardt}, {Blanc}, {Fabricius},
  {Hill}, {Murphy}, {van den Bosch}  \& {van de Ven}}{{Adams}
  et~al.}{2012}]{Adams2012}
{Adams} J.~J.,  {Gebhardt} K.,  {Blanc} G.~A.,  {Fabricius} M.~H.,  {Hill}
  G.~J.,  {Murphy} J.~D.,  {van den Bosch} R. C.~E.,   {van de Ven} G.,  2012,
  \mn@doi [\apj] {10.1088/0004-637X/745/1/92}, \href
  {https://ui.adsabs.harvard.edu/abs/2012ApJ...745...92A} {745, 92}

\bibitem[\protect\citeauthoryear{{Ag{\"u}ero}, {D{\'\i}az}  \&
  {Bajaja}}{{Ag{\"u}ero} et~al.}{2004}]{Aguero2004}
{Ag{\"u}ero} E.~L.,  {D{\'\i}az} R.~J.,   {Bajaja} E.,  2004, \mn@doi [\aap]
  {10.1051/0004-6361:20031644}, \href
  {https://ui.adsabs.harvard.edu/abs/2004A&A...414..453A} {414, 453}

\bibitem[\protect\citeauthoryear{{Aguerri}, {M{\'e}ndez-Abreu}  \&
  {Corsini}}{{Aguerri} et~al.}{2009}]{Aguerri2009}
{Aguerri} J.~A.~L.,  {M{\'e}ndez-Abreu} J.,   {Corsini} E.~M.,  2009, \mn@doi
  [\aap] {10.1051/0004-6361:200810931}, \href
  {https://ui.adsabs.harvard.edu/abs/2009A&A...495..491A} {495, 491}

\bibitem[\protect\citeauthoryear{{Ahn} et~al.,}{{Ahn} et~al.}{2012}]{Ahn2012}
{Ahn} C.~P.,  et~al., 2012, \mn@doi [\apjs] {10.1088/0067-0049/203/2/21}, \href
  {https://ui.adsabs.harvard.edu/abs/2012ApJS..203...21A} {203, 21}

\bibitem[\protect\citeauthoryear{{Albareti} et~al.,}{{Albareti}
  et~al.}{2017}]{Albareti2017}
{Albareti} F.~D.,  et~al., 2017, \mn@doi [\apjs] {10.3847/1538-4365/aa8992},
  \href {https://ui.adsabs.harvard.edu/abs/2017ApJS..233...25A} {233, 25}

\bibitem[\protect\citeauthoryear{{Aquino-Ort{\'\i}z}
  et~al.,}{{Aquino-Ort{\'\i}z} et~al.}{2018}]{Aquino-Ortiz2018}
{Aquino-Ort{\'\i}z} E.,  et~al., 2018, \mn@doi [\mnras]
  {10.1093/mnras/sty1522}, \href
  {https://ui.adsabs.harvard.edu/abs/2018MNRAS.479.2133A} {479, 2133}

\bibitem[\protect\citeauthoryear{{Aquino-Ort{\'\i}z}
  et~al.,}{{Aquino-Ort{\'\i}z} et~al.}{2020}]{Aquino-Ortiz2020}
{Aquino-Ort{\'\i}z} E.,  et~al., 2020, \mn@doi [\apj]
  {10.3847/1538-4357/aba94e}, \href
  {https://ui.adsabs.harvard.edu/abs/2020ApJ...900..109A} {900, 109}

\bibitem[\protect\citeauthoryear{{Ashman}}{{Ashman}}{1992}]{Ashman1992}
{Ashman} K.~M.,  1992, \mn@doi [\pasp] {10.1086/133099}, \href
  {https://ui.adsabs.harvard.edu/abs/1992PASP..104.1109A} {104, 1109}

\bibitem[\protect\citeauthoryear{{Athanassoula}}{{Athanassoula}}{1992}]{Athanassoula1992}
{Athanassoula} E.,  1992, \mn@doi [\mnras] {10.1093/mnras/259.2.328}, \href
  {https://ui.adsabs.harvard.edu/abs/1992MNRAS.259..328A} {259, 328}

\bibitem[\protect\citeauthoryear{{Avila-Reese}, {Zavala}, {Firmani}  \&
  {Hern{\'a}ndez-Toledo}}{{Avila-Reese} et~al.}{2008}]{Avila-Reese2008}
{Avila-Reese} V.,  {Zavala} J.,  {Firmani} C.,   {Hern{\'a}ndez-Toledo} H.~M.,
  2008, \mn@doi [\aj] {10.1088/0004-6256/136/3/1340}, \href
  {https://ui.adsabs.harvard.edu/abs/2008AJ....136.1340A} {136, 1340}

\bibitem[\protect\citeauthoryear{{Barrera-Ballesteros}
  et~al.,}{{Barrera-Ballesteros} et~al.}{2014}]{Barrera-Ballesteros2014AA}
{Barrera-Ballesteros} J.~K.,  et~al., 2014, \mn@doi [\aap]
  {10.1051/0004-6361/201423488}, \href
  {https://ui.adsabs.harvard.edu/abs/2014A&A...568A..70B} {568, A70}

\bibitem[\protect\citeauthoryear{{Bekerait{\.{e}}}}{{Bekerait{\.{e}}}}{2017}]{Bekeraite2017}
{Bekerait{\.{e}}} S.,  2017, PhD thesis, University of Potsdam, Germany

\bibitem[\protect\citeauthoryear{{Bekerait{\'e}} et~al.,}{{Bekerait{\'e}}
  et~al.}{2016a}]{Bekeraite2016a}
{Bekerait{\'e}} S.,  et~al., 2016a, \mn@doi [\aap]
  {10.1051/0004-6361/201527405}, \href
  {https://ui.adsabs.harvard.edu/abs/2016A&A...593A.114B} {593, A114}

\bibitem[\protect\citeauthoryear{{Bekerait{\'e}} et~al.,}{{Bekerait{\'e}}
  et~al.}{2016b}]{Bekeraite2016b}
{Bekerait{\'e}} S.,  et~al., 2016b, \mn@doi [\apjl]
  {10.3847/2041-8205/827/2/L36}, \href
  {https://ui.adsabs.harvard.edu/abs/2016ApJ...827L..36B} {827, L36}

\bibitem[\protect\citeauthoryear{{Bing} et~al.,}{{Bing}
  et~al.}{2019}]{Bing2019}
{Bing} L.,  et~al., 2019, \mn@doi [\mnras] {10.1093/mnras/sty2662}, \href
  {https://ui.adsabs.harvard.edu/abs/2019MNRAS.482..194B} {482, 194}

\bibitem[\protect\citeauthoryear{{Binney} \& {Tremaine}}{{Binney} \&
  {Tremaine}}{1987}]{Binney}
{Binney} J.,  {Tremaine} S.,  1987, {Galactic dynamics}

\bibitem[\protect\citeauthoryear{{Binney}, {Gerhard}, {Stark}, {Bally}  \&
  {Uchida}}{{Binney} et~al.}{1991}]{Binney1991}
{Binney} J.,  {Gerhard} O.~E.,  {Stark} A.~A.,  {Bally} J.,   {Uchida} K.~I.,
  1991, \mn@doi [\mnras] {10.1093/mnras/252.2.210}, \href
  {https://ui.adsabs.harvard.edu/abs/1991MNRAS.252..210B} {252, 210}

\bibitem[\protect\citeauthoryear{{Bolatto} et~al.,}{{Bolatto}
  et~al.}{2017}]{Bolatto2017}
{Bolatto} A.~D.,  et~al., 2017, \mn@doi [\apj] {10.3847/1538-4357/aa86aa},
  \href {https://ui.adsabs.harvard.edu/abs/2017ApJ...846..159B} {846, 159}

\bibitem[\protect\citeauthoryear{{Bosma}}{{Bosma}}{1996}]{Bosma1996}
{Bosma} A.,  1996, in {Buta} R.,  {Crocker} D.~A.,   {Elmegreen} B.~G.,  eds,
  Astronomical Society of the Pacific Conference Series Vol. 91, IAU Colloq.
  157: Barred Galaxies. p.~132

\bibitem[\protect\citeauthoryear{{Bovy}}{{Bovy}}{2015}]{Bovy2015}
{Bovy} J.,  2015, \mn@doi [\apjs] {10.1088/0067-0049/216/2/29}, \href
  {https://ui.adsabs.harvard.edu/abs/2015ApJS..216...29B} {216, 29}

\bibitem[\protect\citeauthoryear{{Bundy} et~al.,}{{Bundy}
  et~al.}{2015}]{Bundy2015}
{Bundy} K.,  et~al., 2015, \mn@doi [\apj] {10.1088/0004-637X/798/1/7}, \href
  {https://ui.adsabs.harvard.edu/abs/2015ApJ...798....7B} {798, 7}

\bibitem[\protect\citeauthoryear{{Burbidge}, {Burbidge}  \&
  {Prendergast}}{{Burbidge} et~al.}{1959}]{Burbidge1959}
{Burbidge} E.~M.,  {Burbidge} G.~R.,   {Prendergast} K.~H.,  1959, \mn@doi
  [\apj] {10.1086/146765}, \href
  {https://ui.adsabs.harvard.edu/abs/1959ApJ...130..739B} {130, 739}

\bibitem[\protect\citeauthoryear{{Burbidge}, {Burbidge}  \&
  {Prendergast}}{{Burbidge} et~al.}{1961}]{Burbidge1961}
{Burbidge} E.~M.,  {Burbidge} G.~R.,   {Prendergast} K.~H.,  1961, \mn@doi
  [\apj] {10.1086/147215}, \href
  {https://ui.adsabs.harvard.edu/abs/1961ApJ...134..874B} {134, 874}

\bibitem[\protect\citeauthoryear{{Bureau} \& {Carignan}}{{Bureau} \&
  {Carignan}}{2002}]{Bureau2002}
{Bureau} M.,  {Carignan} C.,  2002, \mn@doi [\aj] {10.1086/338899}, \href
  {https://ui.adsabs.harvard.edu/abs/2002AJ....123.1316B} {123, 1316}

\bibitem[\protect\citeauthoryear{{Burton} \& {Liszt}}{{Burton} \&
  {Liszt}}{1993}]{Burton1993}
{Burton} W.~B.,  {Liszt} H.~S.,  1993, \aap, \href
  {https://ui.adsabs.harvard.edu/abs/1993A&A...274..765B} {274, 765}

\bibitem[\protect\citeauthoryear{{Cappellari}}{{Cappellari}}{2017}]{Cappellari2017MNRAS}
{Cappellari} M.,  2017, \mn@doi [\mnras] {10.1093/mnras/stw3020}, \href
  {https://ui.adsabs.harvard.edu/abs/2017MNRAS.466..798C} {466, 798}

\bibitem[\protect\citeauthoryear{{Cappellari} \& {Emsellem}}{{Cappellari} \&
  {Emsellem}}{2004}]{Cappellari2004PASP}
{Cappellari} M.,  {Emsellem} E.,  2004, \mn@doi [\pasp] {10.1086/381875}, \href
  {https://ui.adsabs.harvard.edu/abs/2004PASP..116..138C} {116, 138}

\bibitem[\protect\citeauthoryear{{Cherinka} et~al.,}{{Cherinka}
  et~al.}{2019}]{Marvinsoft2019}
{Cherinka} B.,  et~al., 2019, \mn@doi [\aj] {10.3847/1538-3881/ab2634}, \href
  {https://ui.adsabs.harvard.edu/abs/2019AJ....158...74C} {158, 74}

\bibitem[\protect\citeauthoryear{{Ciotti}}{{Ciotti}}{2021}]{Ciotti2021}
{Ciotti} L.,  2021, {Introduction to Stellar Dynamics},
  \mn@doi{10.1017/9780511736117.
}

\bibitem[\protect\citeauthoryear{{Ciotti}}{{Ciotti}}{2022}]{Ciotti2022}
{Ciotti} L.,  2022, \mn@doi [\apj] {10.3847/1538-4357/ac82b3}, \href
  {https://ui.adsabs.harvard.edu/abs/2022ApJ...936..180C} {936, 180}

\bibitem[\protect\citeauthoryear{{Courteau}}{{Courteau}}{1997a}]{Corteau1997AJ}
{Courteau} S.,  1997a, \mn@doi [\aj] {10.1086/118656}, \href
  {https://ui.adsabs.harvard.edu/abs/1997AJ....114.2402C} {114, 2402}

\bibitem[\protect\citeauthoryear{{Courteau}}{{Courteau}}{1997b}]{Courteau1997}
{Courteau} S.,  1997b, \mn@doi [\aj] {10.1086/118656}, \href
  {https://ui.adsabs.harvard.edu/abs/1997AJ....114.2402C} {114, 2402}

\bibitem[\protect\citeauthoryear{{Croom} et~al.,}{{Croom}
  et~al.}{2012}]{Croom2012MNRAS}
{Croom} S.~M.,  et~al., 2012, \mn@doi [\mnras]
  {10.1111/j.1365-2966.2011.20365.x}, \href
  {https://ui.adsabs.harvard.edu/abs/2012MNRAS.421..872C} {421, 872}

\bibitem[\protect\citeauthoryear{{Daod} \& {Zeki}}{{Daod} \&
  {Zeki}}{2019}]{Daod2019}
{Daod} N.~A.,  {Zeki} M.~K.,  2019, \mn@doi [\apj] {10.3847/1538-4357/aaf57b},
  \href {https://ui.adsabs.harvard.edu/abs/2019ApJ...870..107D} {870, 107}

\bibitem[\protect\citeauthoryear{{Davis}, {Graham}  \& {Cameron}}{{Davis}
  et~al.}{2018}]{Davis2018}
{Davis} B.~L.,  {Graham} A.~W.,   {Cameron} E.,  2018, \mn@doi [\apj]
  {10.3847/1538-4357/aae820}, \href
  {https://ui.adsabs.harvard.edu/abs/2018ApJ...869..113D} {869, 113}

\bibitem[\protect\citeauthoryear{{Di Teodoro}, {Posti}, {Ogle}, {Fall}  \&
  {Jarrett}}{{Di Teodoro} et~al.}{2021}]{DiTeodoro2021}
{Di Teodoro} E.~M.,  {Posti} L.,  {Ogle} P.~M.,  {Fall} S.~M.,   {Jarrett} T.,
  2021, \mn@doi [\mnras] {10.1093/mnras/stab2549}, \href
  {https://ui.adsabs.harvard.edu/abs/2021MNRAS.507.5820D} {507, 5820}

\bibitem[\protect\citeauthoryear{{Di Teodoro} et~al.,}{{Di Teodoro}
  et~al.}{2023}]{DiTeodoro2023}
{Di Teodoro} E.~M.,  et~al., 2023, \mn@doi [\mnras] {10.1093/mnras/stac3424},
  \href {https://ui.adsabs.harvard.edu/abs/2023MNRAS.518.6340D} {518, 6340}

\bibitem[\protect\citeauthoryear{{D{\'\i}az}, {Carranza}, {Dottori}  \&
  {Goldes}}{{D{\'\i}az} et~al.}{1999}]{Diaz1999}
{D{\'\i}az} R.,  {Carranza} G.,  {Dottori} H.,   {Goldes} G.,  1999, \mn@doi
  [\apj] {10.1086/306781}, \href
  {https://ui.adsabs.harvard.edu/abs/1999ApJ...512..623D} {512, 623}

\bibitem[\protect\citeauthoryear{{Edelson}, {Malkan}  \& {Rieke}}{{Edelson}
  et~al.}{1987}]{Edelson1987}
{Edelson} R.~A.,  {Malkan} M.~A.,   {Rieke} G.~H.,  1987, \mn@doi [\apj]
  {10.1086/165627}, \href
  {https://ui.adsabs.harvard.edu/abs/1987ApJ...321..233E} {321, 233}

\bibitem[\protect\citeauthoryear{{Falc{\'o}n-Barroso}
  et~al.,}{{Falc{\'o}n-Barroso} et~al.}{2017}]{Falcon-Barroso2017AA}
{Falc{\'o}n-Barroso} J.,  et~al., 2017, \mn@doi [\aap]
  {10.1051/0004-6361/201628625}, \href
  {https://ui.adsabs.harvard.edu/abs/2017A&A...597A..48F} {597, A48}

\bibitem[\protect\citeauthoryear{{Ferrero}, {Navarro}, {Abadi}, {Benavides}  \&
  {Mast}}{{Ferrero} et~al.}{2021}]{Ferrero2021AA}
{Ferrero} I.,  {Navarro} J.~F.,  {Abadi} M.~G.,  {Benavides} J.~A.,   {Mast}
  D.,  2021, \mn@doi [\aap] {10.1051/0004-6361/202039839}, \href
  {https://ui.adsabs.harvard.edu/abs/2021A&A...648A.124F} {648, A124}

\bibitem[\protect\citeauthoryear{{Garc{\'\i}a-Lorenzo}
  et~al.,}{{Garc{\'\i}a-Lorenzo} et~al.}{2015}]{Garcia-Lorenzo2015AA}
{Garc{\'\i}a-Lorenzo} B.,  et~al., 2015, \mn@doi [\aap]
  {10.1051/0004-6361/201423485}, \href
  {https://ui.adsabs.harvard.edu/abs/2015A&A...573A..59G} {573, A59}

\bibitem[\protect\citeauthoryear{{Garma-Oehmichen}, {Cano-D{\'\i}az},
  {Hern{\'a}ndez-Toledo}, {Aquino-Ort{\'\i}z}, {Valenzuela}, {Aguerri},
  {S{\'a}nchez}  \& {Merrifield}}{{Garma-Oehmichen}
  et~al.}{2020}]{Garma-Oehmichen2020}
{Garma-Oehmichen} L.,  {Cano-D{\'\i}az} M.,  {Hern{\'a}ndez-Toledo} H.,
  {Aquino-Ort{\'\i}z} E.,  {Valenzuela} O.,  {Aguerri} J.~A.~L.,  {S{\'a}nchez}
  S.~F.,   {Merrifield} M.,  2020, \mn@doi [\mnras] {10.1093/mnras/stz3101},
  \href {https://ui.adsabs.harvard.edu/abs/2020MNRAS.491.3655G} {491, 3655}

\bibitem[\protect\citeauthoryear{{Granados}, {Torres}, {Casta{\~n}eda},
  {Henao-O.}  \& {Vanegas}}{{Granados} et~al.}{2017}]{Granados2017}
{Granados} A.,  {Torres} D.,  {Casta{\~n}eda} L.,  {Henao-O.} J. L.,
  {Vanegas} S.,  2017, arXiv e-prints, \href
  {https://ui.adsabs.harvard.edu/abs/2017arXiv170501665G} {p. arXiv:1705.01665}

\bibitem[\protect\citeauthoryear{{Greene} \& {Ho}}{{Greene} \&
  {Ho}}{2007}]{Greene2007}
{Greene} J.~E.,  {Ho} L.~C.,  2007, \mn@doi [\apj] {10.1086/520497}, \href
  {https://ui.adsabs.harvard.edu/abs/2007ApJ...667..131G} {667, 131}

\bibitem[\protect\citeauthoryear{{Gunn} et~al.,}{{Gunn}
  et~al.}{2006}]{Gunn2006}
{Gunn} J.~E.,  et~al., 2006, \mn@doi [\aj] {10.1086/500975}, \href
  {https://ui.adsabs.harvard.edu/abs/2006AJ....131.2332G} {131, 2332}

\bibitem[\protect\citeauthoryear{{Guo}, {Mao}, {Athanassoula}, {Li}, {Ge},
  {Long}, {Merrifield}  \& {Masters}}{{Guo} et~al.}{2019}]{Guo2019}
{Guo} R.,  {Mao} S.,  {Athanassoula} E.,  {Li} H.,  {Ge} J.,  {Long} R.~J.,
  {Merrifield} M.,   {Masters} K.,  2019, \mn@doi [\mnras]
  {10.1093/mnras/sty2715}, \href
  {https://ui.adsabs.harvard.edu/abs/2019MNRAS.482.1733G} {482, 1733}

\bibitem[\protect\citeauthoryear{{Hernandez}, {Carignan}, {Amram}, {Chemin}  \&
  {Daigle}}{{Hernandez} et~al.}{2005}]{Hernandez2005}
{Hernandez} O.,  {Carignan} C.,  {Amram} P.,  {Chemin} L.,   {Daigle} O.,
  2005, \mn@doi [\mnras] {10.1111/j.1365-2966.2005.09125.x}, \href
  {https://ui.adsabs.harvard.edu/abs/2005MNRAS.360.1201H} {360, 1201}

\bibitem[\protect\citeauthoryear{{Ilha} et~al.,}{{Ilha}
  et~al.}{2019}]{Ilha2019}
{Ilha} G.~S.,  et~al., 2019, \mn@doi [\mnras] {10.1093/mnras/sty3373}, \href
  {https://ui.adsabs.harvard.edu/abs/2019MNRAS.484..252I} {484, 252}

\bibitem[\protect\citeauthoryear{{Ja{\l}ocha}, {Bratek}, {Kutschera}  \&
  {Skindzier}}{{Ja{\l}ocha} et~al.}{2010}]{Jalocha2010}
{Ja{\l}ocha} J.,  {Bratek} {\L}.,  {Kutschera} M.,   {Skindzier} P.,  2010,
  \mn@doi [\mnras] {10.1111/j.1365-2966.2010.16887.x}, \href
  {https://ui.adsabs.harvard.edu/abs/2010MNRAS.406.2805J} {406, 2805}

\bibitem[\protect\citeauthoryear{{Kalinova}, {van de Ven}, {Lyubenova},
  {Falc{\'o}n-Barroso}, {Colombo}  \& {Rosolowsky}}{{Kalinova}
  et~al.}{2017a}]{Kalinova2017a}
{Kalinova} V.,  {van de Ven} G.,  {Lyubenova} M.,  {Falc{\'o}n-Barroso} J.,
  {Colombo} D.,   {Rosolowsky} E.,  2017a, \mn@doi [\mnras]
  {10.1093/mnras/stw2448}, \href
  {https://ui.adsabs.harvard.edu/abs/2017MNRAS.464.1903K} {464, 1903}

\bibitem[\protect\citeauthoryear{{Kalinova} et~al.,}{{Kalinova}
  et~al.}{2017b}]{Kalinova2017b}
{Kalinova} V.,  et~al., 2017b, \mn@doi [\mnras] {10.1093/mnras/stx901}, \href
  {https://ui.adsabs.harvard.edu/abs/2017MNRAS.469.2539K} {469, 2539}

\bibitem[\protect\citeauthoryear{{Krumm} \& {Salpeter}}{{Krumm} \&
  {Salpeter}}{1977}]{Krumm1977}
{Krumm} N.,  {Salpeter} E.~E.,  1977, \aap, \href
  {https://ui.adsabs.harvard.edu/abs/1977A&A....56..465K} {56, 465}

\bibitem[\protect\citeauthoryear{{Krut}, {Arg{\"u}elles}, {Chavanis}, {Rueda}
  \& {Ruffini}}{{Krut} et~al.}{2023}]{Krut2023}
{Krut} A.,  {Arg{\"u}elles} C.~R.,  {Chavanis} P.~H.,  {Rueda} J.~A.,
  {Ruffini} R.,  2023, \mn@doi [\apj] {10.3847/1538-4357/acb8bd}, \href
  {https://ui.adsabs.harvard.edu/abs/2023ApJ...945....1K} {945, 1}

\bibitem[\protect\citeauthoryear{{Laine}, {Kenney}, {Yun}  \&
  {Gottesman}}{{Laine} et~al.}{1999}]{Laine1999}
{Laine} S.,  {Kenney} J.~D.~P.,  {Yun} M.~S.,   {Gottesman} S.~T.,  1999,
  \mn@doi [\apj] {10.1086/306709}, \href
  {https://ui.adsabs.harvard.edu/abs/1999ApJ...511..709L} {511, 709}

\bibitem[\protect\citeauthoryear{{Lang} et~al.,}{{Lang}
  et~al.}{2020}]{Lang2020}
{Lang} P.,  et~al., 2020, \mn@doi [\apj] {10.3847/1538-4357/ab9953}, \href
  {https://ui.adsabs.harvard.edu/abs/2020ApJ...897..122L} {897, 122}

\bibitem[\protect\citeauthoryear{{Lapi}, {Salucci}  \& {Danese}}{{Lapi}
  et~al.}{2018}]{Lapi2018}
{Lapi} A.,  {Salucci} P.,   {Danese} L.,  2018, \mn@doi [\apj]
  {10.3847/1538-4357/aabf35}, \href
  {https://ui.adsabs.harvard.edu/abs/2018ApJ...859....2L} {859, 2}

\bibitem[\protect\citeauthoryear{{Lelli}, {McGaugh}, {Schombert}  \&
  {Pawlowski}}{{Lelli} et~al.}{2016}]{Lelli2016}
{Lelli} F.,  {McGaugh} S.~S.,  {Schombert} J.~M.,   {Pawlowski} M.~S.,  2016,
  \mn@doi [\apjl] {10.3847/2041-8205/827/1/L19}, \href
  {https://ui.adsabs.harvard.edu/abs/2016ApJ...827L..19L} {827, L19}

\bibitem[\protect\citeauthoryear{{Leung} et~al.,}{{Leung}
  et~al.}{2018}]{Leung2018}
{Leung} G. Y.~C.,  et~al., 2018, \mn@doi [\mnras] {10.1093/mnras/sty288}, \href
  {https://ui.adsabs.harvard.edu/abs/2018MNRAS.477..254L} {477, 254}

\bibitem[\protect\citeauthoryear{{Levy} et~al.,}{{Levy}
  et~al.}{2018}]{Levy2018}
{Levy} R.~C.,  et~al., 2018, \mn@doi [\apj] {10.3847/1538-4357/aac2e5}, \href
  {https://ui.adsabs.harvard.edu/abs/2018ApJ...860...92L} {860, 92}

\bibitem[\protect\citeauthoryear{{Makarov}, {Burenkov}  \& {Tyurina}}{{Makarov}
  et~al.}{2001}]{Makarov2001}
{Makarov} D.~I.,  {Burenkov} A.~N.,   {Tyurina} N.~V.,  2001, \mn@doi
  [Astronomy Letters] {10.1134/1.1358377}, \href
  {https://ui.adsabs.harvard.edu/abs/2001AstL...27..213M} {27, 213}

\bibitem[\protect\citeauthoryear{{Marinova} \& {Jogee}}{{Marinova} \&
  {Jogee}}{2007}]{Marinova2007}
{Marinova} I.,  {Jogee} S.,  2007, \mn@doi [\apj] {10.1086/512355}, \href
  {https://ui.adsabs.harvard.edu/abs/2007ApJ...659.1176M} {659, 1176}

\bibitem[\protect\citeauthoryear{{M{\'a}rquez}, {Masegosa}, {Moles}, {Varela},
  {Bettoni}  \& {Galletta}}{{M{\'a}rquez} et~al.}{2002}]{Marquez2002}
{M{\'a}rquez} I.,  {Masegosa} J.,  {Moles} M.,  {Varela} J.,  {Bettoni} D.,
  {Galletta} G.,  2002, \mn@doi [\aap] {10.1051/0004-6361:20021036}, \href
  {https://ui.adsabs.harvard.edu/abs/2002A&A...393..389M} {393, 389}

\bibitem[\protect\citeauthoryear{{Martinez-Medina}, {Pichardo}  \&
  {Peimbert}}{{Martinez-Medina} et~al.}{2020}]{Martinez-Medina2020}
{Martinez-Medina} L.~A.,  {Pichardo} B.,   {Peimbert} A.,  2020, \mn@doi
  [\mnras] {10.1093/mnras/staa1677}, \href
  {https://ui.adsabs.harvard.edu/abs/2020MNRAS.496.1845M} {496, 1845}

\bibitem[\protect\citeauthoryear{{Martinsson}, {Verheijen}, {Westfall},
  {Bershady}, {Schechtman-Rook}, {Andersen}  \& {Swaters}}{{Martinsson}
  et~al.}{2013}]{Martinsson2013}
{Martinsson} T. P.~K.,  {Verheijen} M. A.~W.,  {Westfall} K.~B.,  {Bershady}
  M.~A.,  {Schechtman-Rook} A.,  {Andersen} D.~R.,   {Swaters} R.~A.,  2013,
  \mn@doi [\aap] {10.1051/0004-6361/201220515}, \href
  {https://ui.adsabs.harvard.edu/abs/2013A&A...557A.130M} {557, A130}

\bibitem[\protect\citeauthoryear{{Mathewson}, {Ford}  \&
  {Buchhorn}}{{Mathewson} et~al.}{1992}]{Mathewson1992}
{Mathewson} D.~S.,  {Ford} V.~L.,   {Buchhorn} M.,  1992, \mn@doi [\apjs]
  {10.1086/191700}, \href
  {https://ui.adsabs.harvard.edu/abs/1992ApJS...81..413M} {81, 413}

\bibitem[\protect\citeauthoryear{{Miyamoto} \& {Nagai}}{{Miyamoto} \&
  {Nagai}}{1975}]{Miyamoto-Nagai1975}
{Miyamoto} M.,  {Nagai} R.,  1975, Publications of the Astronomical Society of
  Japan, \href {https://ui.adsabs.harvard.edu/#abs/1975PASJ...27..533M} {27,
  533}

\bibitem[\protect\citeauthoryear{{Pence}}{{Pence}}{1981}]{Pence1981}
{Pence} W.~D.,  1981, \mn@doi [\apj] {10.1086/159056}, \href
  {https://ui.adsabs.harvard.edu/abs/1981ApJ...247..473P} {247, 473}

\bibitem[\protect\citeauthoryear{{Persic} \& {Salucci}}{{Persic} \&
  {Salucci}}{1995}]{Persic1995}
{Persic} M.,  {Salucci} P.,  1995, \mn@doi [\apjs] {10.1086/192195}, \href
  {https://ui.adsabs.harvard.edu/abs/1995ApJS...99..501P} {99, 501}

\bibitem[\protect\citeauthoryear{{Pilyugin}, {Grebel}, {Zinchenko}, {Nefedyev}
  \& {V{\'\i}lchez}}{{Pilyugin} et~al.}{2019}]{Pilyugin2019}
{Pilyugin} L.~S.,  {Grebel} E.~K.,  {Zinchenko} I.~A.,  {Nefedyev} Y.~A.,
  {V{\'\i}lchez} J.~M.,  2019, \mn@doi [\aap] {10.1051/0004-6361/201834239},
  \href {https://ui.adsabs.harvard.edu/abs/2019A&A...623A.122P} {623, A122}

\bibitem[\protect\citeauthoryear{{Pizagno} et~al.,}{{Pizagno}
  et~al.}{2007}]{Pizagno2007}
{Pizagno} J.,  et~al., 2007, \mn@doi [\aj] {10.1086/519522}, \href
  {https://ui.adsabs.harvard.edu/abs/2007AJ....134..945P} {134, 945}

\bibitem[\protect\citeauthoryear{{Regan} \& {Vogel}}{{Regan} \&
  {Vogel}}{1994}]{Regan1994}
{Regan} M.~W.,  {Vogel} S.~N.,  1994, \mn@doi [\apj] {10.1086/174755}, \href
  {https://ui.adsabs.harvard.edu/abs/1994ApJ...434..536R} {434, 536}

\bibitem[\protect\citeauthoryear{{Rembold} et~al.,}{{Rembold}
  et~al.}{2017}]{Rembold2017}
{Rembold} S.~B.,  et~al., 2017, \mn@doi [\mnras] {10.1093/mnras/stx2264}, \href
  {https://ui.adsabs.harvard.edu/abs/2017MNRAS.472.4382R} {472, 4382}

\bibitem[\protect\citeauthoryear{{Rubin}, {Burbidge}, {Burbidge}  \&
  {Prendergast}}{{Rubin} et~al.}{1964}]{Rubin1964}
{Rubin} V.~C.,  {Burbidge} E.~M.,  {Burbidge} G.~R.,   {Prendergast} K.~H.,
  1964, \mn@doi [\apj] {10.1086/147894}, \href
  {https://ui.adsabs.harvard.edu/abs/1964ApJ...140...80R} {140, 80}

\bibitem[\protect\citeauthoryear{{Rubin}, {Ford}  \& {Thonnard}}{{Rubin}
  et~al.}{1978}]{Rubin1978}
{Rubin} V.~C.,  {Ford} W.~K. J.,   {Thonnard} N.,  1978, \mn@doi [\apjl]
  {10.1086/182804}, \href
  {https://ui.adsabs.harvard.edu/abs/1978ApJ...225L.107R} {225, L107}

\bibitem[\protect\citeauthoryear{{Rubin}, {Ford}  \& {Thonnard}}{{Rubin}
  et~al.}{1980}]{Rubin1980}
{Rubin} V.~C.,  {Ford} W.~K. J.,   {Thonnard} N.,  1980, \mn@doi [\apj]
  {10.1086/158003}, \href
  {https://ui.adsabs.harvard.edu/abs/1980ApJ...238..471R} {238, 471}

\bibitem[\protect\citeauthoryear{{Salucci}, {Yegorova}  \& {Drory}}{{Salucci}
  et~al.}{2008}]{Salucci2008b}
{Salucci} P.,  {Yegorova} I.~A.,   {Drory} N.,  2008, \mn@doi [\mnras]
  {10.1111/j.1365-2966.2008.13295.x}, \href
  {https://ui.adsabs.harvard.edu/abs/2008MNRAS.388..159S} {388, 159}

\bibitem[\protect\citeauthoryear{{S{\'a}nchez} et~al.,}{{S{\'a}nchez}
  et~al.}{2012}]{Sanchez2012AA}
{S{\'a}nchez} S.~F.,  et~al., 2012, \mn@doi [\aap]
  {10.1051/0004-6361/201117353}, \href
  {https://ui.adsabs.harvard.edu/abs/2012A&A...538A...8S} {538, A8}

\bibitem[\protect\citeauthoryear{{S{\'a}nchez} et~al.,}{{S{\'a}nchez}
  et~al.}{2016a}]{Sanchez2016a}
{S{\'a}nchez} S.~F.,  et~al., 2016a, \rmxaa, \href
  {https://ui.adsabs.harvard.edu/abs/2016RMxAA..52...21S} {52, 21}

\bibitem[\protect\citeauthoryear{{S{\'a}nchez} et~al.,}{{S{\'a}nchez}
  et~al.}{2016b}]{Pipe3d}
{S{\'a}nchez} S.~F.,  et~al., 2016b, \rmxaa, \href
  {https://ui.adsabs.harvard.edu/abs/2016RMxAA..52..171S} {52, 171}

\bibitem[\protect\citeauthoryear{{S{\'a}nchez} et~al.,}{{S{\'a}nchez}
  et~al.}{2016c}]{Sanchez2016b}
{S{\'a}nchez} S.~F.,  et~al., 2016c, \rmxaa, \href
  {https://ui.adsabs.harvard.edu/abs/2016RMxAA..52..171S} {52, 171}

\bibitem[\protect\citeauthoryear{{S{\'a}nchez} et~al.,}{{S{\'a}nchez}
  et~al.}{2018a}]{Pipe3dII}
{S{\'a}nchez} S.~F.,  et~al., 2018a, \rmxaa, \href
  {https://ui.adsabs.harvard.edu/abs/2018RMxAA..54..217S} {54, 217}

\bibitem[\protect\citeauthoryear{{S{\'a}nchez} et~al.,}{{S{\'a}nchez}
  et~al.}{2018b}]{Sanchez2018}
{S{\'a}nchez} S.~F.,  et~al., 2018b, \mn@doi [\rmxaa]
  {10.48550/arXiv.1709.05438}, \href
  {https://ui.adsabs.harvard.edu/abs/2018RMxAA..54..217S} {54, 217}

\bibitem[\protect\citeauthoryear{{Schmidt}, {Ferreiro}, {Vega Neme}  \&
  {Oio}}{{Schmidt} et~al.}{2016}]{Schmidt2016}
{Schmidt} E.~O.,  {Ferreiro} D.,  {Vega Neme} L.,   {Oio} G.~A.,  2016, \mn@doi
  [\aap] {10.1051/0004-6361/201629343}, \href
  {https://ui.adsabs.harvard.edu/abs/2016A&A...596A..95S} {596, A95}

\bibitem[\protect\citeauthoryear{{Schmidt}, {Mast}, {D{\'\i}az}, {Ag{\"u}ero},
  {G{\"u}nthardt}, {Gimeno}, {Oio}  \& {Gaspar}}{{Schmidt}
  et~al.}{2019}]{Schmidt2019}
{Schmidt} E.~O.,  {Mast} D.,  {D{\'\i}az} R.~J.,  {Ag{\"u}ero} M.~P.,
  {G{\"u}nthardt} G.,  {Gimeno} G.,  {Oio} G.,   {Gaspar} G.,  2019, \mn@doi
  [\aj] {10.3847/1538-3881/ab2882}, \href
  {https://ui.adsabs.harvard.edu/abs/2019AJ....158...60S} {158, 60}

\bibitem[\protect\citeauthoryear{{Sheth} et~al.,}{{Sheth}
  et~al.}{2008}]{Sheth2008}
{Sheth} K.,  et~al., 2008, \mn@doi [\apj] {10.1086/524980}, \href
  {https://ui.adsabs.harvard.edu/abs/2008ApJ...675.1141S} {675, 1141}

\bibitem[\protect\citeauthoryear{{Shetty}, {Vogel}, {Ostriker}  \&
  {Teuben}}{{Shetty} et~al.}{2007}]{Shetty2007}
{Shetty} R.,  {Vogel} S.~N.,  {Ostriker} E.~C.,   {Teuben} P.~J.,  2007,
  \mn@doi [\apj] {10.1086/520037}, \href
  {https://ui.adsabs.harvard.edu/abs/2007ApJ...665.1138S} {665, 1138}

\bibitem[\protect\citeauthoryear{{Sofue}}{{Sofue}}{2016}]{Sofue2016}
{Sofue} Y.,  2016, \mn@doi [\pasj] {10.1093/pasj/psv103}, \href
  {https://ui.adsabs.harvard.edu/abs/2016PASJ...68....2S} {68, 2}

\bibitem[\protect\citeauthoryear{{Sofue}}{{Sofue}}{2017}]{Sofue2017}
{Sofue} Y.,  2017, \mn@doi [\pasj] {10.1093/pasj/psw103}, \href
  {https://ui.adsabs.harvard.edu/abs/2017PASJ...69R...1S} {69, R1}

\bibitem[\protect\citeauthoryear{{Sofue} \& {Rubin}}{{Sofue} \&
  {Rubin}}{2001}]{Sofue2001}
{Sofue} Y.,  {Rubin} V.,  2001, \mn@doi [\araa]
  {10.1146/annurev.astro.39.1.137}, \href
  {https://ui.adsabs.harvard.edu/abs/2001ARA&A..39..137S} {39, 137}

\bibitem[\protect\citeauthoryear{{Sofue}, {Tutui}, {Honma}, {Tomita},
  {Takamiya}, {Koda}  \& {Takeda}}{{Sofue} et~al.}{1999}]{Sofue1999}
{Sofue} Y.,  {Tutui} Y.,  {Honma} M.,  {Tomita} A.,  {Takamiya} T.,  {Koda} J.,
    {Takeda} Y.,  1999, \mn@doi [\apj] {10.1086/307731}, \href
  {https://ui.adsabs.harvard.edu/abs/1999ApJ...523..136S} {523, 136}

\bibitem[\protect\citeauthoryear{{Spano}, {Marcelin}, {Amram}, {Carignan},
  {Epinat}  \& {Hernandez}}{{Spano} et~al.}{2008}]{Spano2008}
{Spano} M.,  {Marcelin} M.,  {Amram} P.,  {Carignan} C.,  {Epinat} B.,
  {Hernandez} O.,  2008, \mn@doi [\mnras] {10.1111/j.1365-2966.2007.12545.x},
  \href {https://ui.adsabs.harvard.edu/abs/2008MNRAS.383..297S} {383, 297}

\bibitem[\protect\citeauthoryear{{Stern} \& {Laor}}{{Stern} \&
  {Laor}}{2012}]{Stern2012}
{Stern} J.,  {Laor} A.,  2012, \mn@doi [\mnras]
  {10.1111/j.1365-2966.2012.20901.x}, \href
  {https://ui.adsabs.harvard.edu/abs/2012MNRAS.423..600S} {423, 600}

\bibitem[\protect\citeauthoryear{{Toba} et~al.,}{{Toba}
  et~al.}{2014}]{Toba2014}
{Toba} Y.,  et~al., 2014, \mn@doi [\apj] {10.1088/0004-637X/788/1/45}, \href
  {https://ui.adsabs.harvard.edu/abs/2014ApJ...788...45T} {788, 45}

\bibitem[\protect\citeauthoryear{{Tully} \& {Fisher}}{{Tully} \&
  {Fisher}}{1977}]{Tully1977}
{Tully} R.~B.,  {Fisher} J.~R.,  1977, \aap, \href
  {https://ui.adsabs.harvard.edu/abs/1977A&A....54..661T} {54, 661}

\bibitem[\protect\citeauthoryear{{Vaughan}}{{Vaughan}}{1989}]{Vaughan1989}
{Vaughan} J.~M.,  1989, {The Fabry-Perot interferometer. History, theory,
  practice and applications}

\bibitem[\protect\citeauthoryear{{Vogt}, {Haynes}, {Giovanelli}  \&
  {Herter}}{{Vogt} et~al.}{2004a}]{Vogt2004a}
{Vogt} N.~P.,  {Haynes} M.~P.,  {Giovanelli} R.,   {Herter} T.,  2004a, \mn@doi
  [\aj] {10.1086/420702}, \href
  {https://ui.adsabs.harvard.edu/abs/2004AJ....127.3300V} {127, 3300}

\bibitem[\protect\citeauthoryear{{Vogt}, {Haynes}, {Giovanelli}  \&
  {Herter}}{{Vogt} et~al.}{2004b}]{Vogt2004b}
{Vogt} N.~P.,  {Haynes} M.~P.,  {Giovanelli} R.,   {Herter} T.,  2004b, \mn@doi
  [\aj] {10.1086/420703}, \href
  {https://ui.adsabs.harvard.edu/abs/2004AJ....127.3325V} {127, 3325}

\bibitem[\protect\citeauthoryear{{Williams}, {Bureau}  \&
  {Cappellari}}{{Williams} et~al.}{2010}]{Williams2010}
{Williams} M.~J.,  {Bureau} M.,   {Cappellari} M.,  2010, \mn@doi [\mnras]
  {10.1111/j.1365-2966.2010.17406.x}, \href
  {https://ui.adsabs.harvard.edu/abs/2010MNRAS.409.1330W} {409, 1330}

\bibitem[\protect\citeauthoryear{{Yoon}, {Park}, {Chung}  \& {Zhang}}{{Yoon}
  et~al.}{2021}]{Yoon2021ApJ}
{Yoon} Y.,  {Park} C.,  {Chung} H.,   {Zhang} K.,  2021, \mn@doi [\apj]
  {10.3847/1538-4357/ac2302}, \href
  {https://ui.adsabs.harvard.edu/abs/2021ApJ...922..249Y} {922, 249}

\bibitem[\protect\citeauthoryear{{Yu} et~al.,}{{Yu} et~al.}{2022}]{YU2022AA}
{Yu} S.-Y.,  et~al., 2022, \mn@doi [\aap] {10.1051/0004-6361/202244306}, \href
  {https://ui.adsabs.harvard.edu/abs/2022A&A...666A.175Y} {666, A175}

\makeatother
\end{thebibliography}




\appendix

\section{Fitted rotation curves of the 46 barred galaxies}


\begin{figure*} 
\centering
\begin{tabular}{cc}
    \includegraphics[width=0.44\linewidth]{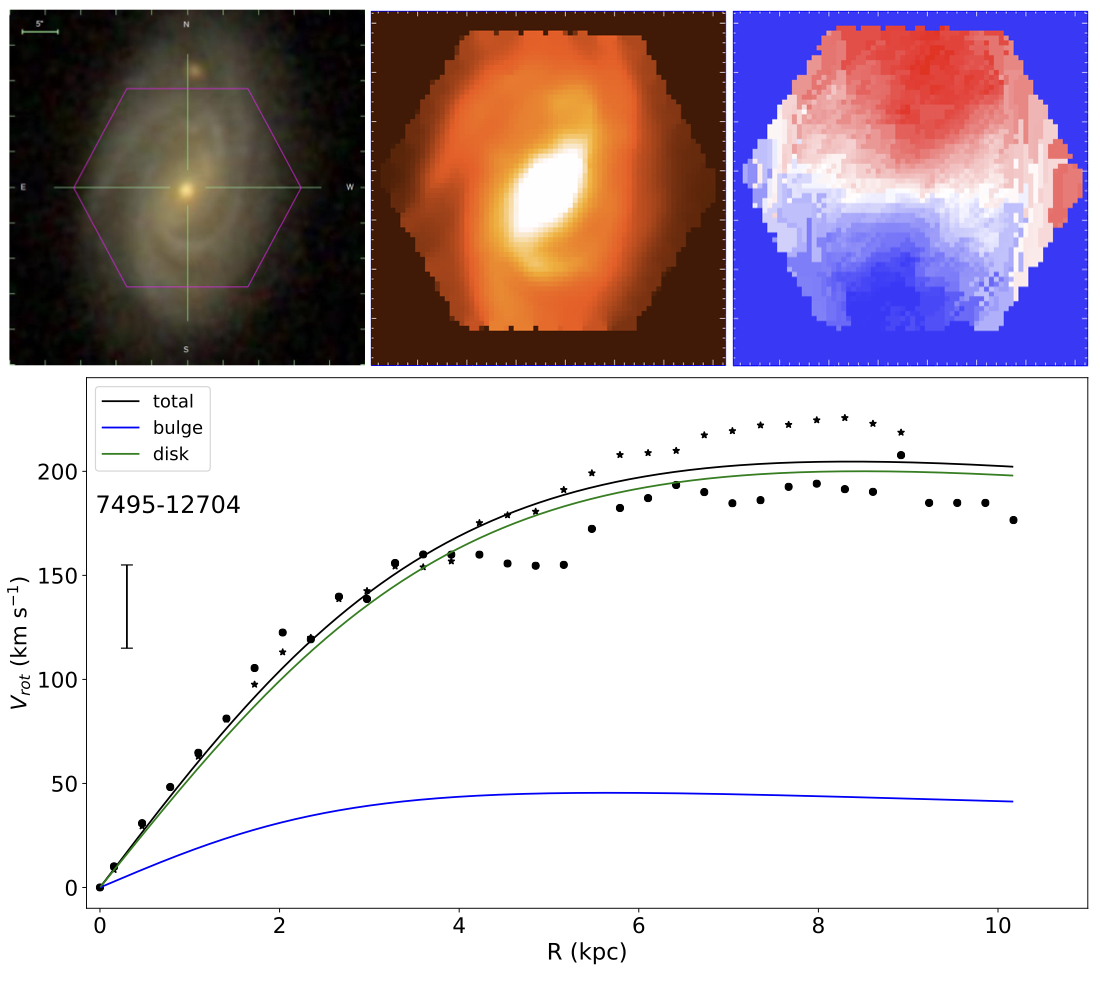}&
    \includegraphics[width=0.44\linewidth]{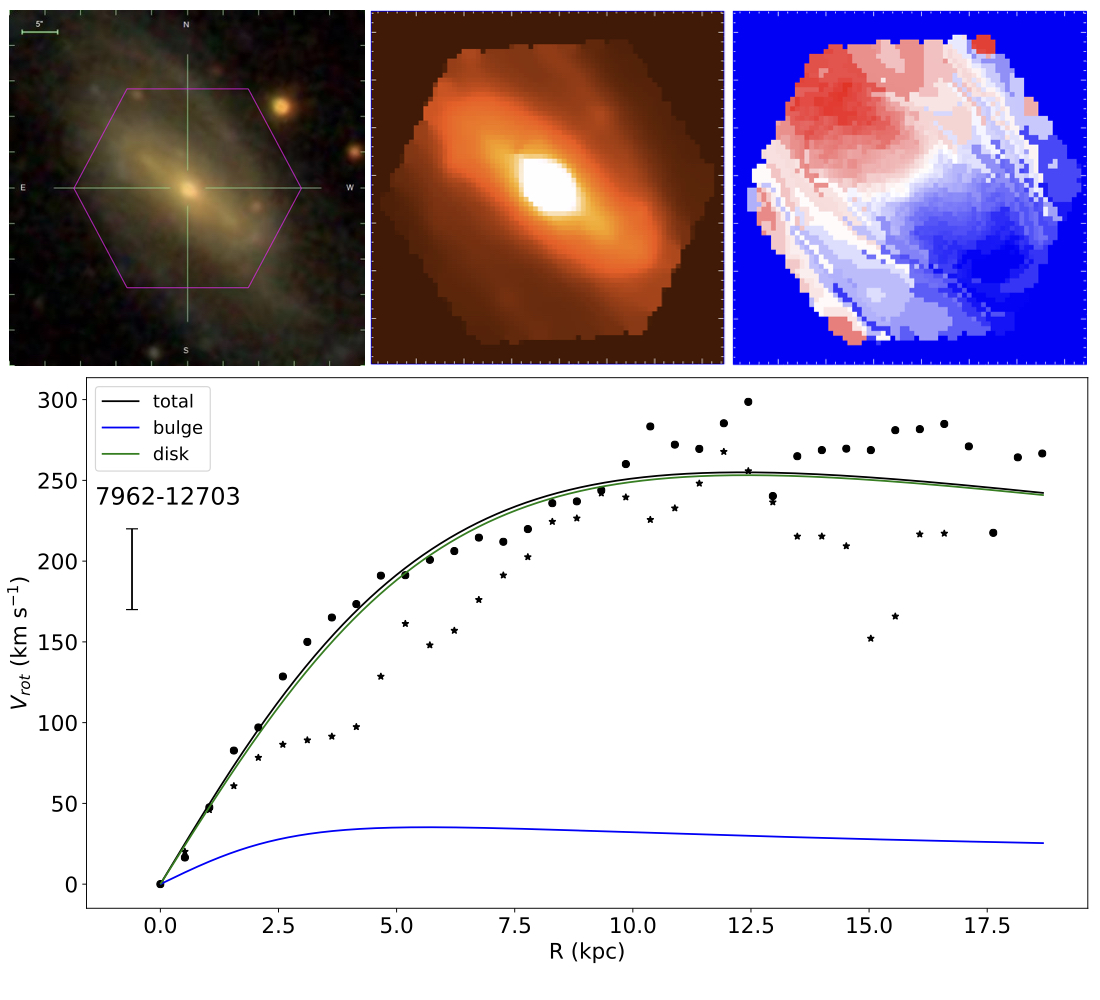}\\[2\tabcolsep]
    \includegraphics[width=0.44\linewidth]{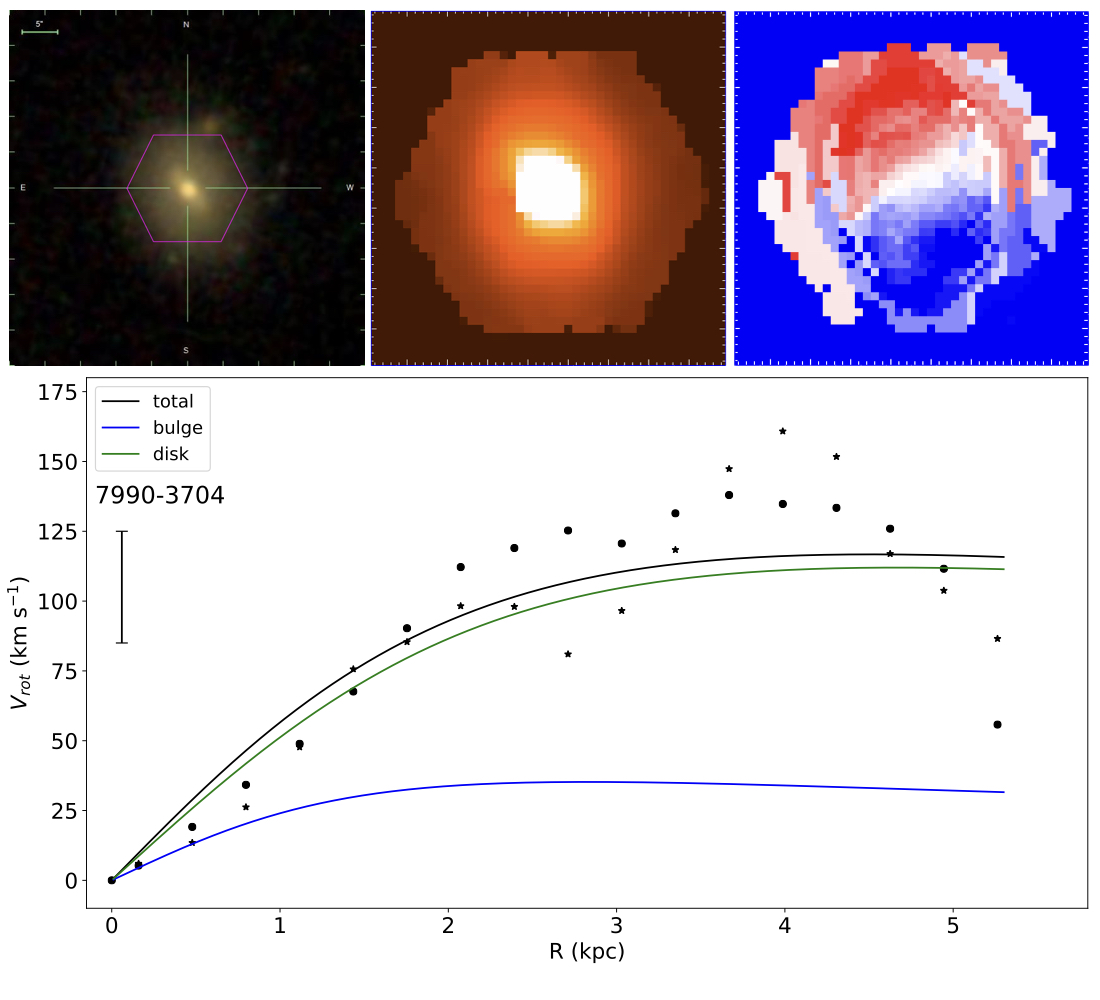}&
    \includegraphics[width=0.44\linewidth]{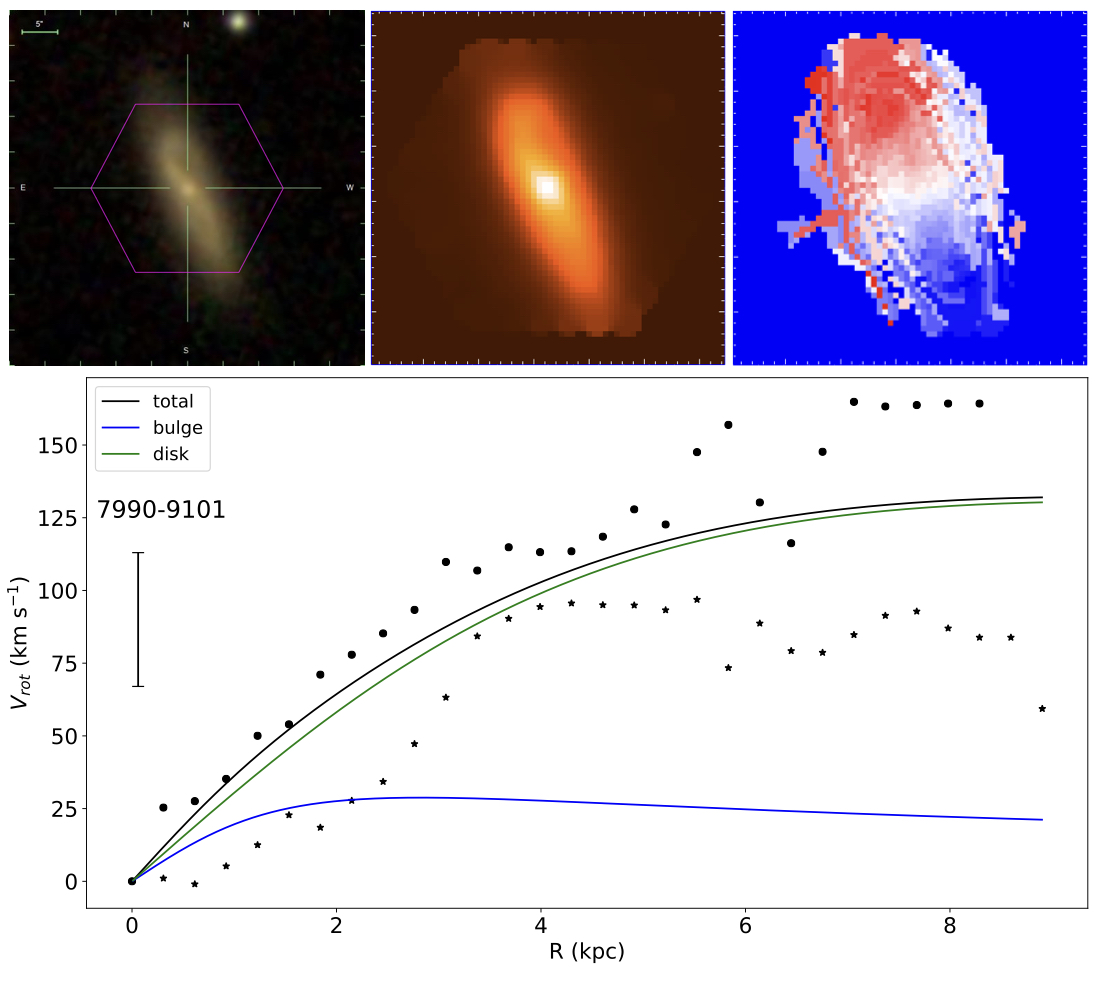}\\
    \includegraphics[width=0.44\linewidth]{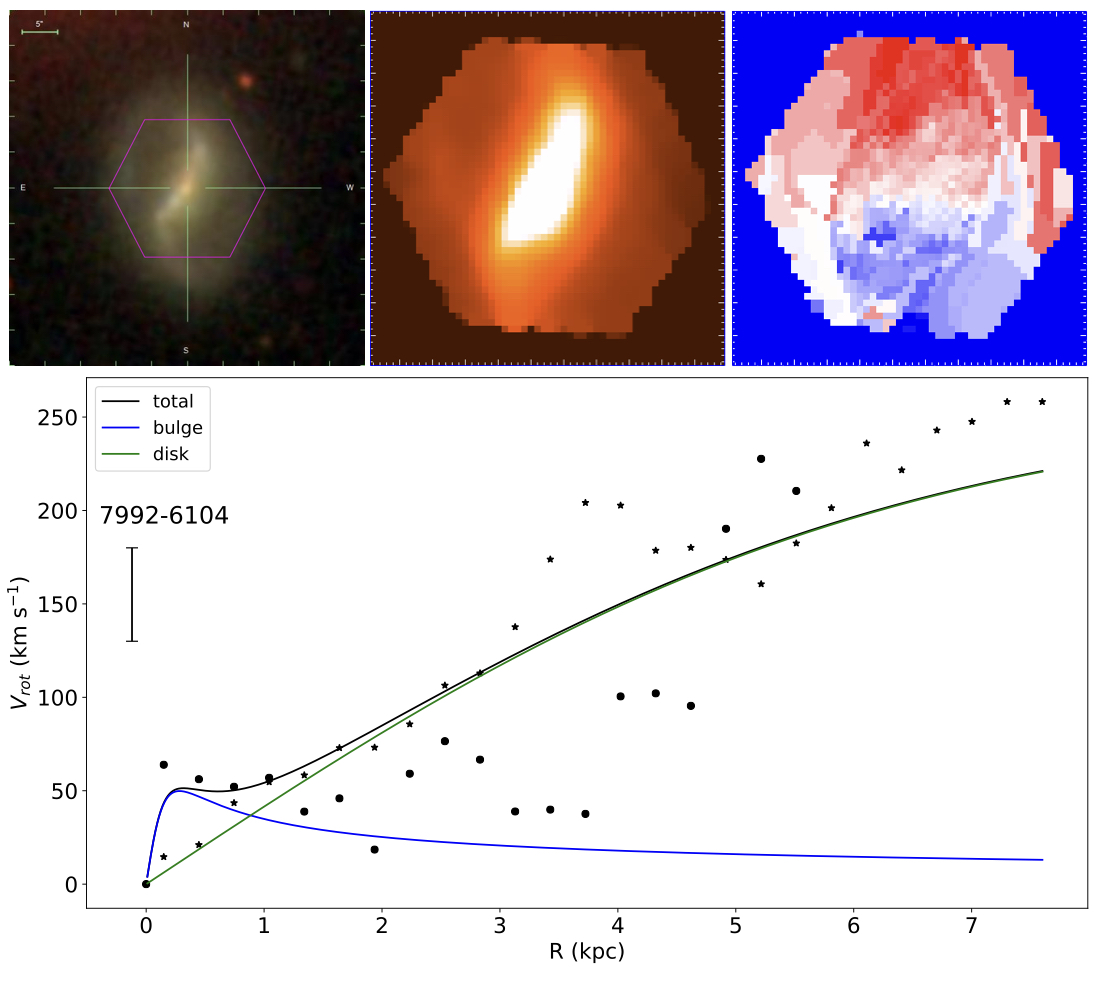}&
    \includegraphics[width=0.44\linewidth]{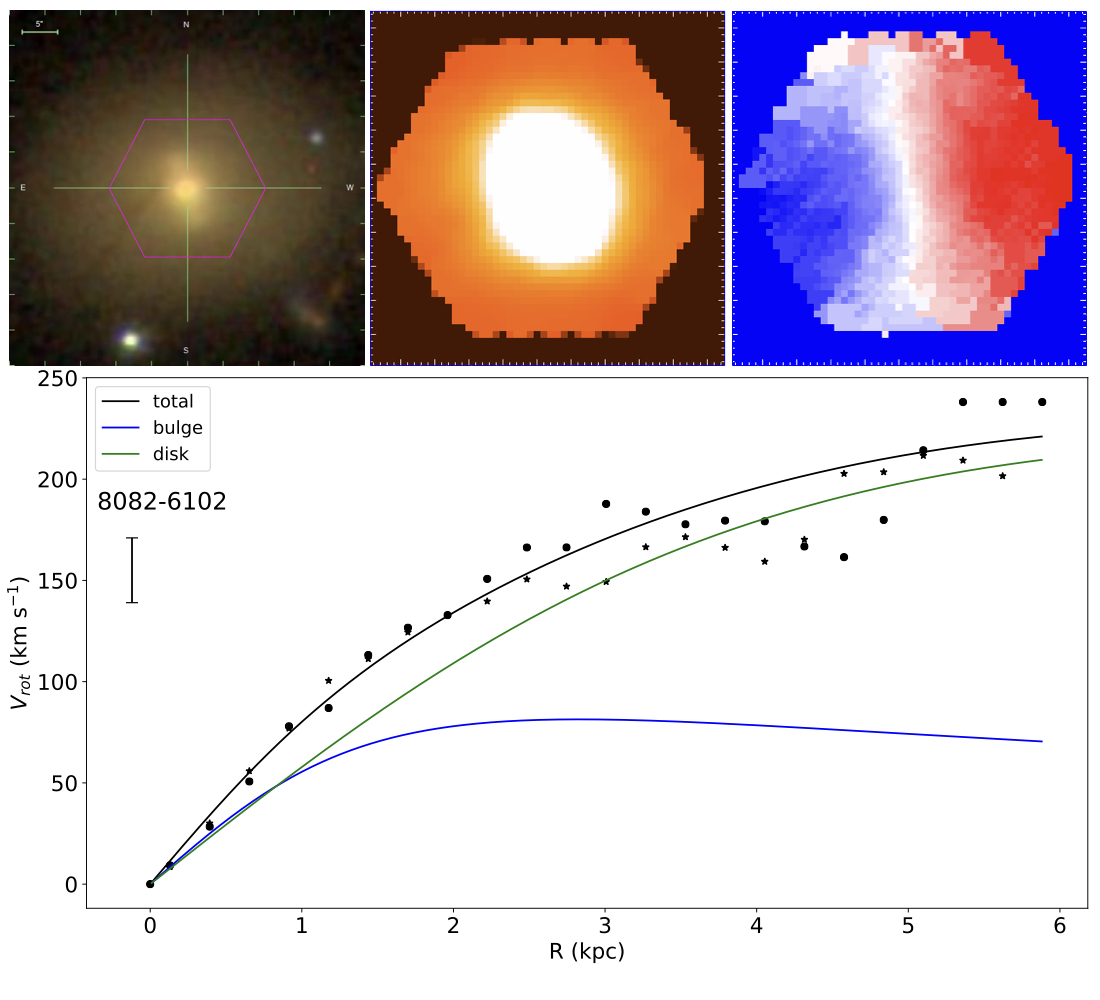}\\
\end{tabular}
\caption{Each panel shows from left to right and top to bottom: MaNGA FoV depicted over $gri$ false-color images from SDSS, pseudo-V-band image extracted from the datacubes, the velocity field with red color representing positive radial velocities and blue representing negative radial velocities with respect to the systemic velocity (see text for details), and the rotation curve fitted with two mass components corresponding to an axisymetric Miyamoto$-$Nagai gravitational potential. In all the maps and images, North is up and East is to the left. Note that the maps are not at the same scale as the false color image. Ticksmarks on the maps are separated each 0.5 arcsec. Circles and stars represent different sides of the velocity field. A typical error bar is shown in the top left corner.}
\label{fig:curves9}

\end{figure*}

\begin{figure*} 
\ContinuedFloat
\centering
\begin{tabular}{cc}
    \includegraphics[width=0.47\linewidth]{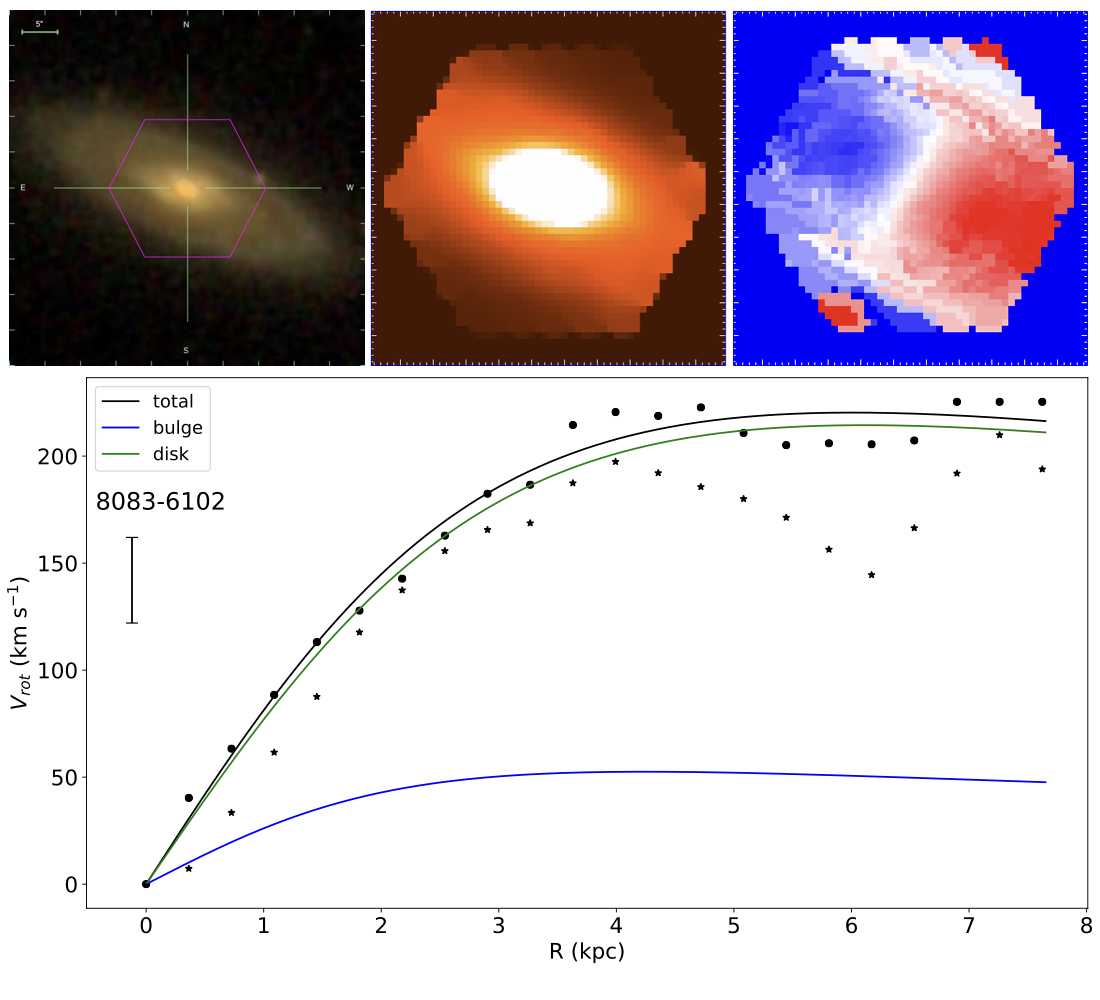}&
    \includegraphics[width=0.47\linewidth]{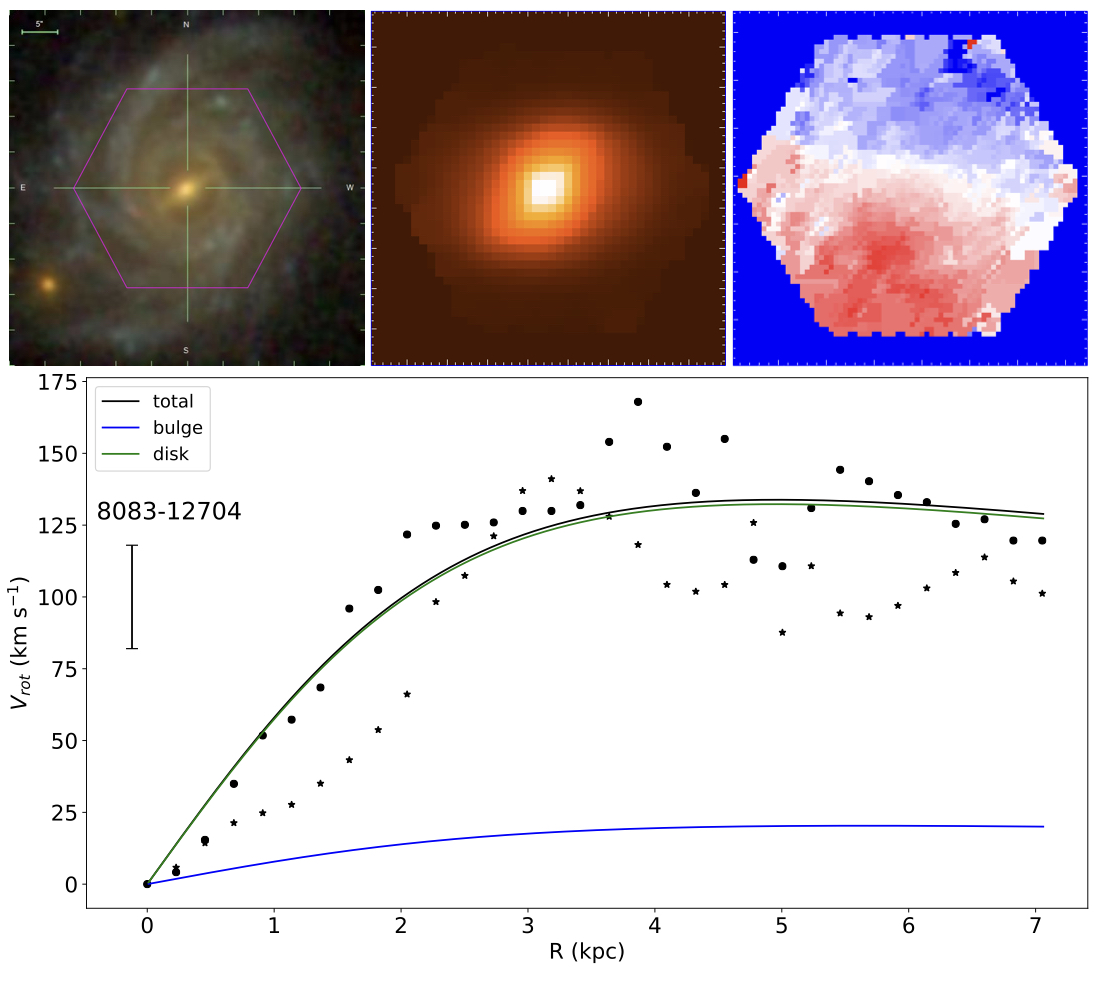}\\[2\tabcolsep]
    \includegraphics[width=0.47\linewidth]{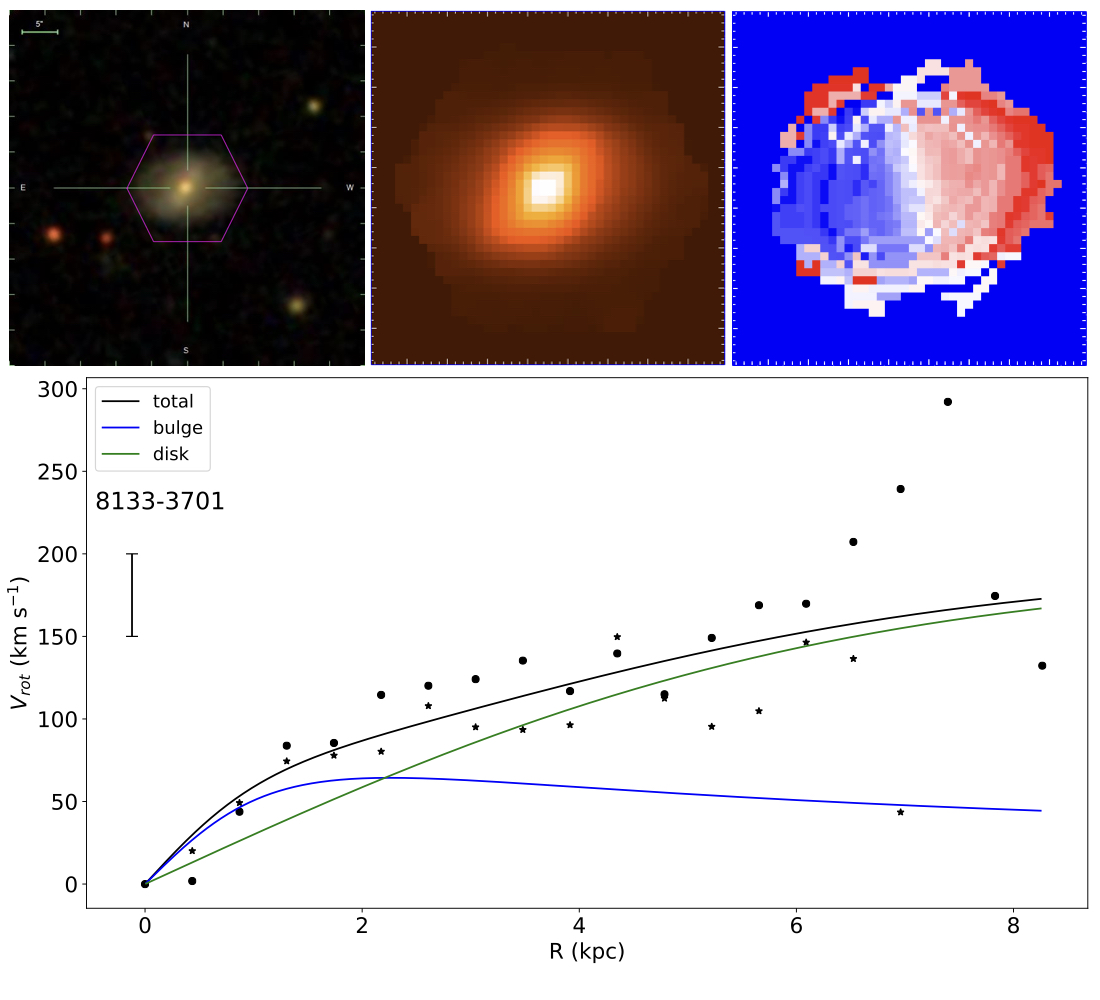}&
    \includegraphics[width=0.47\linewidth]{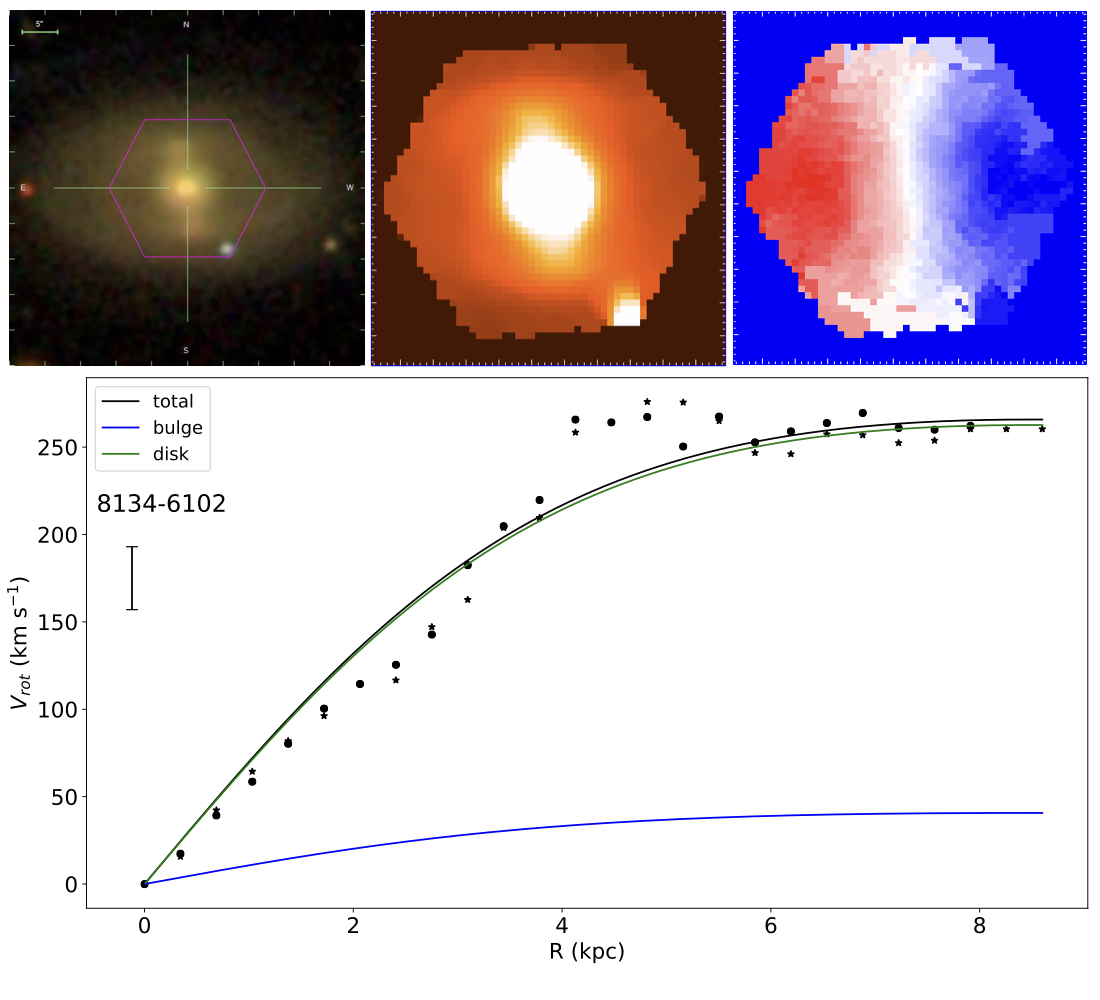}\\
    \includegraphics[width=0.47\linewidth]{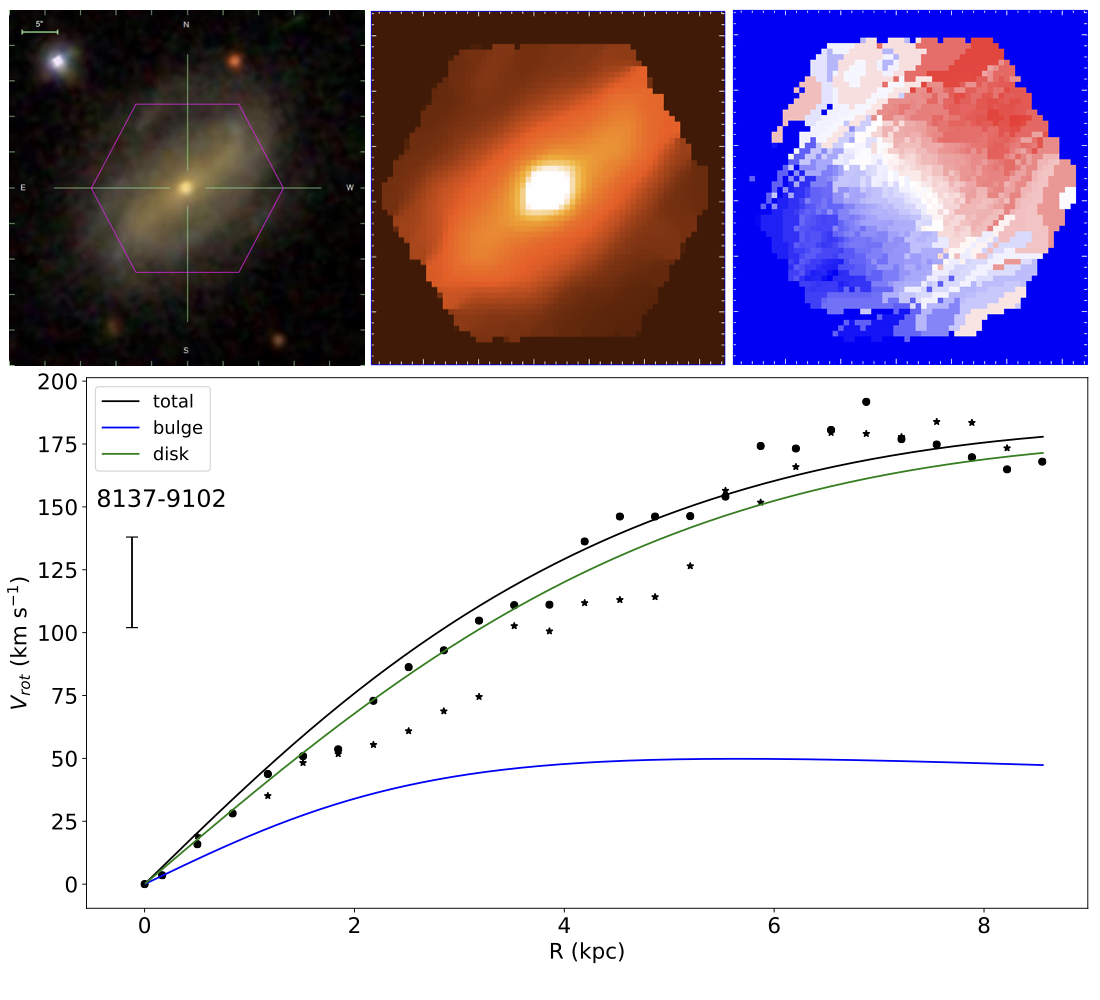}&
    \includegraphics[width=0.47\linewidth]{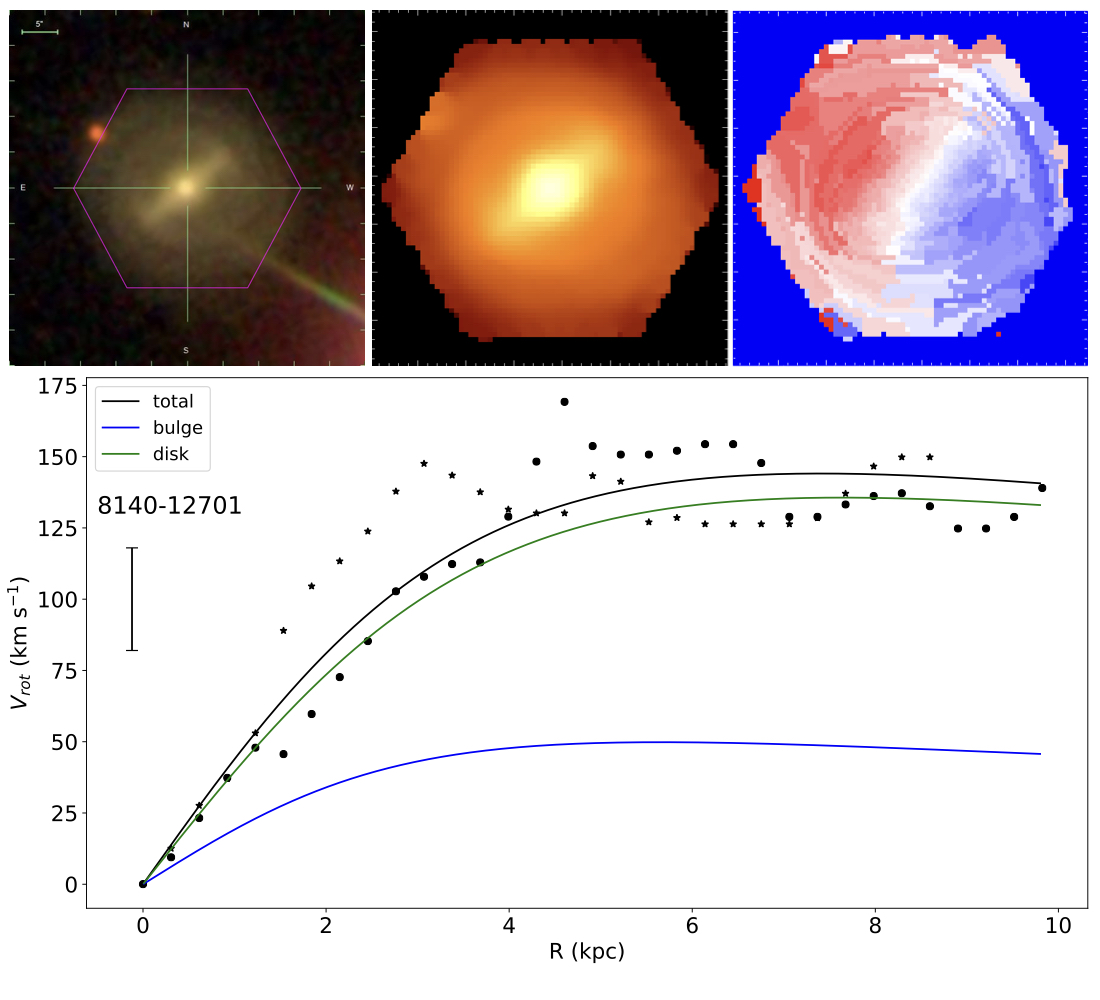}\\
\end{tabular}
\caption{continued}
\label{fig:curves8}
\end{figure*}

\begin{figure*} 
\ContinuedFloat
\centering
\begin{tabular}{cc}
    \includegraphics[width=0.47\linewidth]{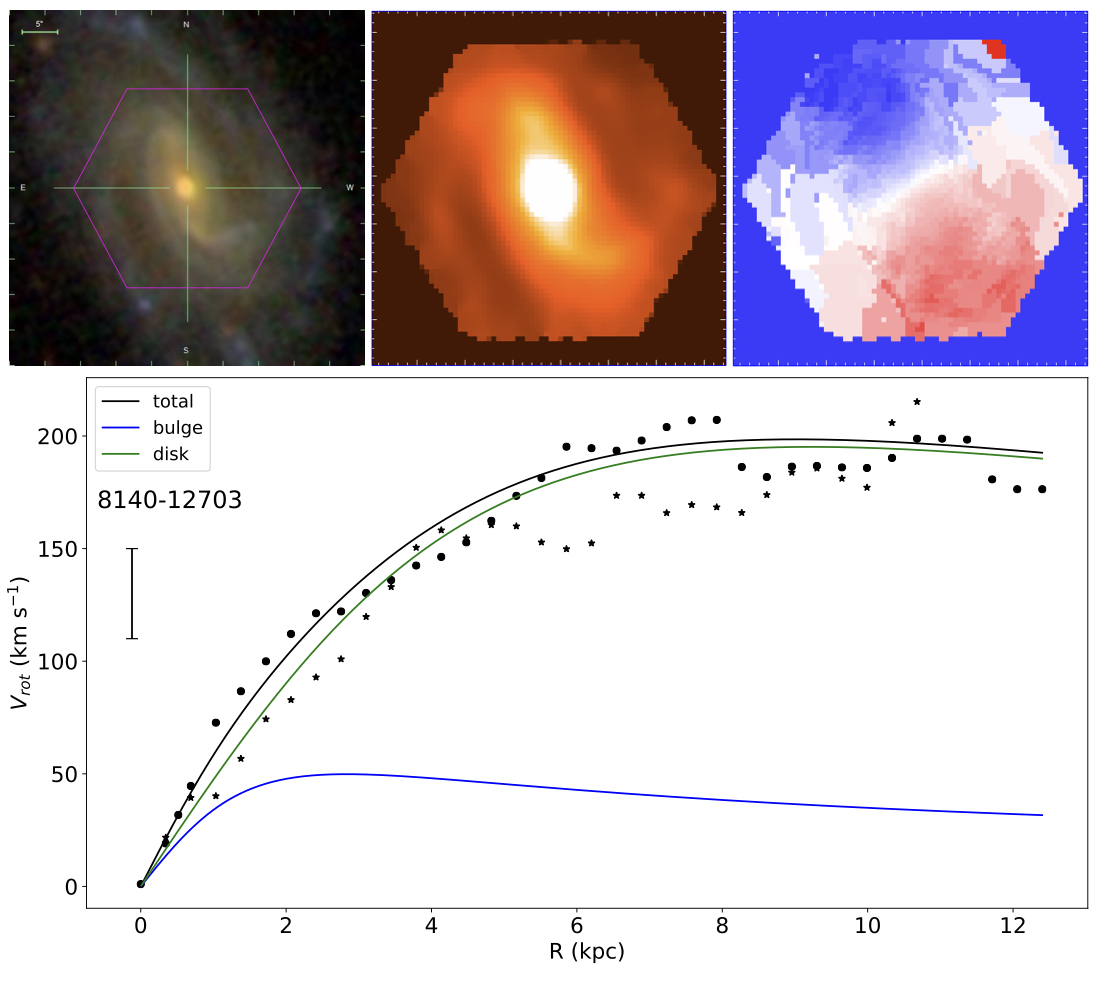}&
    \includegraphics[width=0.47\linewidth]{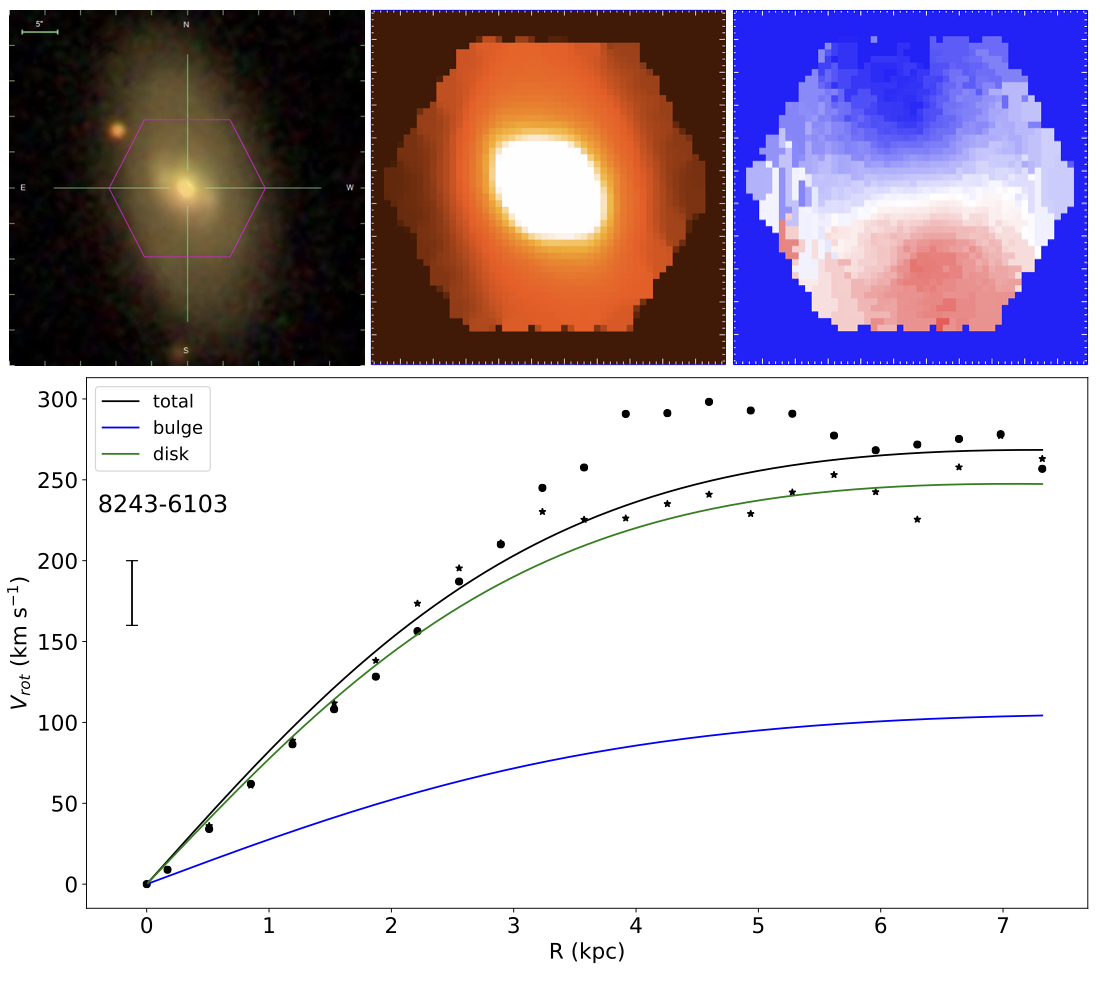}\\[2\tabcolsep]
    \includegraphics[width=0.47\linewidth]{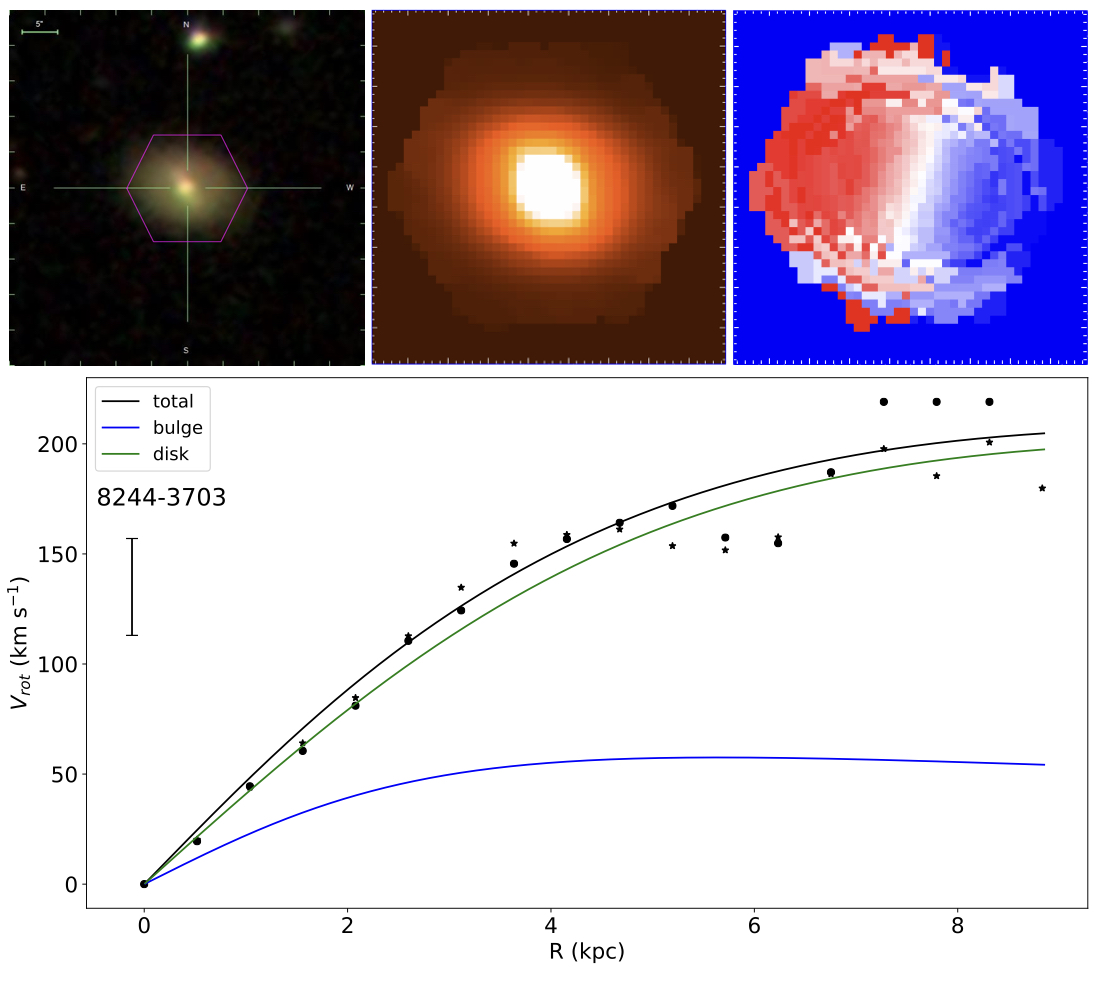}&
    \includegraphics[width=0.47\linewidth]{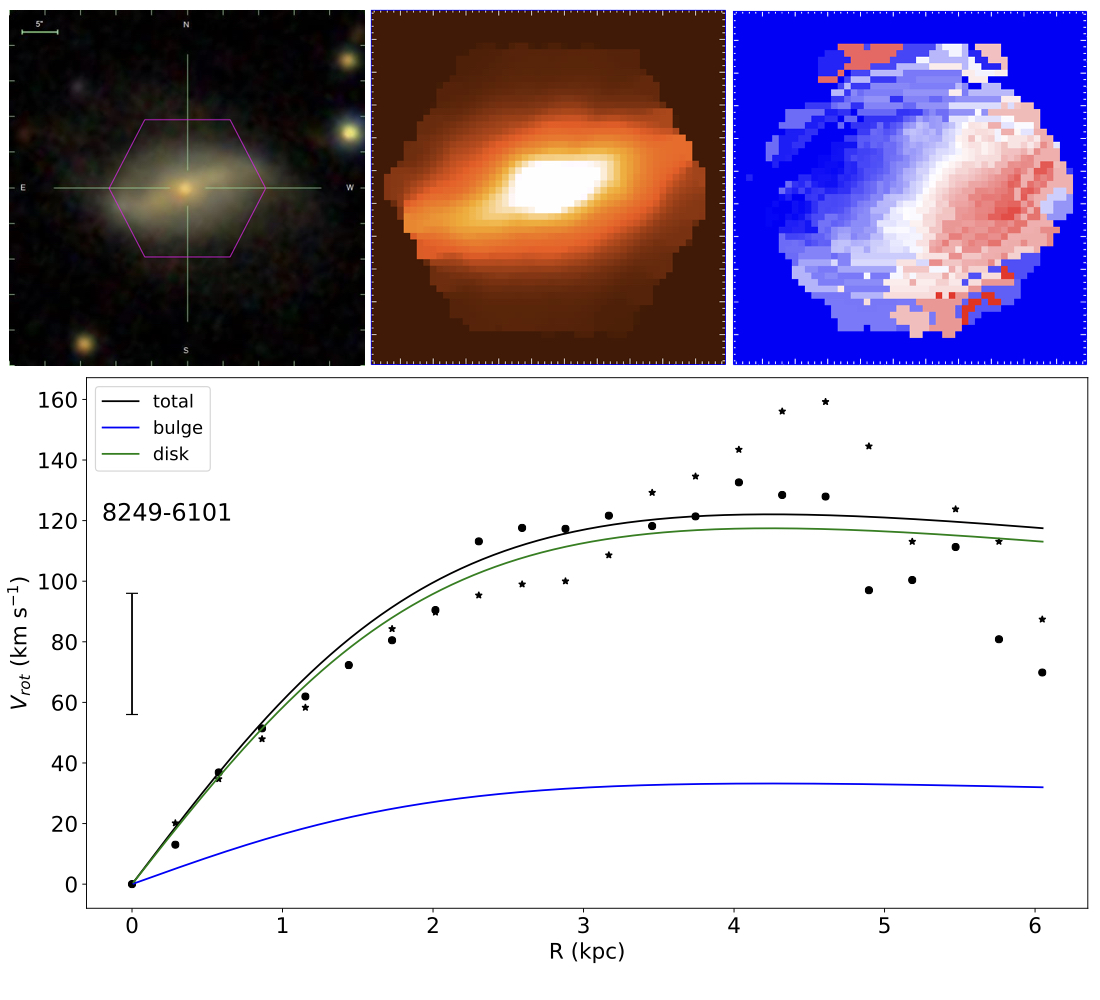}\\
    \includegraphics[width=0.47\linewidth]{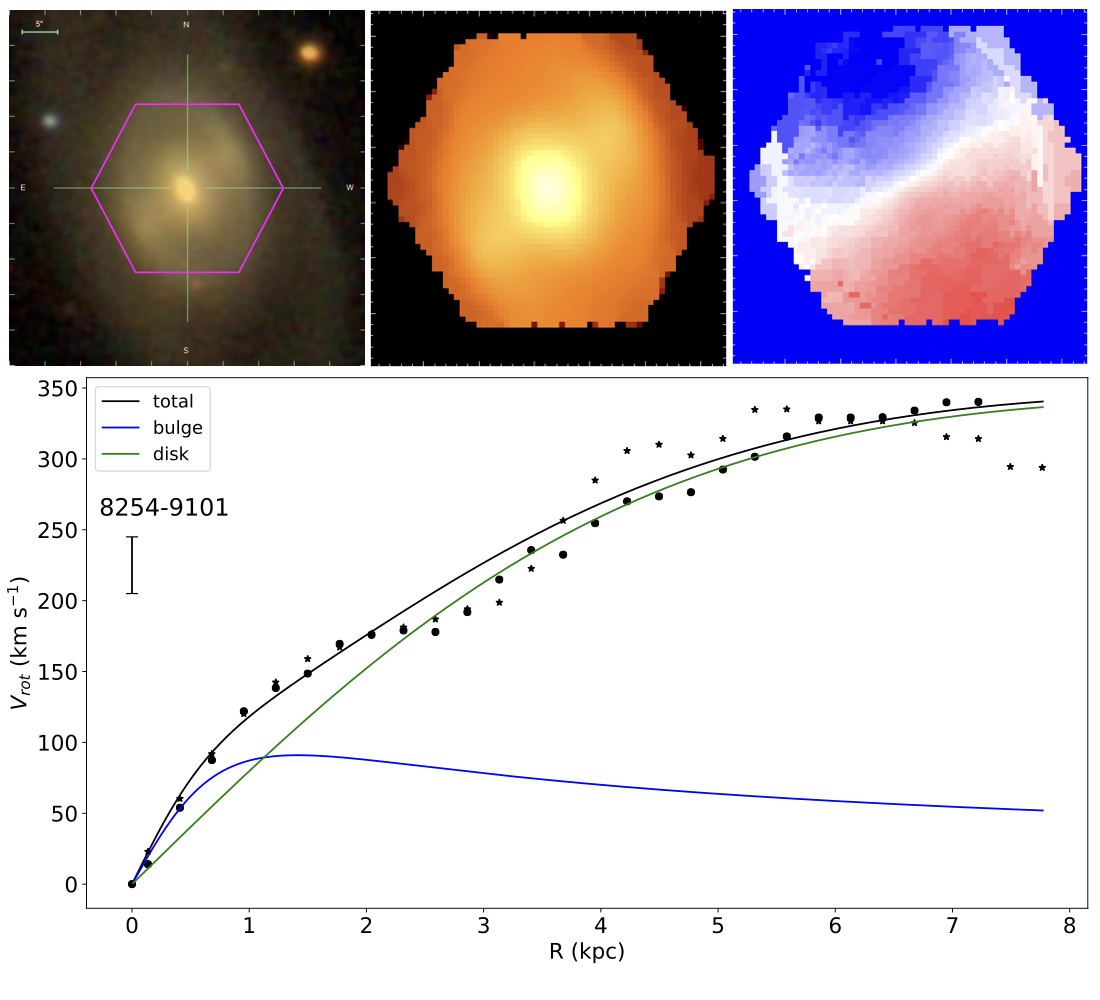}&
    \includegraphics[width=0.47\linewidth]{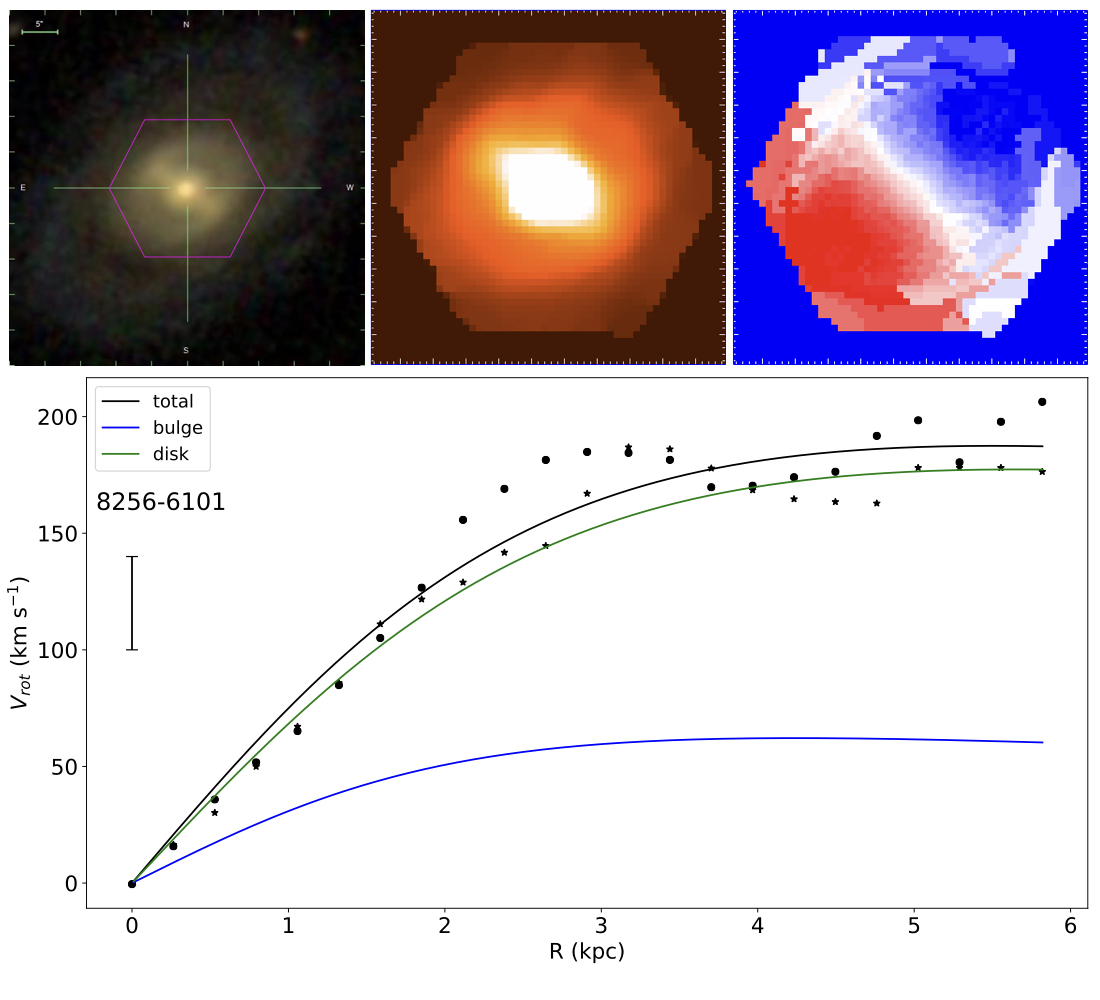}\\
\end{tabular}
\caption{continued}
\label{fig:curves}
\end{figure*}

\begin{figure*} 
\ContinuedFloat
\centering
\begin{tabular}{cc}
    \includegraphics[width=0.47\linewidth]{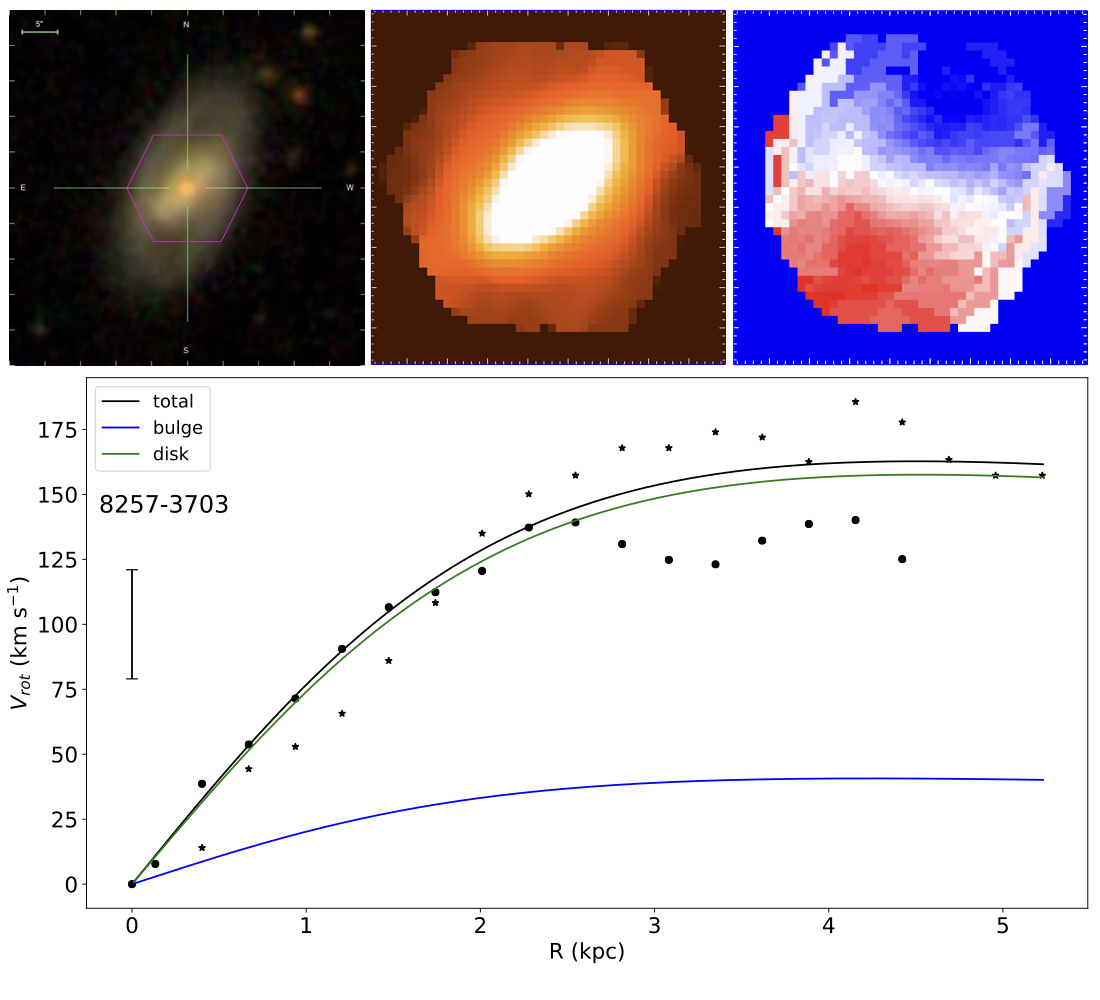}&
    \includegraphics[width=0.47\linewidth]{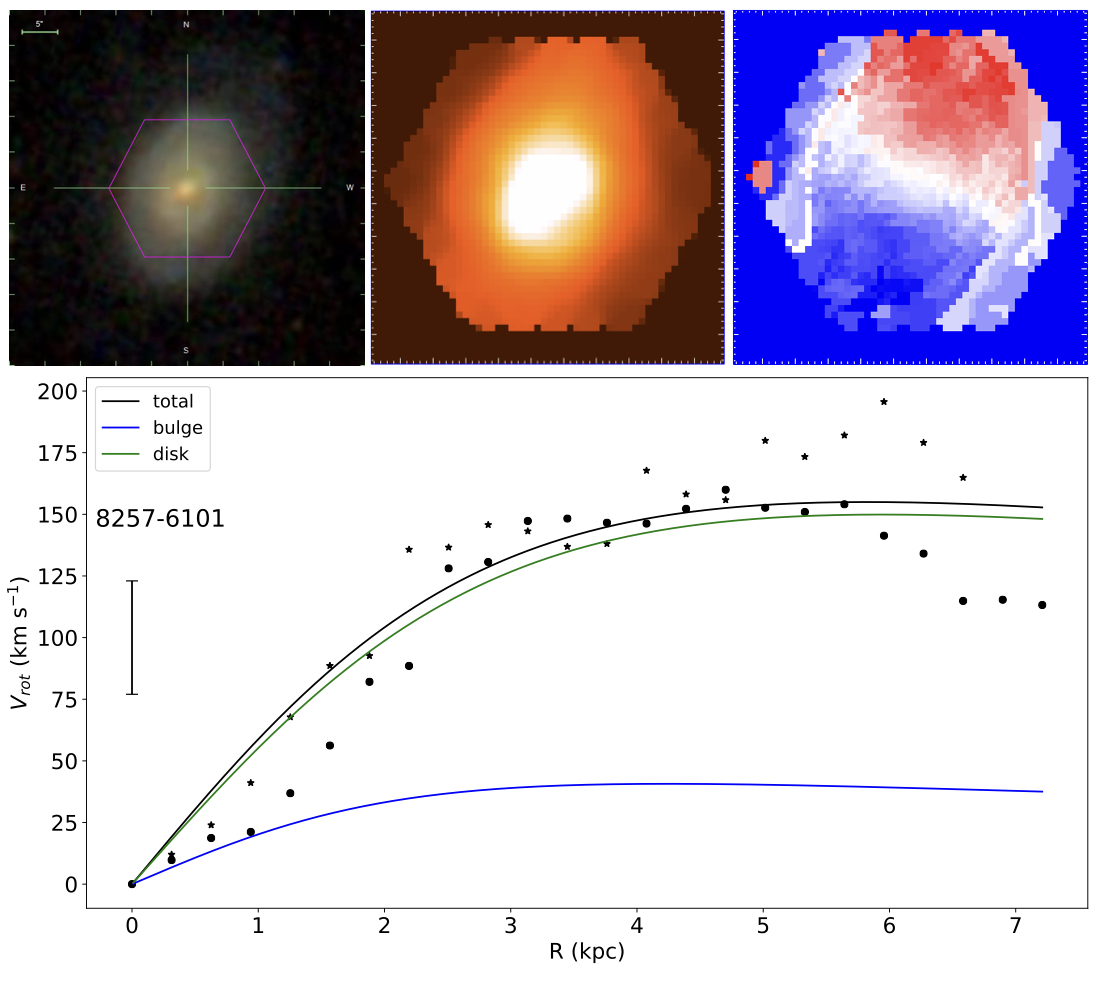}\\[2\tabcolsep]
    \includegraphics[width=0.47\linewidth]{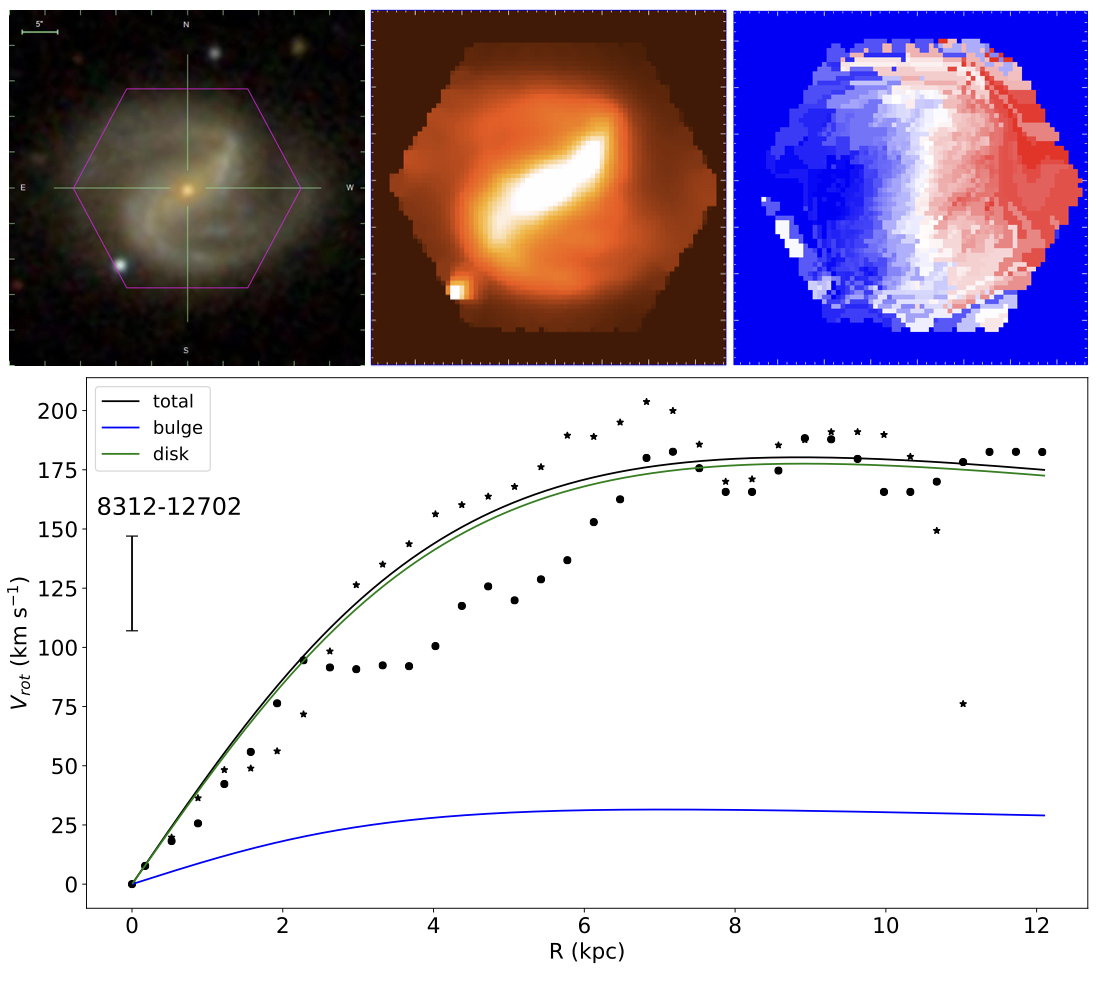}&
    \includegraphics[width=0.47\linewidth]{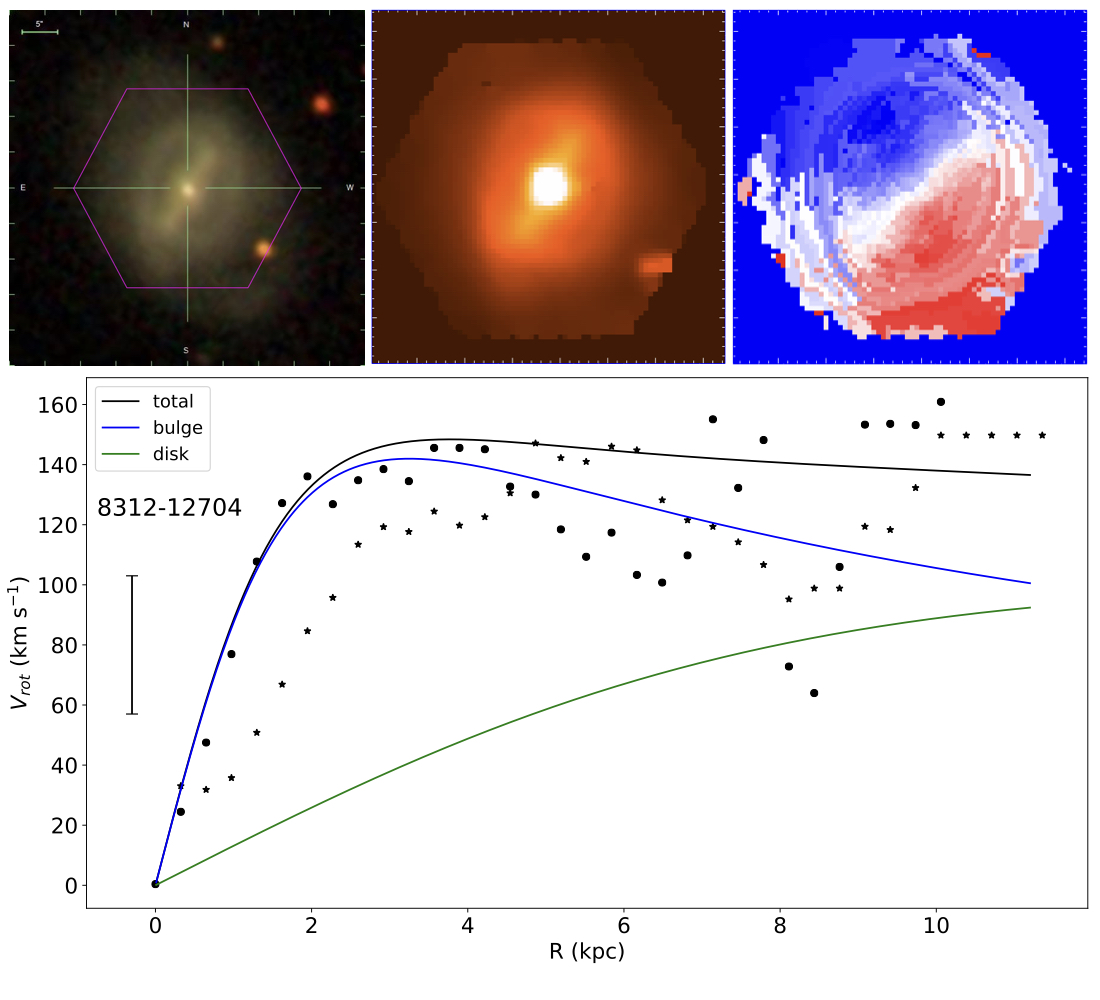}\\
    \includegraphics[width=0.47\linewidth]{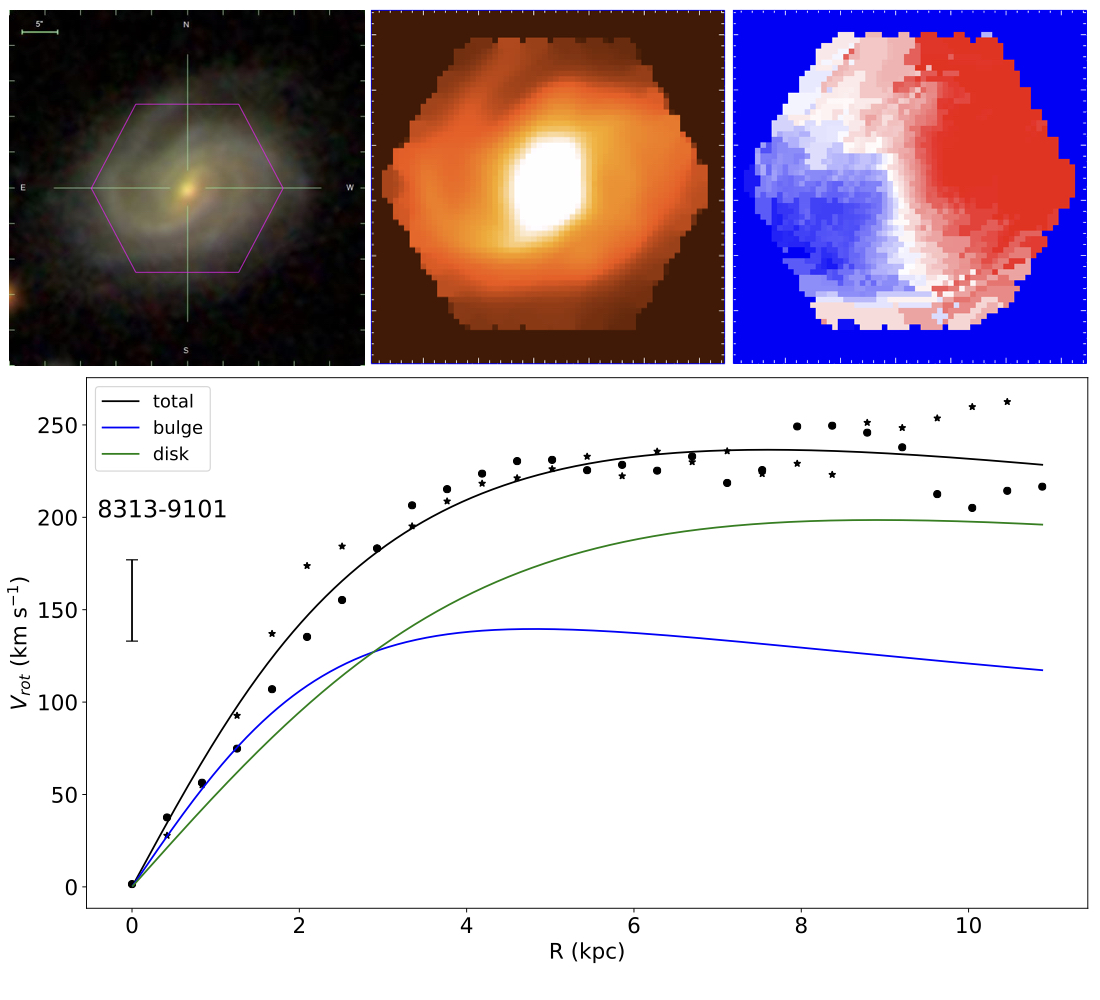}&
    \includegraphics[width=0.47\linewidth]{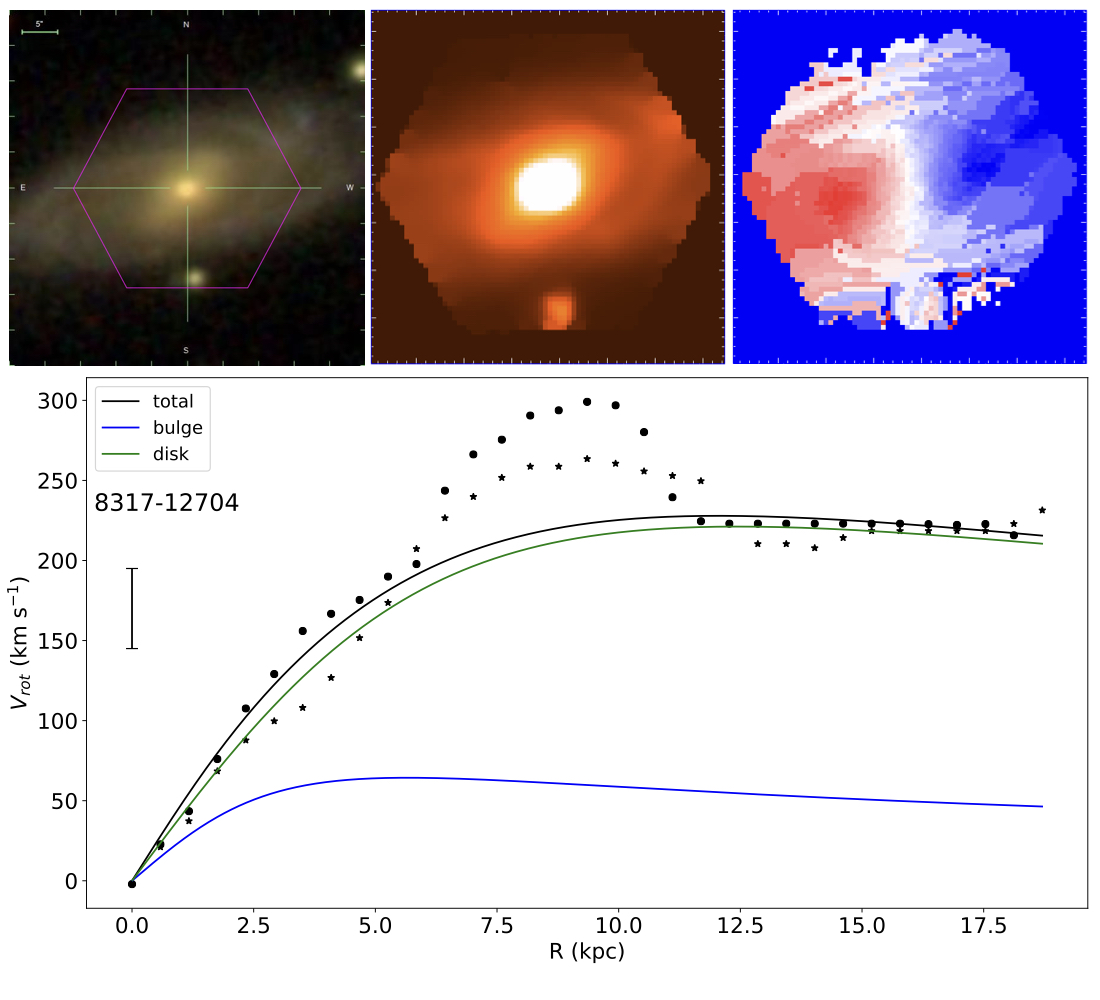}\\
\end{tabular}
\caption{continued}
\label{fig:curves2}
\end{figure*}

\begin{figure*} 
\ContinuedFloat
\centering
\begin{tabular}{cc}
    \includegraphics[width=0.47\linewidth]{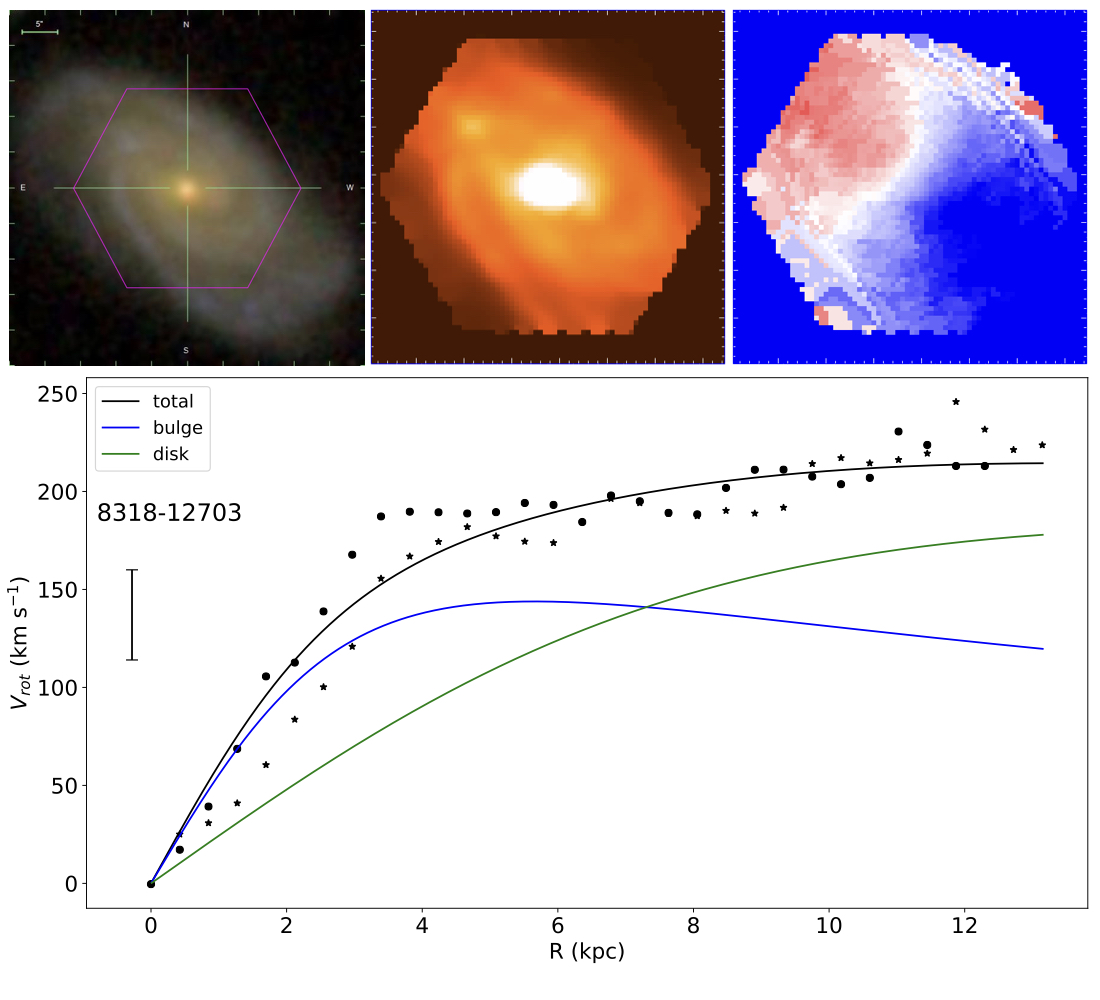}&
    \includegraphics[width=0.47\linewidth]{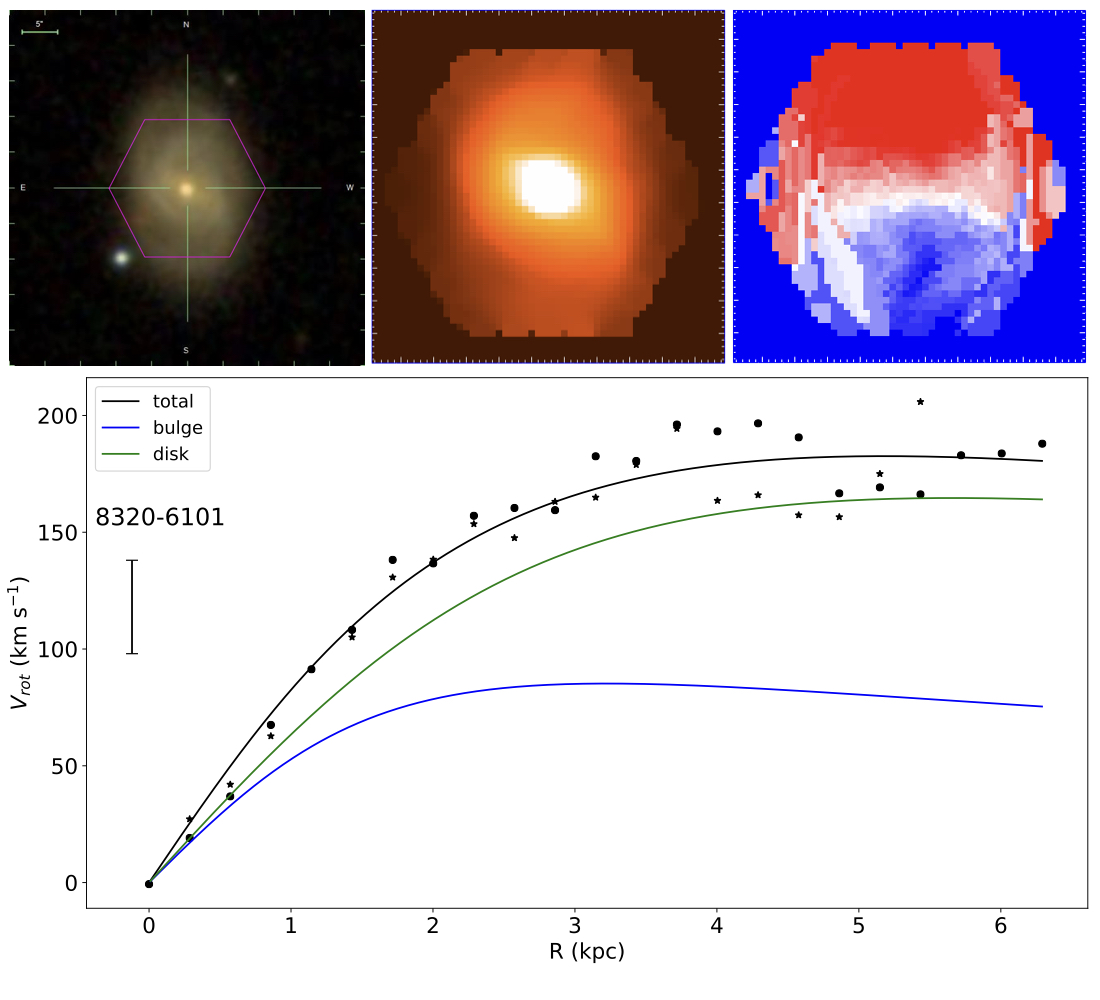}\\[2\tabcolsep]
    \includegraphics[width=0.47\linewidth]{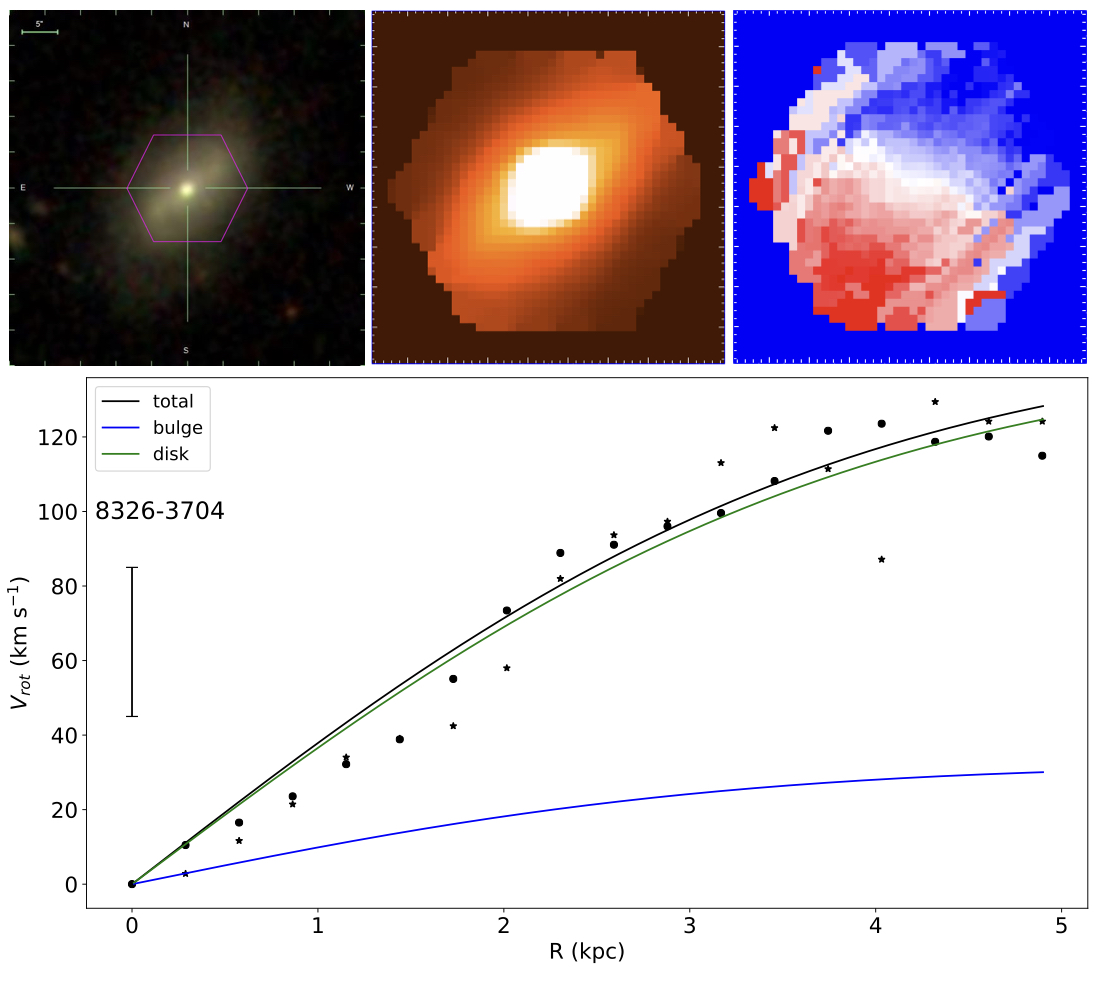}&
    \includegraphics[width=0.47\linewidth]{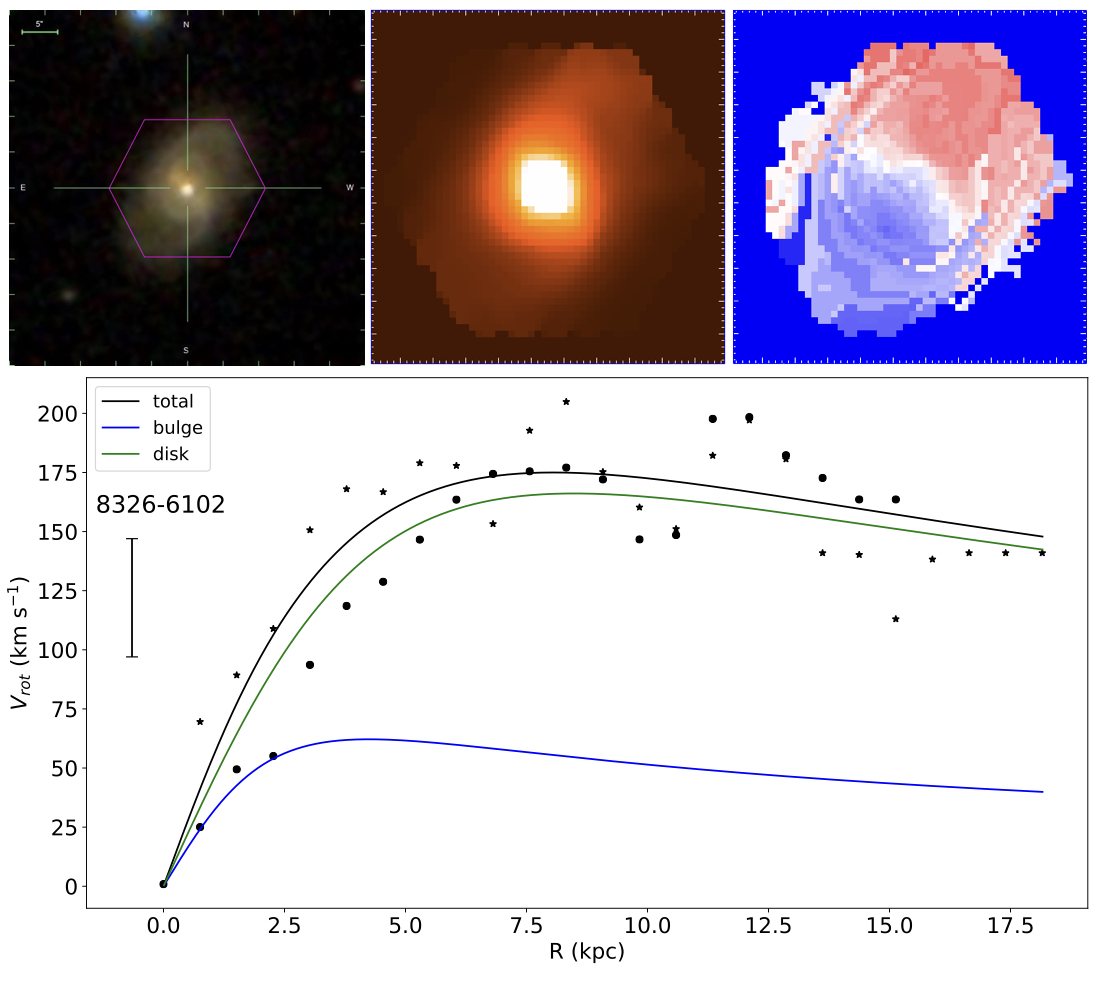}\\
    \includegraphics[width=0.47\linewidth]{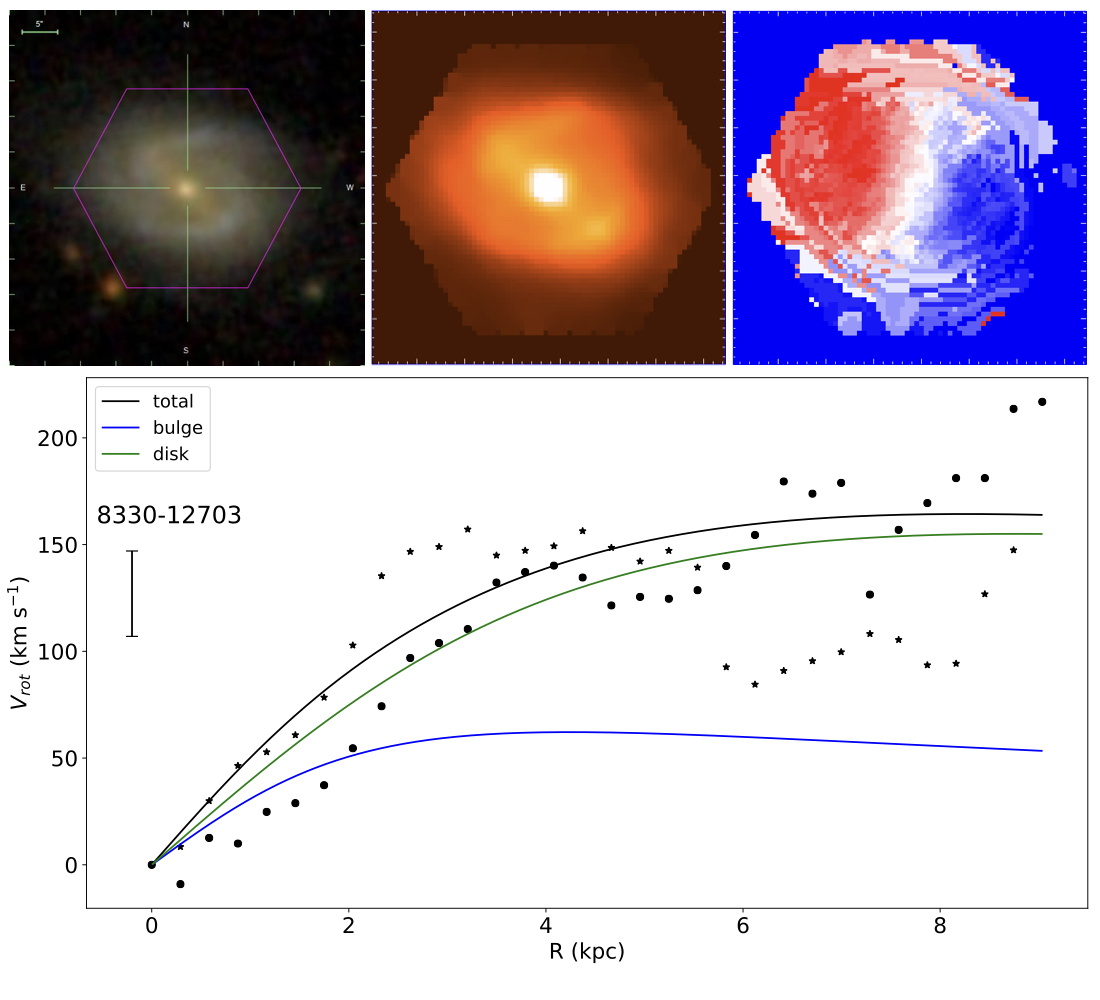}&
    \includegraphics[width=0.47\linewidth]{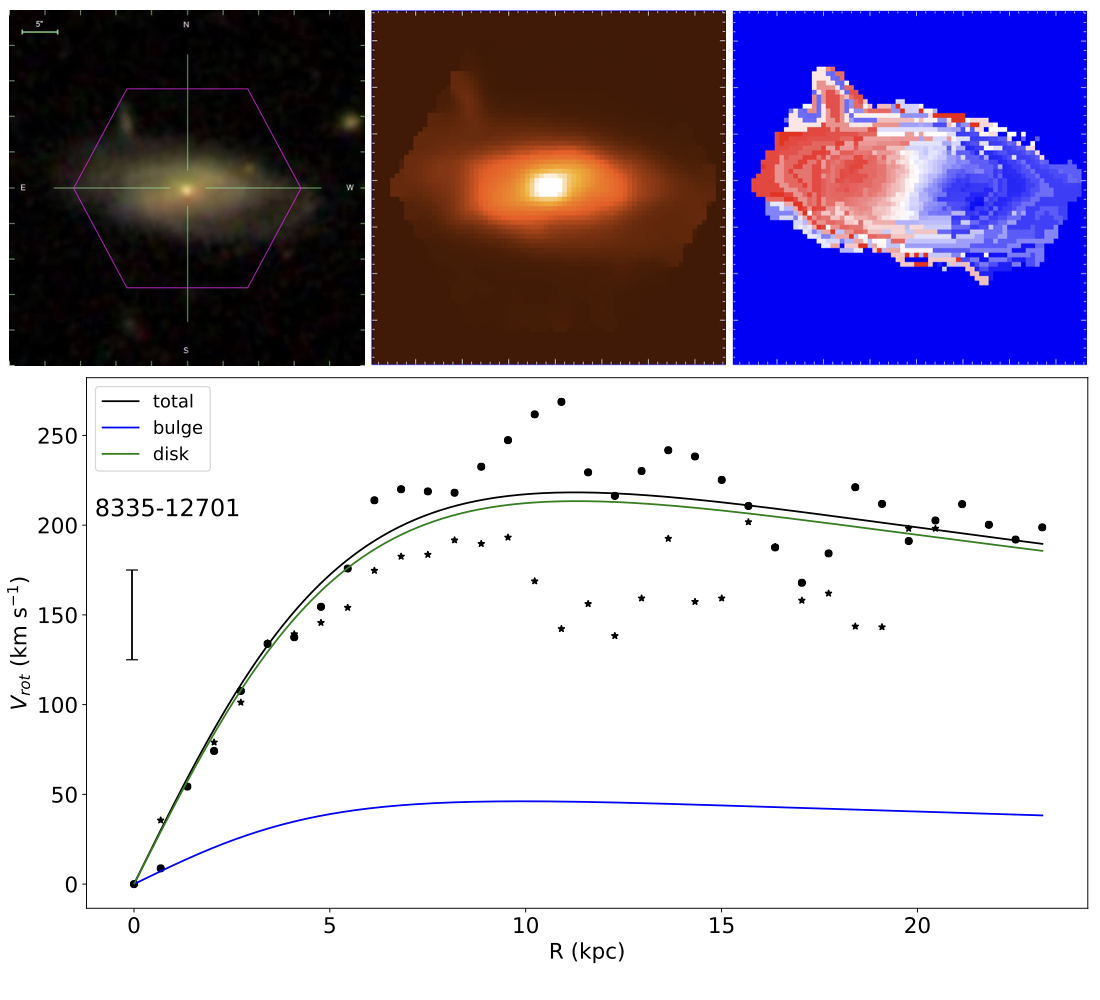}\\
\end{tabular}
\caption{continued}
\label{fig:curves3}
\end{figure*}

\begin{figure*} 
\ContinuedFloat
\centering
\begin{tabular}{cc}
    \includegraphics[width=0.47\linewidth]{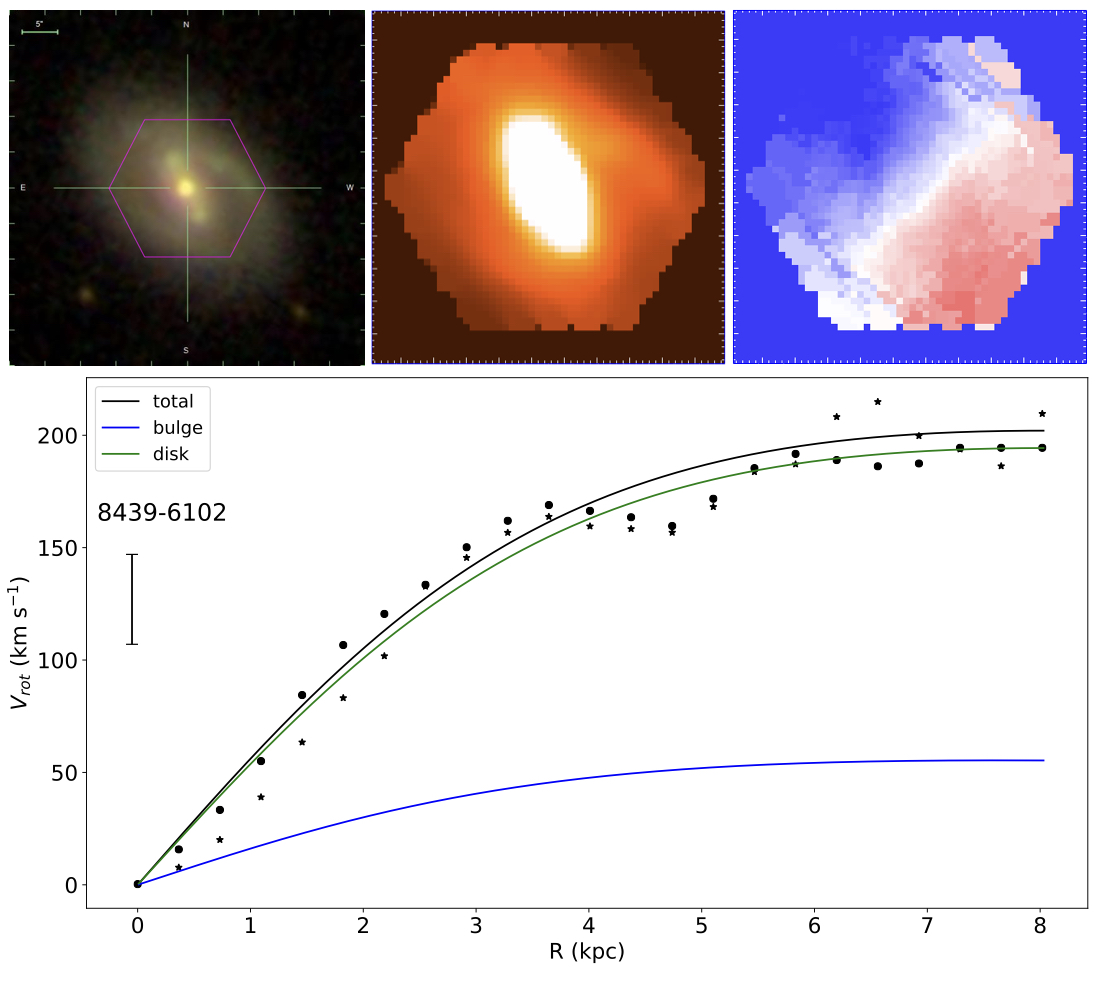}&
    \includegraphics[width=0.47\linewidth]{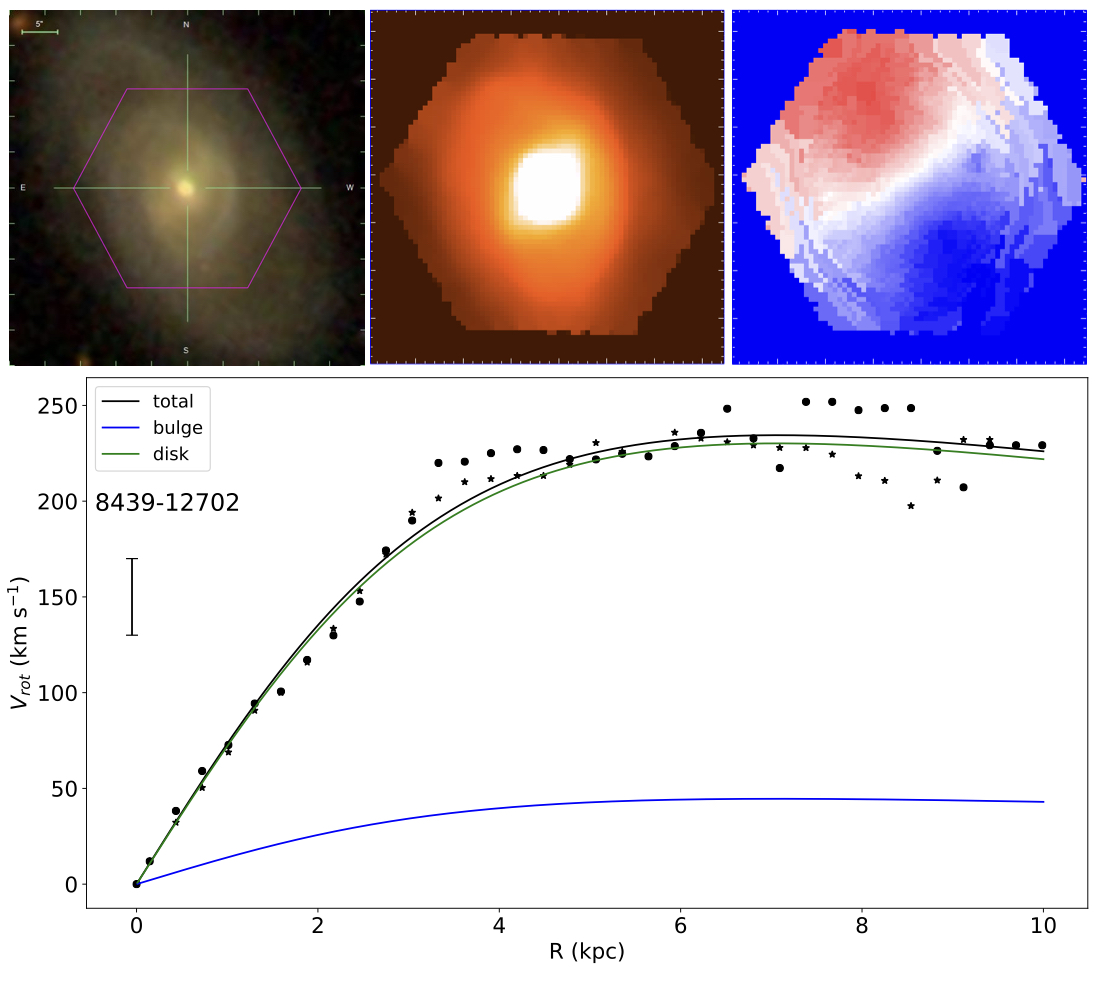}\\[2\tabcolsep]
    \includegraphics[width=0.47\linewidth]{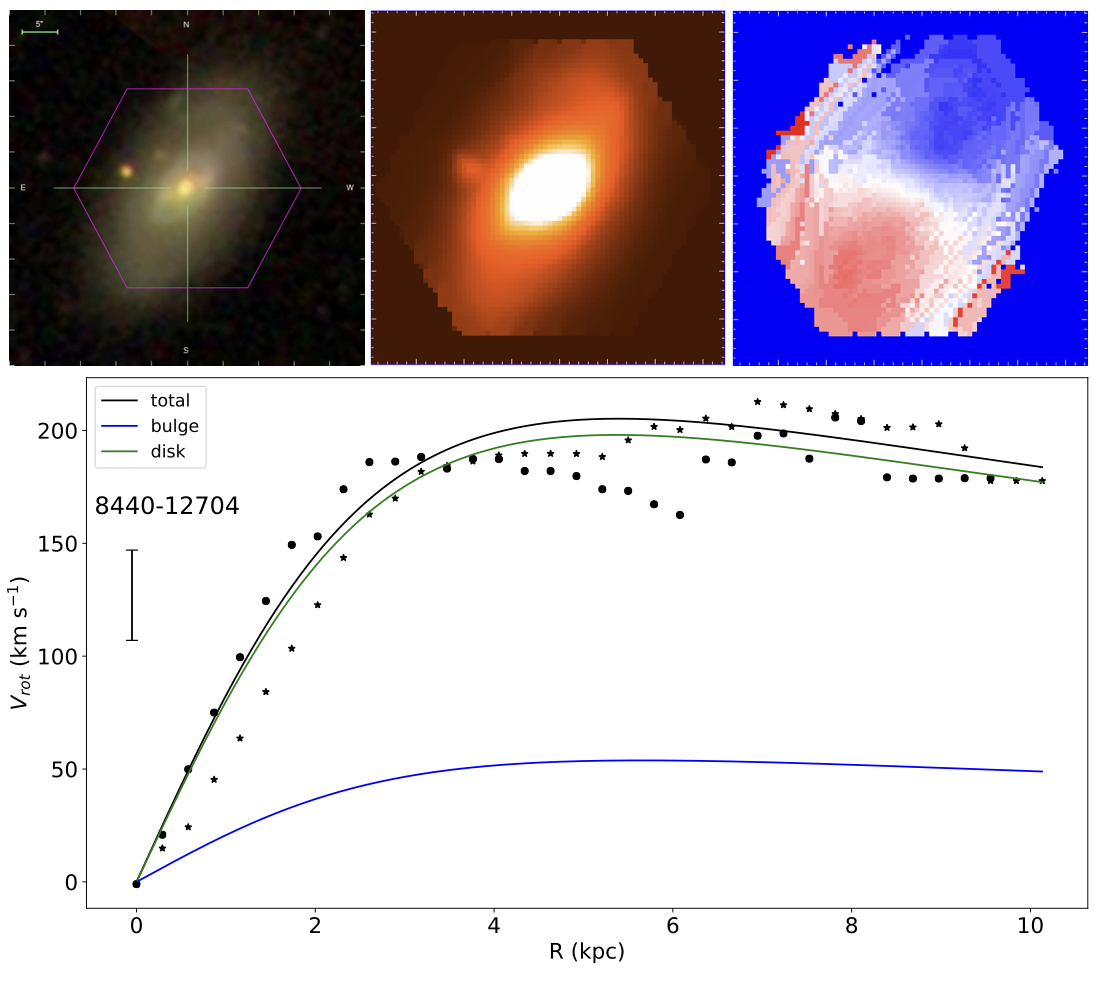}&
    \includegraphics[width=0.47\linewidth]{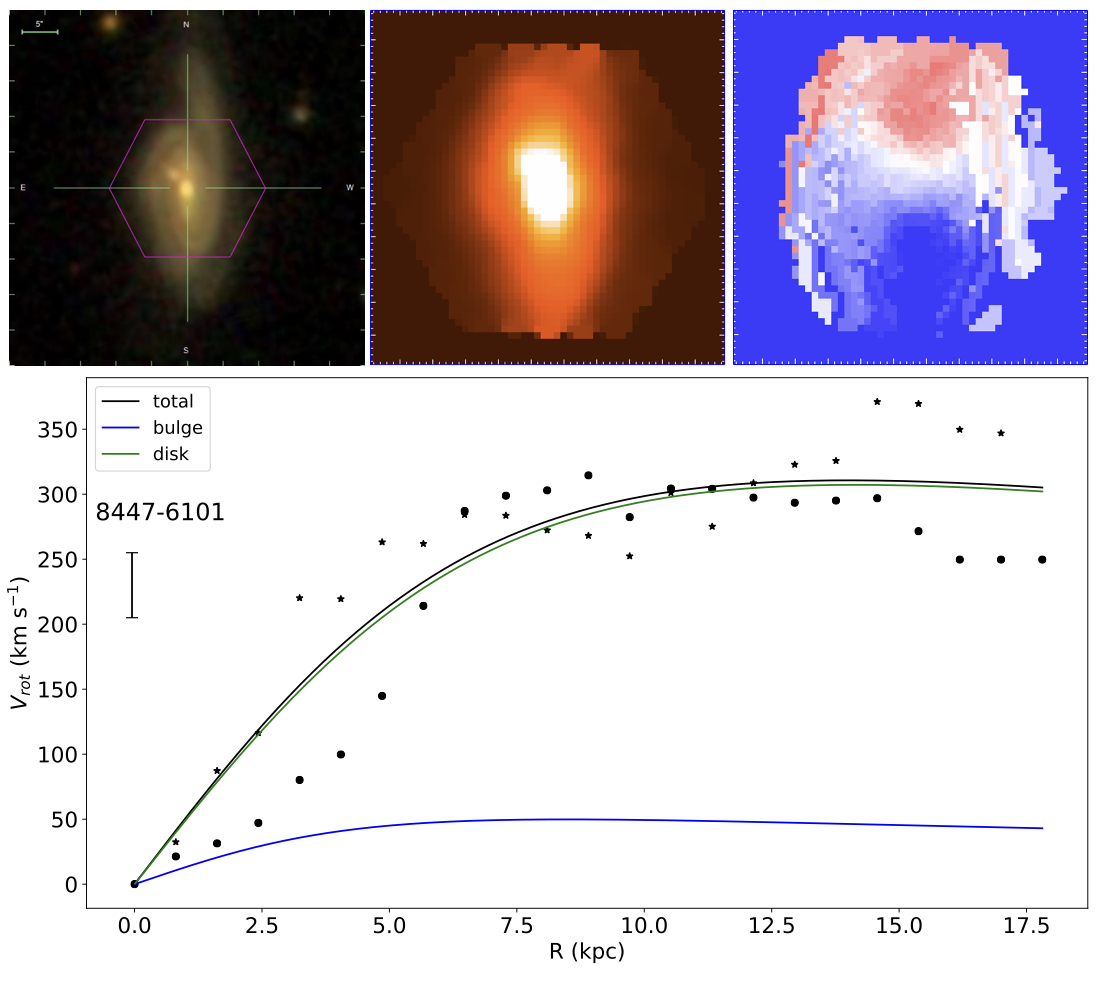}\\
    \includegraphics[width=0.47\linewidth]{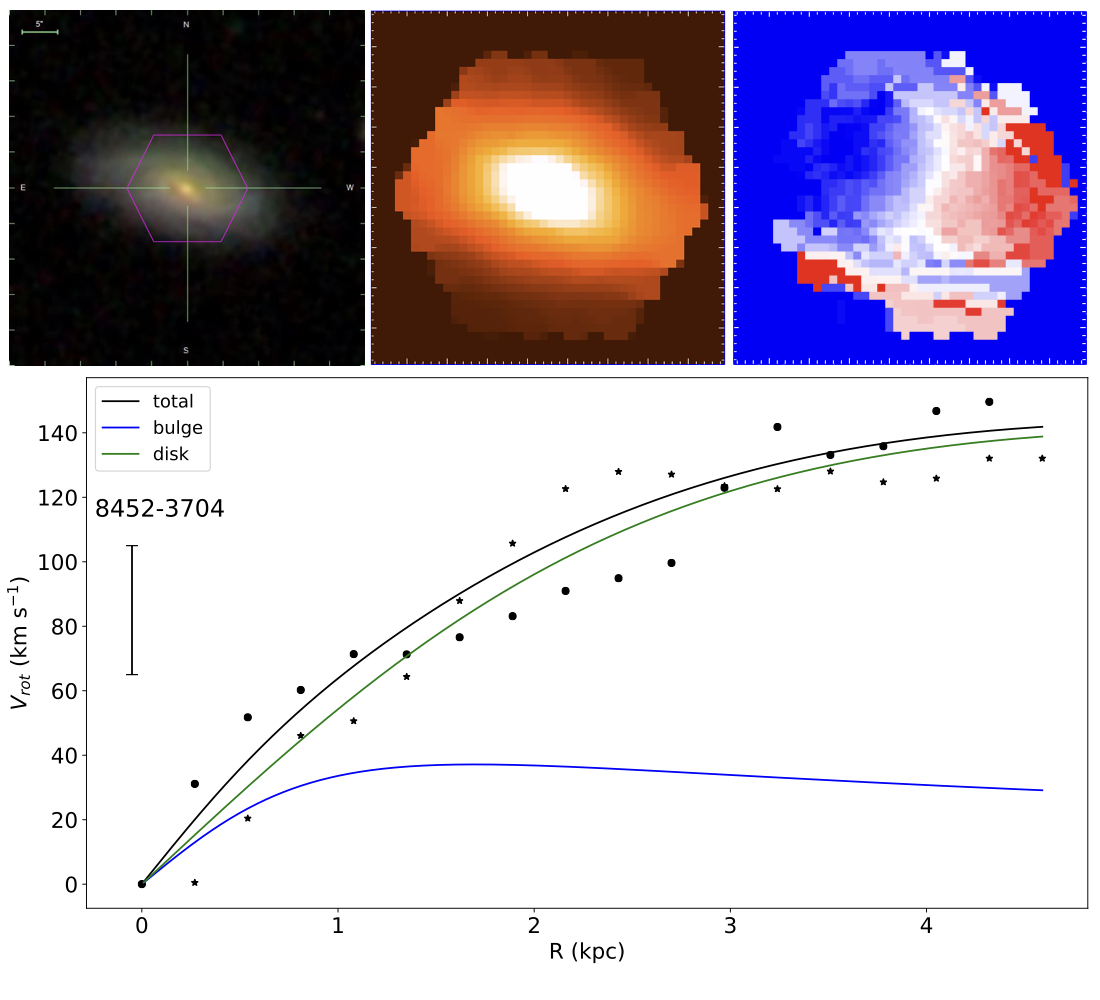}&
    \includegraphics[width=0.47\linewidth]{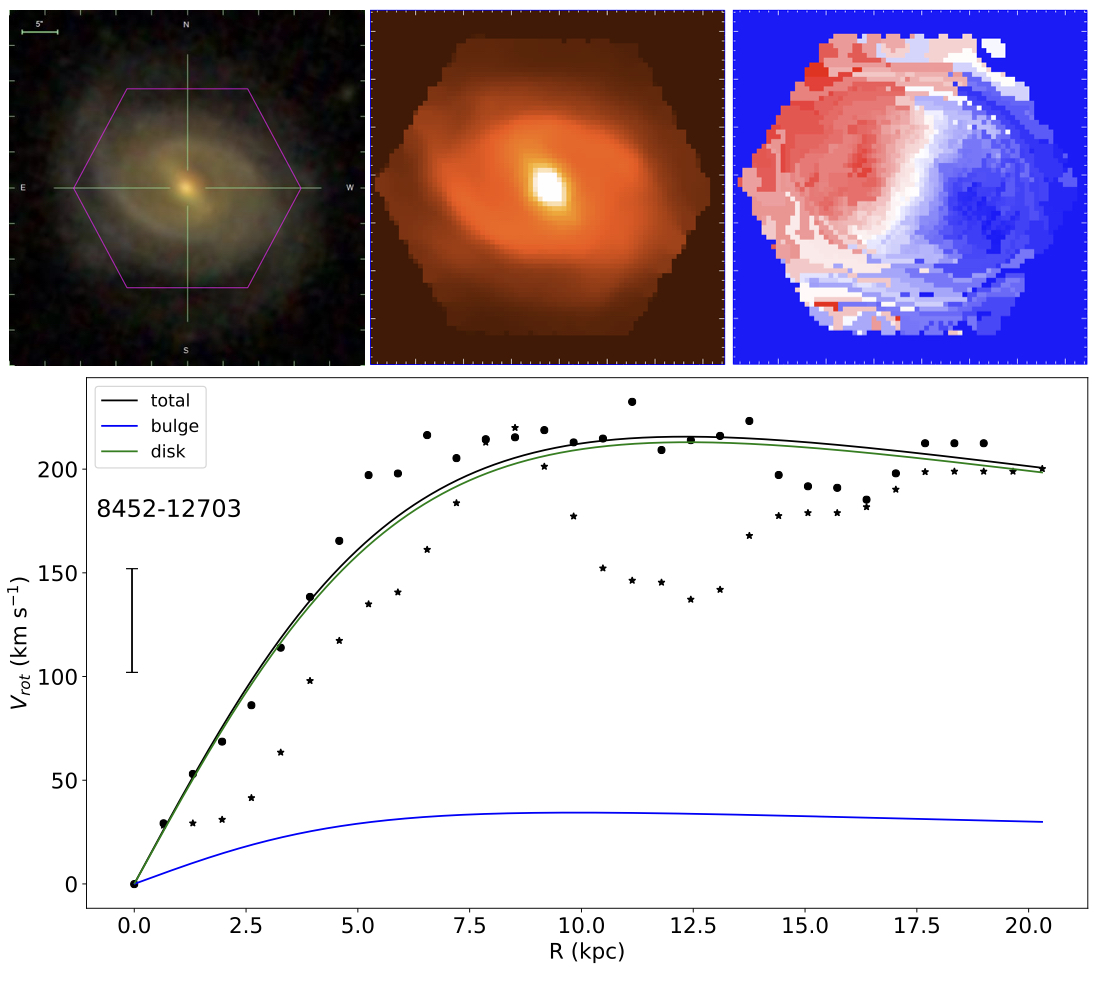}\\
\end{tabular}
\caption{continued}
\label{fig:curves4}
\end{figure*}

\begin{figure*} 
\ContinuedFloat
\centering
\begin{tabular}{cc}
    \includegraphics[width=0.47\linewidth]{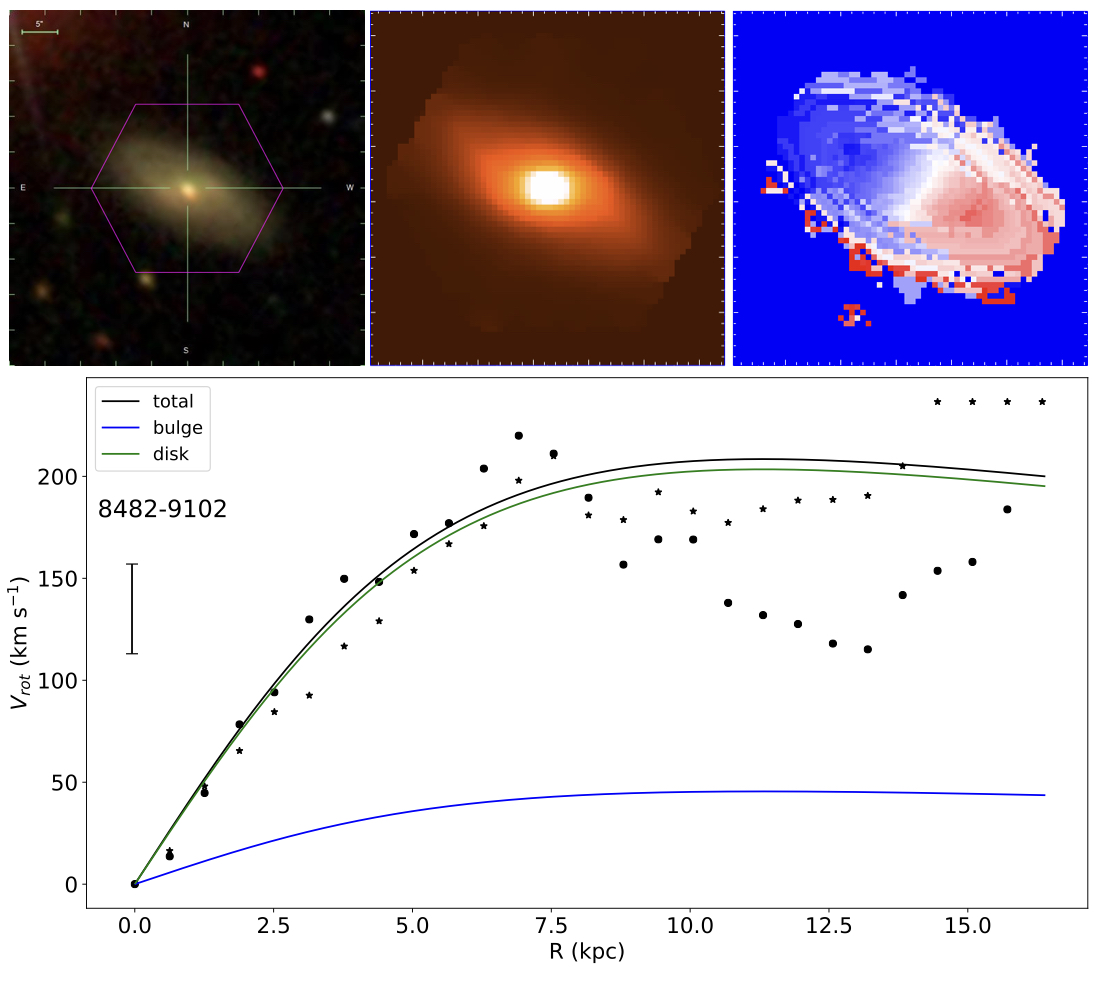}&
    \includegraphics[width=0.47\linewidth]{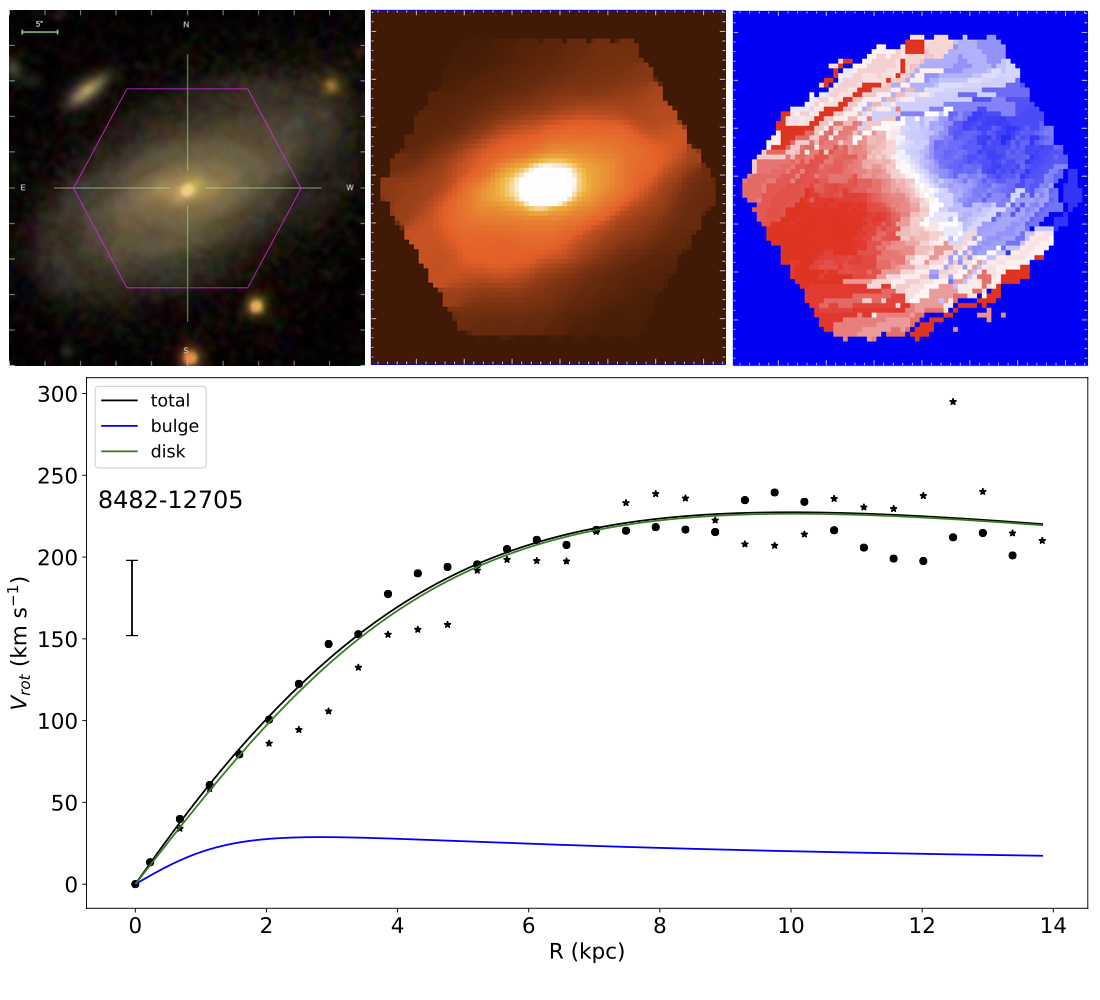}\\[2\tabcolsep]
    \includegraphics[width=0.47\linewidth]{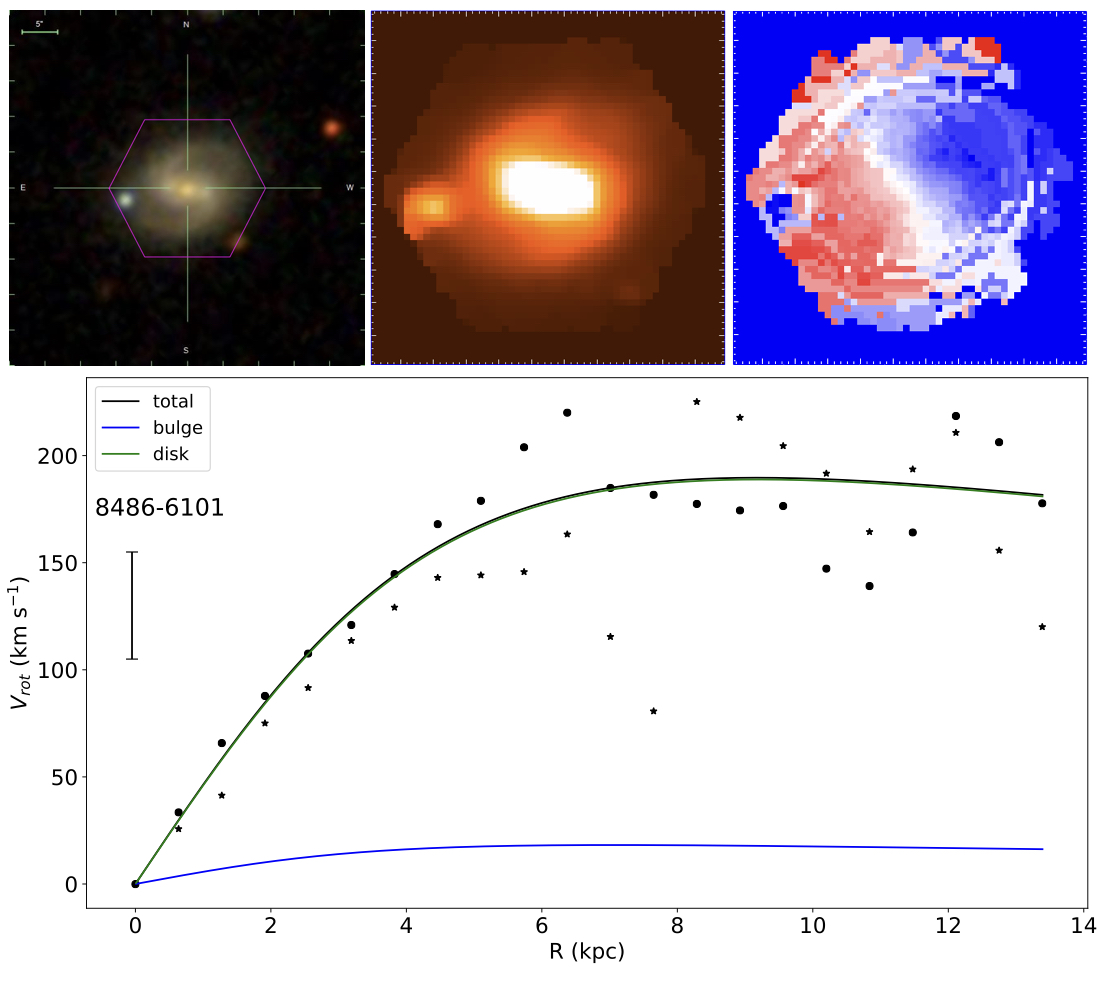}&
    \includegraphics[width=0.47\linewidth]{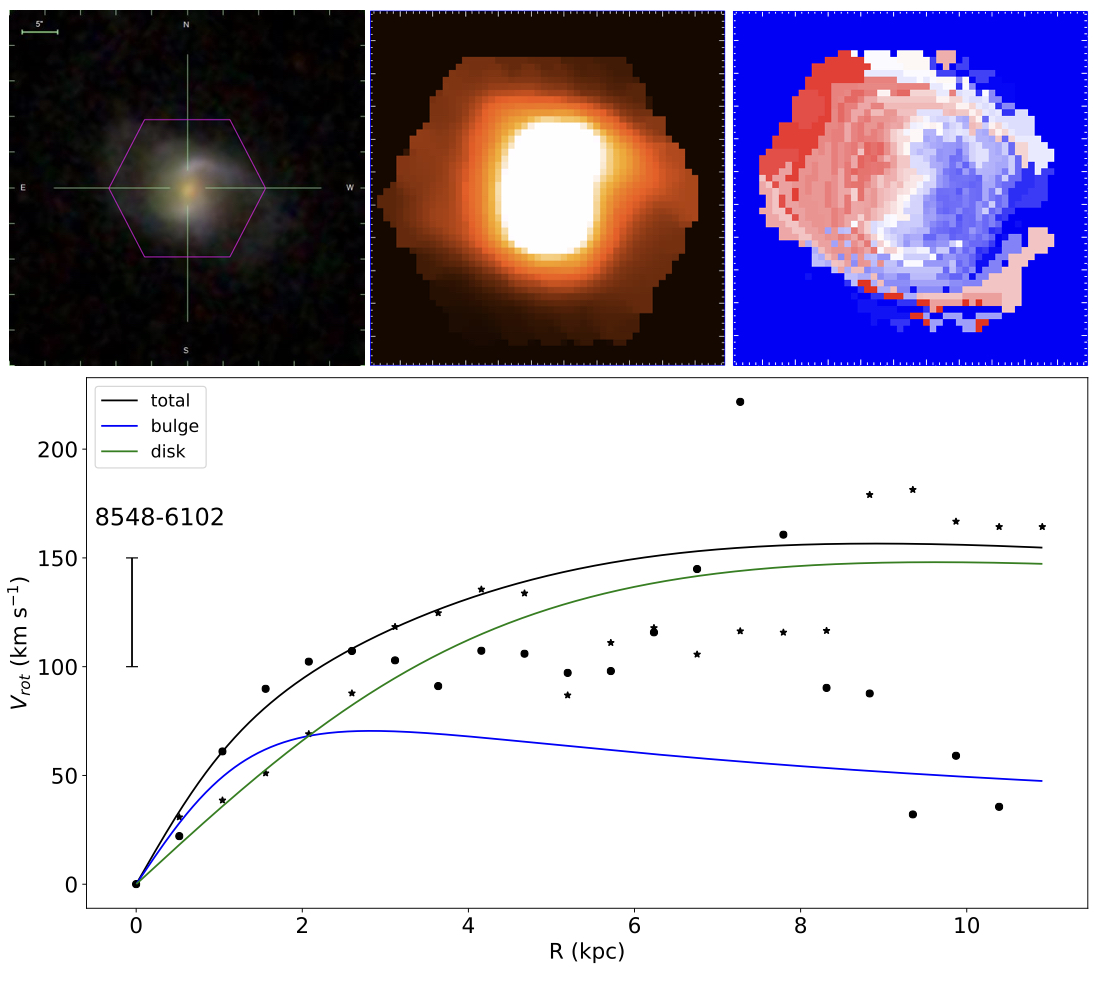}\\
    \includegraphics[width=0.47\linewidth]{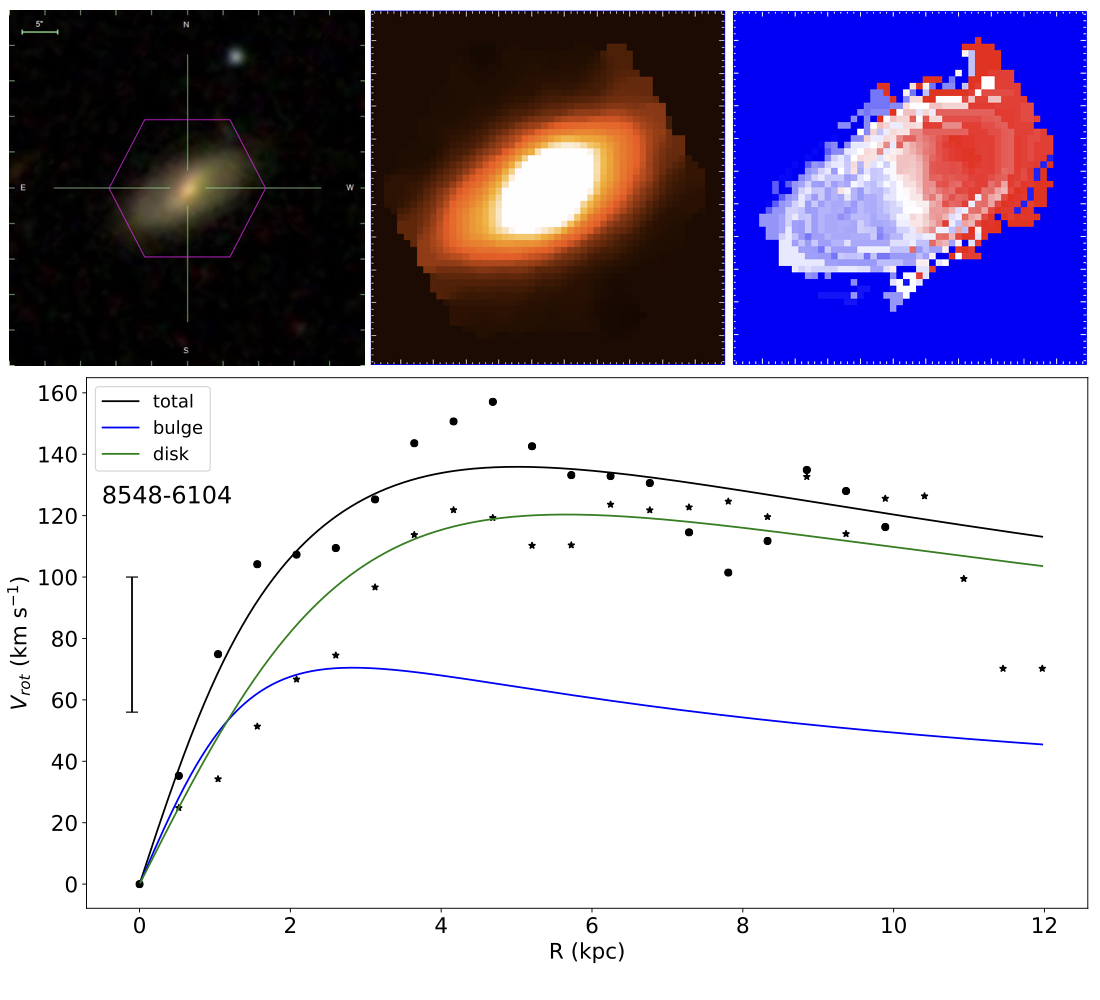}&
    \includegraphics[width=0.47\linewidth]{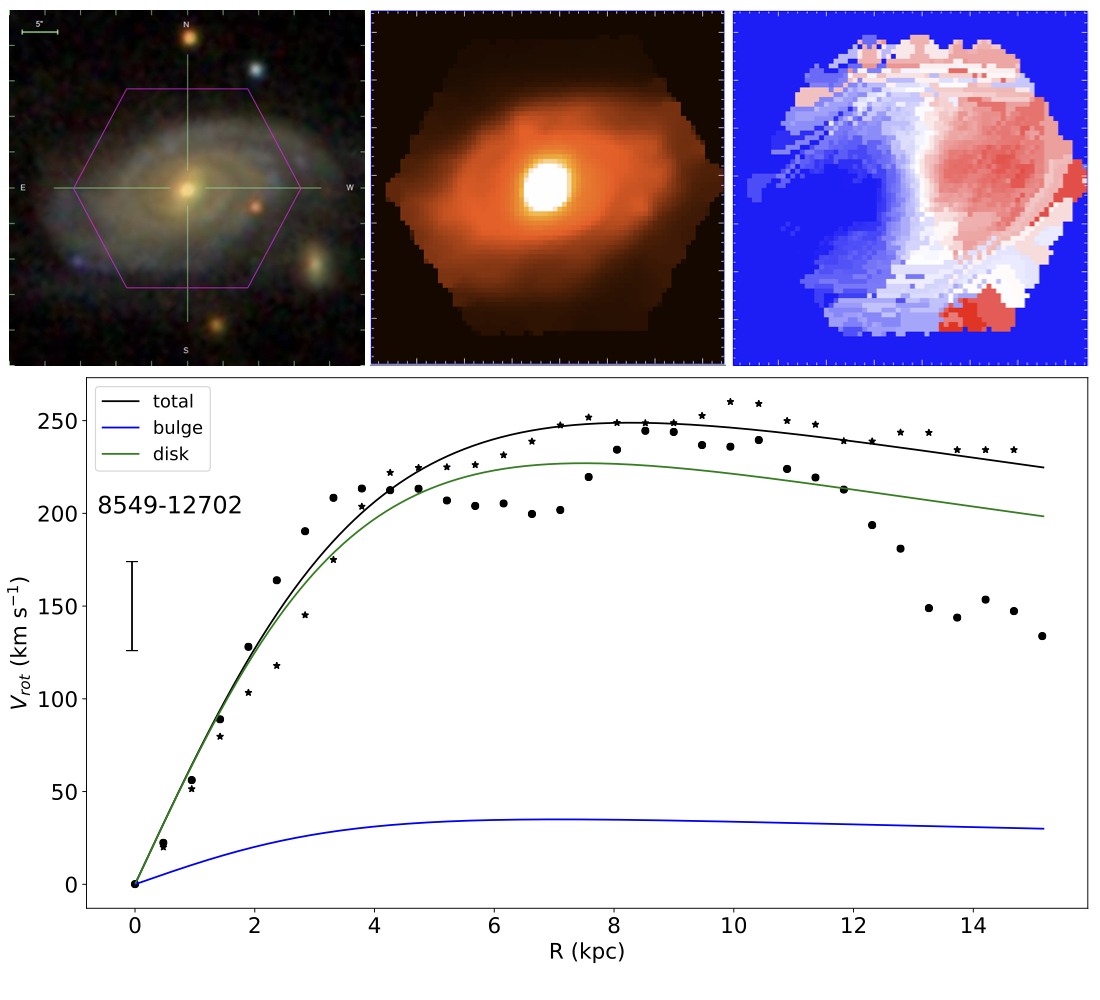}\\
\end{tabular}
\caption{continued}
\label{fig:curves5}
\end{figure*}

\begin{figure*} 
\ContinuedFloat
\centering
\begin{tabular}{cc}
    \includegraphics[width=0.47\linewidth]{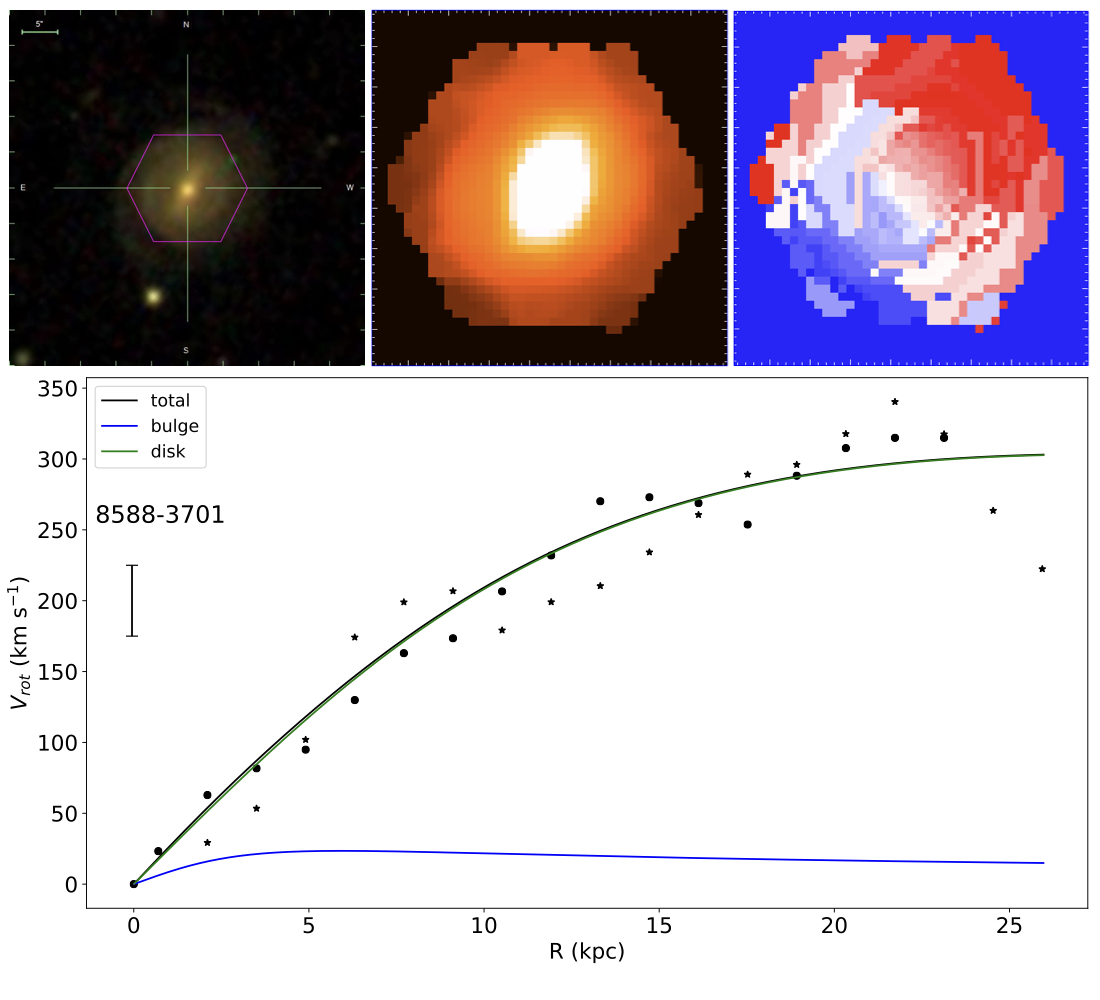}&
    \includegraphics[width=0.47\linewidth]{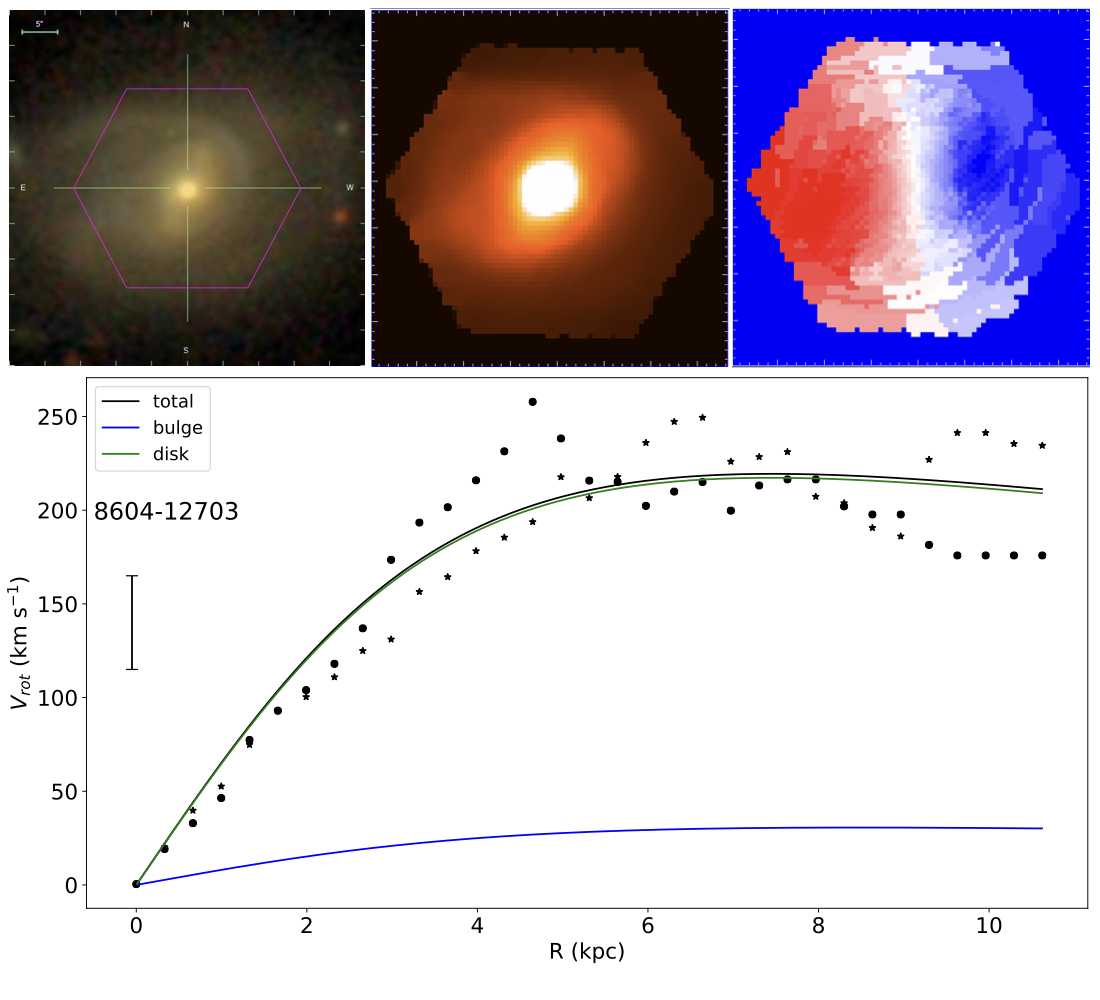}\\[2\tabcolsep]
    \includegraphics[width=0.47\linewidth]{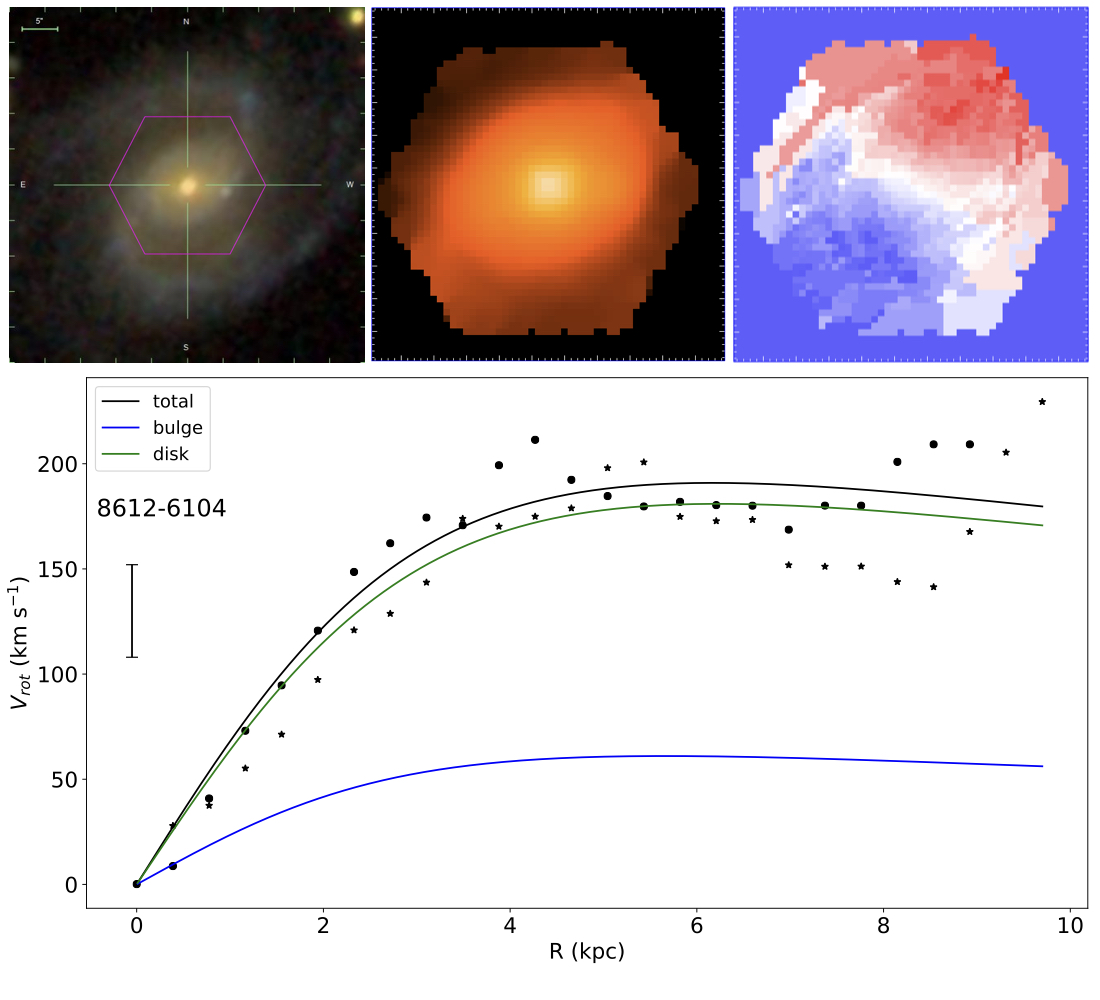}&
    \includegraphics[width=0.47\linewidth]{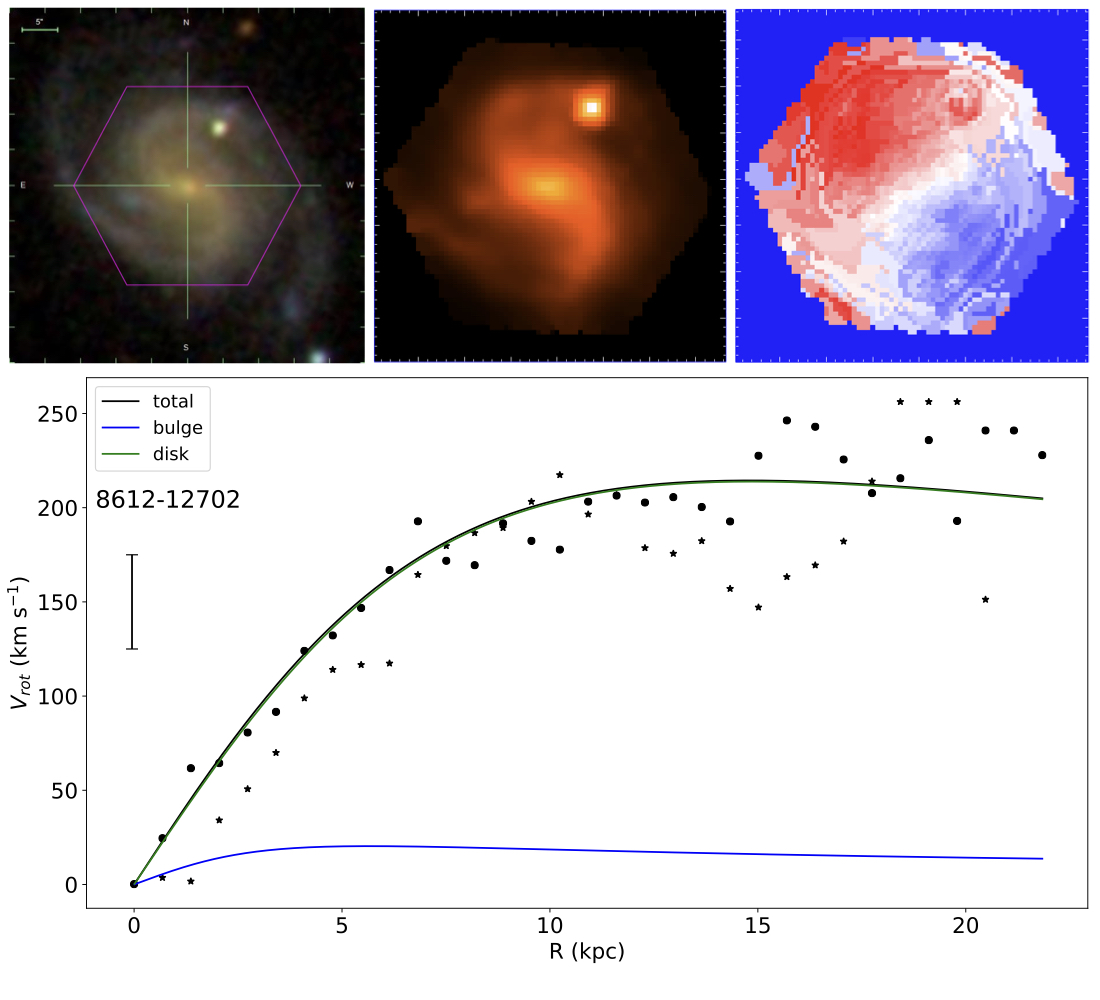}\\
\end{tabular}
\caption{continued}
\label{fig:curves6}
\end{figure*}


\section{Dynamical parameters of the galaxies}

\begin{table*}
	\centering
\caption{\label{table:params} Dynamical parameters of the galaxies. Columns: (1) MaNGA ID; (2) stellar mass; (3) and (4) shape parameters and dynamical mass of the component 1 and 2 of the fitted Miyamoto–Nagai model, respectively; (5) maximun rotation velocity measured over each rotation curve; (6) total dynamical mass given by the sum of M$_{1}$ and M$_{2}$. All data were obtained in Sect. \ref{sec: results}. As mentioned in the text, errors in stellar masses are of $\sim$ 1 \%. These errors, V$_{max}$, and M$_{dyn}$ errors only include uncertainties in our measurement process (see Sect. \ref{sub:Mstar}, \ref{sub:rotation_curves}, and \ref{sub:Mdyn} respectively for details).}
\begin{tabular}{l c c c c c}
\hline
Plate-IFU     &  M$_{star}$/ M$_{\odot}$   &    Component 1                              &       Component 2                            & V$_{max}$ (km s$^{-1}$)   & M$_{dyn}$/ M$_{\odot}$               \\ 
\hline
7495-12704    &  8.04 $\times$ 10$^{10}$   & a$_{1}=$ 2.0 kpc                            &     a$_{2}=$ 4.3 kpc                         &   205 $\pm$ 7  &  (1.45 $\pm$ 0.07) $\times$ 10$^{11}$     \\   
              &                            & b$_{1}$/a$_{1} =$ 1.00                      & b$_{2}$/a$_{2} =$0.39                        &                &                             \\  
              &                            & M$_{1}=$ 5.0 $\times$ 10$^{9}$ M$_{\odot}$  & M$_{2}=$ 1.4 $\times$ 10$^{11}$ M$_{\odot}$  &                &                             \\
\hline
7962-12703    &  2.25 $\times$ 10$^{11}$   & a$_{1}=$ 2.0 kpc                            & a$_{2}=$ 6.4 kpc                             &  255 $\pm$ 13  &  (3.4 $\pm$ 0.3) $\times$ 10$^{11}$     \\   
              &                            & b$_{1}$/a$_{1} =$ 1.00                      & b$_{2}$/a$_{2} =$0.38                        &                                                  \\  
              &                            & M$_{1}=$ 3.0 $\times$ 10$^{9}$ M$_{\odot}$  & M$_{2}=$ 3.4 $\times$ 10$^{11}$ M$_{\odot}$  &                &                                 \\   
\hline
7990-3704     &  2.08 $\times$ 10$^{10}$   & a$_{1}=$ 1.0 kpc                            & a$_{2}=$ 2.4 kpc                             &  117 $\pm$ 7   &  (2.70 $\pm$ 0.24) $\times$ 10$^{10}$    \\   
              &                            & b$_{1}$/a$_{1} =$ 1.00                      & b$_{2}$/a$_{2} =$0.37                        &                &                                 \\  
              &                            & M$_{1}=$ 1.5 $\times$ 10$^{9}$ M$_{\odot}$  & M$_{2}=$ 2.5 $\times$ 10$^{10}$ M$_{\odot}$  &                &                                  \\
\hline
7990-9101     &  2.47 $\times$ 10$^{10}$   & a$_{1}=$ 1.0 kpc                            & a$_{2}=$ 4.3 kpc                             &  132 $\pm$ 8   &  (7.10 $\pm$ 0.64) $\times$ 10$^{10}$     \\   
              &                            & b$_{1}$/a$_{1} =$ 1.00                      & b$_{2}$/a$_{2} =$0.58                        &                &                                  \\  
              &                            & M$_{1}=$ 1.0 $\times$ 10$^{9}$ M$_{\odot}$  & M$_{2}=$ 7.0 $\times$ 10$^{10}$ M$_{\odot}$  &                &                                  \\
\hline
7992-6104     &  3.59 $\times$ 10$^{10}$   & a$_{1}=$ 0.1 kpc                            & a$_{2}=$ 6.1 kpc                             &  221 $\pm$ 11  &   (3.60 $\pm$ 0.22) $\times$ 10$^{11}$    \\   
              &                            & b$_{1}$/a$_{1} =$ 1.00                      & b$_{2}$/a$_{2} =$0.57                        &                &                                  \\  
              &                            & M$_{1}=$ 3.0 $\times$ 10$^{8}$ M$_{\odot}$  & M$_{2}=$ 3.6 $\times$ 10$^{11}$ M$_{\odot}$  &                &                                  \\
\hline
8082-6102     & 8.02 $\times$ 10$^{10}$    & a$_{1}=$ 1.0 kpc                            & a$_{2}=$ 4.4 kpc                             &  221 $\pm$ 9   &   (2.7 $\pm$ 0.1) $\times$ 10$^{10}$    \\   
              &                            & b$_{1}$/a$_{1} =$ 1.00                      & b$_{2}$/a$_{2} =$0.36                        &                &                                  \\  
              &                            & M$_{1}=$ 8.0 $\times$ 10$^{9}$ M$_{\odot}$  & M$_{2}=$ 1.7 $\times$ 10$^{11}$ M$_{\odot}$  &                &                                  \\
\hline
8083-6102     & 1.29 $\times$ 10$^{11}$    & a$_{1}=$ 1.5 kpc                            & a$_{2}=$ 3.3 kpc                             & 220 $\pm$ 9    &  (1.20 $\pm$ 0.05) $\times$ 10$^{11}$     \\   
              &                            & b$_{1}$/a$_{1} =$ 1.00                      & b$_{2}$/a$_{2} =$ 0.32                       &                &                                  \\  
              &                            & M$_{1}=$ 5.0 $\times$ 10$^{9}$ M$_{\odot}$  & M$_{2}=$ 1.2 $\times$ 10$^{11}$ M$_{\odot}$  &                &                                  \\
\hline
8083-12704    & 3.66 $\times$ 10$^{10}$    & a$_{1}=$ 2.0 kpc                            & a$_{2}=$ 2.2 kpc                             & 134 $\pm$ 8    &   (3.8 $\pm$ 0.3) $\times$ 10$^{10}$    \\   
              &                            & b$_{1}$/a$_{1} =$ 1.00                      & b$_{2}$/a$_{2} =$ 0.59                       &                &                                  \\  
              &                            & M$_{1}=$ 1.0 $\times$ 10$^{9}$ M$_{\odot}$  & M$_{2}=$ 3.7 $\times$ 10$^{10}$ M$_{\odot}$  &                &                                  \\
\hline
8133-3701     & 2.44 $\times$ 10$^{10}$    & a$_{1}=$ 0.8 kpc                            & a$_{2}=$ 6.6 kpc                             & 173 $\pm$ 12   &    (2.0 $\pm$ 0.1) $\times$ 10$^{11}$     \\   
              &                            & b$_{1}$/a$_{1} =$ 1.00                      & b$_{2}$/a$_{2} =$ 0.48                       &                &                                  \\  
              &                            & M$_{1}=$ 4.0 $\times$ 10$^{9}$ M$_{\odot}$  & M$_{2}=$ 2.0 $\times$ 10$^{11}$ M$_{\odot}$  &                &                                  \\
\hline
8134-6102     & 6.06 $\times$ 10$^{10}$    & a$_{1}=$ 3.0 kpc                            & a$_{2}=$ 4.5 kpc                             & 266 $\pm$ 8    &  (2.50 $\pm$ 0.07) $\times$ 10$^{11}$   \\   
              &                            & b$_{1}$/a$_{1} =$ 1.00                      & b$_{2}$/a$_{2} =$ 0.33                       &                &                                  \\  
              &                            & M$_{1}=$ 6.0 $\times$ 10$^{9}$ M$_{\odot}$  & M$_{2}=$ 2.5 $\times$ 10$^{11}$ M$_{\odot}$  &                &                                  \\
\hline
8137-9102     & 4.34 $\times$ 10$^{10}$    & a$_{1}=$ 2.0 kpc                            & a$_{2}=$ 5.7 kpc                             &  178 $\pm$ 5   &  (1.5 $\pm$ 0.5) $\times$ 10$^{11}$     \\   
              &                            & b$_{1}$/a$_{1} =$ 1.00                      & b$_{2}$/a$_{2} =$ 0.40                       &                &                                  \\  
              &                            & M$_{1}=$ 6.0 $\times$ 10$^{9}$ M$_{\odot}$  & M$_{2}=$ 1.5 $\times$ 10$^{11}$ M$_{\odot}$  &                &                                  \\
\hline
8140-12701    & 4.53 $\times$ 10$^{10}$    & a$_{1}=$ 2.0 kpc                            & a$_{2}=$ 4.0 kpc                             &  144 $\pm$ 5   &  (6.6 $\pm$ 0.6) $\times$ 10$^{10}$     \\   
              &                            & b$_{1}$/a$_{1} =$ 1.00                      & b$_{2}$/a$_{2} =$ 0.35                       &                &                                  \\  
              &                            & M$_{1}=$ 6.0 $\times$ 10$^{9}$ M$_{\odot}$  & M$_{2}=$ 6.0 $\times$ 10$^{10}$ M$_{\odot}$  &                &                                  \\
\hline
8140-12703    & 1.07 $\times$ 10$^{11}$    & a$_{1}=$ 1.0 kpc                            & a$_{2}=$ 4.3 kpc                             &  199 $\pm$ 6   &  (1.5 $\pm$ 0.5) $\times$ 10$^{11}$    \\   
              &                            & b$_{1}$/a$_{1} =$ 1.00                      & b$_{2}$/a$_{2} =$ 0.53                       &                &                                  \\  
              &                            & M$_{1}=$ 3.0 $\times$ 10$^{9}$ M$_{\odot}$  & M$_{2}=$ 1.5 $\times$ 10$^{11}$ M$_{\odot}$  &                &                                  \\
\hline
8243-6103     & 5.83 $\times$ 10$^{10}$    & a$_{1}=$ 3.0 kpc                            & a$_{2}=$ 3.5 kpc                             &  269 $\pm$ 8   &  (2.2 $\pm$ 0.1) $\times$ 10$^{11}$     \\   
              &                            & b$_{1}$/a$_{1} =$ 1.00                      & b$_{2}$/a$_{2} =$ 0.42                       &                &                                  \\  
              &                            & M$_{1}=$ 4.0 $\times$ 10$^{10}$ M$_{\odot}$ & M$_{2}=$ 1.8 $\times$ 10$^{11}$ M$_{\odot}$  &                &                                  \\
\hline
8244-3703     & 4.98 $\times$ 10$^{10}$    & a$_{1}=$ 2.0 kpc                            & a$_{2}=$ 5.4 kpc                             &  205 $\pm$ 9   &  (1.90 $\pm$ 0.06) $\times$ 10$^{11}$    \\   
              &                            & b$_{1}$/a$_{1} =$ 1.00                      & b$_{2}$/a$_{2} =$ 0.44                       &                &                                  \\  
              &                            & M$_{1}=$ 8.0 $\times$ 10$^{9}$ M$_{\odot}$  & M$_{2}=$ 1.9 $\times$ 10$^{11}$ M$_{\odot}$  &                &                                  \\
\hline
8249-6101     & 3.16 $\times$ 10$^{10}$    & a$_{1}=$ 1.5 kpc                            & a$_{2}=$ 2.5 kpc                             &  122 $\pm$ 8   &  (2.7 $\pm$ 0.2) $\times$ 10$^{10}$    \\   
              &                            & b$_{1}$/a$_{1} =$ 1.00                      & b$_{2}$/a$_{2} =$ 0.20                       &                &                                  \\  
              &                            & M$_{1}=$ 2.0 $\times$ 10$^{9}$ M$_{\odot}$  & M$_{2}=$ 2.5 $\times$ 10$^{10}$ M$_{\odot}$  &                &                                  \\
\hline              
8254-9101     & 1.31 $\times$ 10$^{11}$    & a$_{1}=$ 0.5 kpc                            & a$_{2}=$ 4.8 kpc                             &  340 $\pm$ 12  &  (4.8 $\pm$ 0.1) $\times$ 10$^{11}$     \\   
              &                            & b$_{1}$/a$_{1} =$ 1.00                      & b$_{2}$/a$_{2} =$ 0.42                       &                &                                  \\  
              &                            & M$_{1}=$ 5.0 $\times$ 10$^{9}$ M$_{\odot}$  & M$_{2}=$ 4.8 $\times$ 10$^{11}$ M$_{\odot}$  &                &                                  \\
\hline
\end{tabular}
 \end{table*}

\begin{table*}
 \contcaption{}
 \label{tab:continued}
 \begin{tabular}{l c c c c c}
  \hline

8256-6101     & 3.40 $\times$ 10$^{10}$    & a$_{1}=$ 1.5 kpc                            & a$_{2}=$ 3.0 kpc                             &   188 $\pm$ 6  &  (8.3 $\pm$ 0.2) $\times$ 10$^{10}$     \\ 
              &                            & b$_{1}$/a$_{1} =$ 1.00                      & b$_{2}$/a$_{2} =$ 0.33                       &                &                                  \\  
              &                            & M$_{1}=$ 7.0 $\times$ 10$^{9}$ M$_{\odot}$  & M$_{2}=$ 7.6 $\times$ 10$^{10}$ M$_{\odot}$  &                &                                  \\
\hline
8257-3703     & 5.92 $\times$ 10$^{10}$    & a$_{1}=$ 1.5 kpc                            & a$_{2}=$ 2.4 kpc                             &   163 $\pm$ 5  &  (5.1 $\pm$ 0.2) $\times$ 10$^{10}$     \\   
              &                            & b$_{1}$/a$_{1} =$ 1.00                      & b$_{2}$/a$_{2} =$ 0.33                       &                &                                  \\  
              &                            & M$_{1}=$ 3.0 $\times$ 10$^{9}$ M$_{\odot}$  & M$_{2}=$ 4.8 $\times$ 10$^{10}$ M$_{\odot}$  &                &                                  \\

\hline
8257-6101     & 3.67 $\times$ 10$^{10}$    & a$_{1}=$ 1.5 kpc                            & a$_{2}=$ 3.2 kpc                             &   155 $\pm$ 9  &  (6.0 $\pm$ 0.3) $\times$ 10$^{10}$     \\   
              &                            & b$_{1}$/a$_{1} =$ 1.00                      & b$_{2}$/a$_{2} =$ 0.31                       &                &                                  \\  
              &                            & M$_{1}=$ 3.0 $\times$ 10$^{9}$ M$_{\odot}$  & M$_{2}=$ 5.7 $\times$ 10$^{10}$ M$_{\odot}$  &                &                                  \\
\hline
8312-12702    & 6.59 $\times$ 10$^{10}$    & a$_{1}=$ 2.5 kpc                            & a$_{2}=$ 4.5 kpc                             &   180 $\pm$ 6  &  (1.20 $\pm$ 0.04) $\times$ 10$^{11}$     \\   
              &                            & b$_{1}$/a$_{1} =$ 1.00                      & b$_{2}$/a$_{2} =$ 0.40                       &                &                                  \\  
              &                            & M$_{1}=$ 3.0 $\times$ 10$^{9}$ M$_{\odot}$  & M$_{2}=$ 1.2 $\times$ 10$^{11}$ M$_{\odot}$  &                &                                  \\
\hline
8312-12704    & 3.58 $\times$ 10$^{10}$    & a$_{1}=$ 1.1 kpc                            & a$_{2}=$ 8.0 kpc                             &   149 $\pm$ 10 &  (9.8 $\pm$ 0.9) $\times$ 10$^{10}$     \\   
              &                            & b$_{1}$/a$_{1} =$ 1.00                      & b$_{2}$/a$_{2} =$ 0.50                       &                &                                  \\  
              &                            & M$_{1}=$ 2.8 $\times$ 10$^{10}$ M$_{\odot}$ & M$_{2}=$ 7.0 $\times$ 10$^{11}$ M$_{\odot}$  &                &                                  \\
\hline
8313-9101     & 1.28 $\times$ 10$^{11}$    & a$_{1}=$ 1.7 kpc                            & a$_{2}=$ 4.5 kpc                             &   236 $\pm$ 7  &  (1.90 $\pm$ 0.05) $\times$ 10$^{11}$    \\   
              &                            & b$_{1}$/a$_{1} =$ 1.00                      & b$_{2}$/a$_{2} =$ 0.40                       &                &                                  \\  
              &                            & M$_{1}=$ 4.0 $\times$ 10$^{10}$ M$_{\odot}$ & M$_{2}=$ 1.5 $\times$ 10$^{11}$ M$_{\odot}$  &                &                                  \\
\hline
8317-12704    & 2.33 $\times$ 10$^{11}$    & a$_{1}=$ 2.0 kpc                            & a$_{2}=$ 6.0 kpc                             &   229 $\pm$ 10 &  (2.7 $\pm$ 0.2) $\times$ 10$^{11}$     \\   
              &                            & b$_{1}$/a$_{1} =$ 1.00                      & b$_{2}$/a$_{2} =$ 0.46                       &                &                                  \\  
              &                            & M$_{1}=$ 1.0 $\times$ 10$^{10}$ M$_{\odot}$ & M$_{2}=$ 2.6 $\times$ 10$^{11}$ M$_{\odot}$  &                &                                  \\
\hline
8318-12703    & 1.3 $\times$ 10$^{11}$     & a$_{1}=$ 2.0 kpc                            & a$_{2}=$ 8.3 kpc                             &   214 $\pm$ 9  &  (2.90 $\pm$ 0.09) $\times$ 10$^{10}$     \\   
              &                            & b$_{1}$/a$_{1} =$ 1.00                      & b$_{2}$/a$_{2} =$ 0.44                       &                &                                  \\  
              &                            & M$_{1}=$ 5.0 $\times$ 10$^{10}$ M$_{\odot}$ & M$_{2}=$ 2.4 $\times$ 10$^{11}$ M$_{\odot}$  &                &                                  \\
\hline
8320-6101     & 3.79 $\times$ 10$^{10}$    & a$_{1}=$ 1.1 kpc                            & a$_{2}=$ 3.2 kpc                             &   183 $\pm$ 5  &  (7.6 $\pm$ 0.3) $\times$ 10$^{10}$     \\   
              &                            & b$_{1}$/a$_{1} =$ 1.00                      & b$_{2}$/a$_{2} =$ 0.25                       &                &                                  \\  
              &                            & M$_{1}=$ 1.0 $\times$ 10$^{10}$ M$_{\odot}$ & M$_{2}=$ 6.6 $\times$ 10$^{10}$ M$_{\odot}$  &                &                                  \\
\hline
8326-3704     & 1.34 $\times$ 10$^{10}$    & a$_{1}=$ 2.5 kpc                            & a$_{2}=$ 4.0 kpc                             &   128 $\pm$ 4  &  (7.3 $\pm$ 0.4) $\times$ 10$^{10}$    \\   
              &                            & b$_{1}$/a$_{1} =$ 1.00                      & b$_{2}$/a$_{2} =$ 0.50                       &                &                                  \\  
              &                            & M$_{1}=$ 3.0 $\times$ 10$^{9}$ M$_{\odot}$  & M$_{2}=$ 7.0 $\times$ 10$^{10}$ M$_{\odot}$  &                &                                  \\
\hline
8326-6102     & 1.72 $\times$ 10$^{11}$    & a$_{1}=$ 1.5 kpc                            & a$_{2}=$ 5.0 kpc                             &   175 $\pm$ 12 &  (1.00 $\pm$ 0.07) $\times$ 10$^{11}$    \\   
              &                            & b$_{1}$/a$_{1} =$ 1.00                      & b$_{2}$/a$_{2} =$ 0.20                       &                &                                  \\  
              &                            & M$_{1}=$ 7.0 $\times$ 10$^{9}$ M$_{\odot}$  & M$_{2}=$ 1.0 $\times$ 10$^{11}$ M$_{\odot}$  &                &                                  \\
\hline
8330-12703    & 3.53 $\times$ 10$^{10}$    & a$_{1}=$ 1.5 kpc                            & a$_{2}=$ 5.2 kpc                             &   165 $\pm$ 13 &  (9.7 $\pm$ 0.9) $\times$ 10$^{10}$     \\   
              &                            & b$_{1}$/a$_{1} =$ 1.00                      & b$_{2}$/a$_{2} =$ 0.19                       &                &                                  \\  
              &                            & M$_{1}=$ 7.0 $\times$ 10$^{9}$ M$_{\odot}$  & M$_{2}=$ 9.0 $\times$ 10$^{10}$ M$_{\odot}$  &                &                                  \\
\hline
8335-12701    & 1.25 $\times$ 10$^{11}$    & a$_{1}=$ 3.5 kpc                            & a$_{2}=$ 6.0 kpc                             &   218 $\pm$ 12 &  (2.2 $\pm$ 0.1) $\times$ 10$^{11}$     \\   
              &                            & b$_{1}$/a$_{1} =$ 1.00                      & b$_{2}$/a$_{2} =$ 0.33                       &                &                                  \\  
              &                            & M$_{1}=$ 9.0 $\times$ 10$^{9}$ M$_{\odot}$  & M$_{2}=$ 2.2 $\times$ 10$^{11}$ M$_{\odot}$  &                &                                  \\
\hline
8439-6102     & 1.06 $\times$ 10$^{11}$    & a$_{1}=$ 2.7 kpc                            & a$_{2}=$ 4.0 kpc                             &   202 $\pm$ 6  &  (1.40 $\pm$ 0.04) $\times$ 10$^{11}$    \\   
              &                            & b$_{1}$/a$_{1} =$ 1.00                      & b$_{2}$/a$_{2} =$ 0.42                       &                &                                  \\  
              &                            & M$_{1}=$ 1.0 $\times$ 10$^{10}$ M$_{\odot}$ & M$_{2}=$ 1.3 $\times$ 10$^{11}$ M$_{\odot}$  &                &                                  \\
\hline
8439-12702    & 6.05 $\times$ 10$^{10}$    & a$_{1}=$ 2.5 kpc                            & a$_{2}=$ 4.0 kpc                             &   235 $\pm$ 7  &  (1.60 $\pm$ 0.05) $\times$ 10$^{11}$    \\   
              &                            & b$_{1}$/a$_{1} =$ 1.00                      & b$_{2}$/a$_{2} =$ 0.25                       &                &                                  \\  
              &                            & M$_{1}=$ 6.0 $\times$ 10$^{9}$ M$_{\odot}$  & M$_{2}=$ 1.6 $\times$ 10$^{11}$ M$_{\odot}$  &                &                                  \\
\hline
8440-12704    & 5.17 $\times$ 10$^{10}$    & a$_{1}=$ 2.0 kpc                            & a$_{2}=$ 2.8 kpc                             &   205 $\pm$ 6  &  (9.7 $\pm$ 0.3) $\times$ 10$^{10}$    \\   
              &                            & b$_{1}$/a$_{1} =$ 1.00                      & b$_{2}$/a$_{2} =$ 0.35                       &                &                                  \\  
              &                            & M$_{1}=$ 7.0 $\times$ 10$^{9}$ M$_{\odot}$  & M$_{2}=$ 9.0 $\times$ 10$^{10}$ M$_{\odot}$  &                &                                  \\
\hline
8447-6101     & 3.66 $\times$ 10$^{11}$    & a$_{1}=$ 3.0 kpc                            & a$_{2}=$ 7.2 kpc                             &   312 $\pm$ 14 &  (5.7 $\pm$ 0.3) $\times$ 10$^{11}$     \\   
              &                            & b$_{1}$/a$_{1} =$ 1.00                      & b$_{2}$/a$_{2} =$ 0.39                       &                &                                  \\  
              &                            & M$_{1}=$ 9.0 $\times$ 10$^{9}$ M$_{\odot}$  & M$_{2}=$ 5.7 $\times$ 10$^{11}$ M$_{\odot}$  &                &                                  \\
\hline

8452-3704     & 1.84 $\times$ 10$^{10}$    & a$_{1}=$ 0.6 kpc                            & a$_{2}=$ 3.0 kpc                             &   142 $\pm$ 4  &  (4.9 $\pm$ 0.2) $\times$ 10$^{10}$     \\   
              &                            & b$_{1}$/a$_{1} =$ 1.00                      & b$_{2}$/a$_{2} =$ 0.33                       &                &                                  \\  
              &                            & M$_{1}=$ 1.0 $\times$ 10$^{9}$ M$_{\odot}$  & M$_{2}=$ 4.8 $\times$ 10$^{10}$ M$_{\odot}$  &                &                                  \\
\hline

\end{tabular}
 \end{table*}

\begin{table*}
 \contcaption{}
 \label{tab:continued}
 \begin{tabular}{l c c c c c}
  \hline

8452-12703    & 2.25 $\times$ 10$^{11}$    & a$_{1}=$ 3.5 kpc                            & a$_{2}=$ 6.1 kpc                             &   216 $\pm$ 11 &  (2.4 $\pm$ 0.1) $\times$ 10$^{11}$    \\   
              &                            & b$_{1}$/a$_{1} =$ 1.00                      & b$_{2}$/a$_{2} =$ 0.44                       &                &                                  \\  
              &                            & M$_{1}=$ 5.0 $\times$ 10$^{9}$ M$_{\odot}$  & M$_{2}=$ 2.4 $\times$ 10$^{11}$ M$_{\odot}$  &                &                                  \\
\hline
8482-9102     & 9.51 $\times$ 10$^{10}$    & a$_{1}=$ 4.0 kpc                            & a$_{2}=$ 5.5 kpc                             &   208 $\pm$ 15 &  (2.1 $\pm$ 0.2) $\times$ 10$^{11}$     \\   
              &                            & b$_{1}$/a$_{1} =$ 1.00                      & b$_{2}$/a$_{2} =$ 0.45                       &                &                                  \\  
              &                            & M$_{1}=$ 1.0 $\times$ 10$^{10}$ M$_{\odot}$ & M$_{2}=$ 2.0 $\times$ 10$^{11}$ M$_{\odot}$  &                &                                  \\
\hline
8482-12705    & 1.26 $\times$ 10$^{11}$    & a$_{1}=$ 1.0 kpc                            & a$_{2}=$ 5.0 kpc                             &   227 $\pm$ 7  &  (2.20 $\pm$ 0.05) $\times$ 10$^{11}$   \\   
              &                            & b$_{1}$/a$_{1} =$ 1.00                      & b$_{2}$/a$_{2} =$ 0.42                       &                &                                  \\  
              &                            & M$_{1}=$ 1.0 $\times$ 10$^{9}$ M$_{\odot}$  & M$_{2}=$ 2.2 $\times$ 10$^{11}$ M$_{\odot}$  &                &                                  \\
\hline
8486-6101     & 1.12 $\times$ 10$^{11}$    & a$_{1}=$ 2.5 kpc    & a$_{2}=$ 5.0 kpc                                                     &   189 $\pm$ 11 &  (1.40 $\pm$ 0.07) $\times$ 10$^{11}$     \\   
              &                            & b$_{1}$/a$_{1} =$ 1.00                      & b$_{2}$/a$_{2} =$ 0.30                       &                &                                  \\  
              &                            & M$_{1}=$ 1.0 $\times$ 10$^{9}$ M$_{\odot}$  & M$_{2}=$ 1.4 $\times$ 10$^{11}$ M$_{\odot}$  &                &                                  \\
              
\hline
8548-6102     & 2.69 $\times$ 10$^{10}$    & a$_{1}=$ 1.0 kpc                            & a$_{2}=$ 4.4 kpc                             &   155 $\pm$ 9  &  (9.6 $\pm$ 0.5) $\times$ 10$^{10}$     \\   
              &                            & b$_{1}$/a$_{1} =$ 1.00                      & b$_{2}$/a$_{2} =$ 0.54                       &                &                                  \\  
              &                            & M$_{1}=$ 6.0 $\times$ 10$^{9}$ M$_{\odot}$  & M$_{2}=$ 9.0 $\times$ 10$^{10}$ M$_{\odot}$  &                &                                  \\
\hline
8548-6104     & 6.04 $\times$ 10$^{10}$    & a$_{1}=$ 1.0 kpc                            & a$_{2}=$ 3.0 kpc                             &   136 $\pm$ 6  &  (4.10 $\pm$ 0.15) $\times$ 10$^{10}$     \\   
              &                            & b$_{1}$/a$_{1} =$ 1.00                      & b$_{2}$/a$_{2} =$ 0.33                       &                &                                  \\  
              &                            & M$_{1}=$ 6.0 $\times$ 10$^{9}$ M$_{\odot}$  & M$_{2}=$ 3.5 $\times$ 10$^{10}$ M$_{\odot}$  &                &                                  \\
\hline
8549-12702    & 1.61 $\times$ 10$^{11}$    & a$_{1}=$ 2.6 kpc                            & a$_{2}=$ 4.5 kpc                             &   249 $\pm$ 7  &  (2.20 $\pm$ 0.07) $\times$ 10$^{11}$     \\   
              &                            & b$_{1}$/a$_{1} =$ 1.00                      & b$_{2}$/a$_{2} =$ 0.36                       &                &                                  \\  
              &                            & M$_{1}=$ 6.0 $\times$ 10$^{10}$ M$_{\odot}$ & M$_{2}=$ 1.6 $\times$ 10$^{11}$ M$_{\odot}$  &                &                                  \\
\hline
8588-3701     & 2.86 $\times$ 10$^{11}$    & a$_{1}=$ 2.1 kpc                            & a$_{2}=$ 12.9 kpc                            &   303 $\pm$ 9  &  (1.10 $\pm$ 0.04) $\times$ 10$^{12}$     \\ 
              &                            & b$_{1}$/a$_{1} =$ 1.00                      & b$_{2}$/a$_{2} =$ 0.53                       &                &                                  \\  
              &                            & M$_{1}=$ 1.4 $\times$ 10$^{9}$ M$_{\odot}$  & M$_{2}=$ 1.1 $\times$ 10$^{12}$ M$_{\odot}$  &                &                                  \\
\hline
8604-12703    & 1.10 $\times$ 10$^{11}$    & a$_{1}=$ 3.0 kpc                            & a$_{2}=$ 3.6 kpc                             &   220 $\pm$ 9  &  (1.50 $\pm$ 0.06) $\times$ 10$^{11}$    \\   
              &                            & b$_{1}$/a$_{1} =$ 1.00                      & b$_{2}$/a$_{2} =$ 0.24                       &                &                                  \\  
              &                            & M$_{1}=$ 3.4 $\times$ 10$^{9}$ M$_{\odot}$  & M$_{2}=$ 1.5 $\times$ 10$^{11}$ M$_{\odot}$  &                &                                  \\
\hline
8612-6104     & 7.90 $\times$ 10$^{10}$    & a$_{1}=$ 2.0 kpc                            & a$_{2}=$ 3.2 kpc                             &   190 $\pm$ 8  &  (9.6 $\pm$ 0.4) $\times$ 10$^{10}$    \\   
              &                            & b$_{1}$/a$_{1} =$ 1.00                      & b$_{2}$/a$_{2} =$ 0.37                       &                &                                  \\  
              &                            & M$_{1}=$ 9.0 $\times$ 10$^{9}$ M$_{\odot}$  & M$_{2}=$ 8.7 $\times$ 10$^{10}$ M$_{\odot}$  &                &                                  \\
\hline
8612-12702    & 3.06 $\times$ 10$^{11}$    & a$_{1}=$ 2.0 kpc                            & a$_{2}=$ 7.0  kpc                            &   214 $\pm$ 11 &  (2.9 $\pm$ 0.2) $\times$ 10$^{11}$    \\   
              &                            & b$_{1}$/a$_{1} =$ 1.00                      & b$_{2}$/a$_{2} =$ 0.50                       &                &                                  \\  
              &                            & M$_{1}=$ 1.0 $\times$ 10$^{9}$ M$_{\odot}$  & M$_{2}=$ 2.9 $\times$ 10$^{11}$ M$_{\odot}$  &                &                                  \\
\hline
\end{tabular}
 \end{table*}


\bsp	
\label{lastpage}
\end{document}